# A Superdimensional Dual-covariant Field Theory

Yaroslav Derbenev
derbenev@jlab.org

Abstract

An approach to *Unified Field Theory* (UFT) is developed as an attempt to establish unification of the Theory of Quantum Fields (QFT) and General Theory of Relativity (GTR) on the background of a *covariant differential calculus*. A notion of a $\mu$-component *matter function* (MF) in a $N$-dimensional *unified manifold* (UM) is introduced based on the *homorphism* principle. In the context of the interpretation, the "extra-dimensions" of UM might be associated with the non-boson fields as the autonomic degrees of freedom of matter. MF can be interpreted as subordinated to a fundamental *differential law* (DL), subject to find out. A dual object consisting of covariant and contravariant $N$-component functions (*dual state vector*, DSV, an extended analog of the *state vector* of QFT), is introduced based on the MF derivatives. It represents *matter* in DL and plays a primary role in the theory based on the *irreducibility* principle. DSV is supposed to transform in a way distinct from that of the differentials of the UM variables though these transformations are supposed to be connected depending on the dynamic law of DSV. Consequently, the notions of *hybrid tensors* and a *hybrid affine tensor* (*unified gauge field,* UGF, an extended analog of both, *Christoffel's symbols* of differential geometry and *gauge fields* of QFT), are introduced. Transformation laws of the hybrid objects include transformations in both MF and UM spaces in product. The *hybrid curvature form* (HCF) is introduced as *covariant derivatives of UGF*. Based on the *extreme action* principle, a system of covariant Euler-Lagrange (EL) equations for DSV, UGF and *triadic* hybrid tensor (*Split Metric matrices*, SM, an extended analog of Dirac's matrices) is derived. A scalar Lagrangian form is composed based on a set of principles suited for UFT in the author's sight, including the *homogeneity* in the UM space, *differential irreducibility,* existence of a conservative *supercurrent*, *scale invariance* and *mini-max* principle. *Grand Metric* tensor (GM) built on binary bundles of SM is introduced for invariant integration of the scalar forms. Scalar Lagrangian consists of a *matter* part and a *geometry* part. The *matter scalar* is structured as a binary form on DSV and its covariant derivatives, using SM. The *geometry scalar* is structured as a second power bundle of the non-contracted HCF using GM tensor. No fundamental constants are introduced. The type of the manifold geometry still unspecified in neither local (signature) nor regional (topology) aspects. Equations for DSV play role of the Schrödinger-Dirac equation in space of UM. By the correspondent EL equations, UGF and SM are connected to DSV and become responsible for the non-linear features of the system i.e. *interactions*. Equations on SM (*metric equations*) together with equations on DSV allow one to directly express SM as function of DSV, UGF and GM. In turn, this relation leads to algebraic equations of the *forth* power on GM as function of DSV and UGF. In this paper we mark breaking of a background paradigm of QFT, the *superposition principle*. The issue of the UM−MF dimensionality will be addressed, and relations to the principles and methodology of QFT and GTR will be discussed.

# Contents

# Preface

This publication presents the motivations and results of the author's studies through a number of years in the area of the foundations and trends of the modern field theory – towards the *unified field theory* (UFT). According to the author's sight (which does not pretend to be exclusively original) on the modern fundamental theory of the elementary particles and their interactions and transformations, it is a differential theory of an object (*state vector*) representing *matter*; so should be the UFT, in general.

There is a general consensus in the high energy physics community that the unified theory should include and explain the phenomenon of *gravitation*. A theory of this class is titled sometimes *theory of everything* (ToE). We prefer use term *UFT*, meaning that the property of *gravitation* should arrive as an intrinsic and imprescriptible consequence of a universal covariant theory of the micro-world.

Reading of the paper does not require a possession of the mathematical techniques beyond the basic methods of the differential calculus, equations, geometry, and tensor calculus. A comprehensive knowledge of the elementary particle physics and modern field theory also is not required for reading and understanding of the texts; but, of course, the professionals in the area have an obvious advantage of a conscious and profound evaluation of the content.

Introduction (*Prolegomena*) to the paper is supplied with a *Synopsis* for a quick reviewing of the contents. A consecutive exposition of the principles and derivations of the edited SFT concept is presented in Chapters 3 through 9. Sections 5.5 and 5.6 in Chapter 5 treat in general terms application of the variation principle and the related background properties of a covariant field theory. Reading the *Prolegomena* is supposed to preceding the reading of the *Synopsis* as well as the whole the text. A comparative discussion of the derived equations is conducted in the conclusive Chapter 10.

# 1. Introduction and Prolegomena

## 1.1. Introduction

*Quantum Field Theory* (QFT) [1-4] is developed and practiced as a theory of the microscopic structure of *matter*, in which matter is represented by a variety of *fields* in a 4-dimensional space-time manifold (STM) with *Minkowski metric* of *Special Theory of Relativity*. In QFT, the space-time geometry is given and is not influenced by matter. It is worth noting, however, that the *quantum fields* are not the same as fields in a rigorous mathematical definition, which are the "classical fields" in terminology of modern theoretical physics. While in *classical field theory* (CFT) the fields are represented in STM by continuous analytical functions $\psi(\vec{r}, t)$, in QFT the field function and its gradients at a point $(\vec{r}, t)$ do not co-exist, just like in the *quantum mechanics* (QM), a theory established by W. Heisenberg, E. Schrödinger and P. Dirac [5,6], the particle coordinate $\vec{r}$ and its derivative with respect to time, $\dot{\vec{r}}$, do not co-exist. In fact, functions $\psi(\vec{r}, t)$ are treat and utilized in the QFT as *operators*, which are applied to affect the *secondary quantization* "wave function" or *State Vector* ϒ (SV). According to the two basic methodological paradigms of QM, the *superposition* and *correspondence* principles, this object is supposed to be subordinate to *Schrödinger equation* with *energy operator* ("quantum Hamiltonian") figured out in a procedure of the so-called *quantization* applied to *energy form* of a relativistic CFT. This form is defined as integral over space of the "time-time" component $T_{00}$ of *energy-momentum tensor* $T_{kl}$ determined based on a suitable *Lagrangian density* form of the *Extreme Action Principle* of CFT [1, 2].

Reviewing the present state of QFT, one recognizes a variety of fields, "particles", group properties, symmetries, interaction models, methods of regularization of diverging integrals and more. There are also great guiding principles and paradigms like *renormalization* of charges and masses as a way to overcome the divergences, the *gauge principle* as a way of structuring the renormalizable QFT, *asymptotic freedom*, *continual integration* and other advances. Based on all these principles and methods, the *Standard Model* had been established in field theory in the -50s through -70s of the previous century. It combines in one dynamic system the *electromagnetic*, *weak* and *strong* interactions and describes with a great effectiveness all known particles and their transformations (generation in collisions, decays) in the TeV energy range [7]. "The interactions of Standard Model…explain virtually all the particles and interactions which have been observed in accelerators. Yet the underlying laws can be summarized in a few lines. … As a theoretical structure, it also explains successfully what might be viewed as mysterious conservation laws: baryon and separate lepton numbers" [10].

But, there still exist in QFT open questions and problems of a critical meaning. Standard Model is not free of internal problems basically due to the large constants of strong interaction, which makes it difficult to calculate or explain some critical processes and phenomena like *confinement*, *hierarchies* and other. After all, mastering the Standard Model certainly does not look as corresponding to sculpturing a *unified theory*. "... it has seventeen numbers – sixteen of which are pure numbers with values which range "all over the map". ...it fails to account for some of the most basic phenomena of the universe: dark matter, dark energy, and the existence of gravity itself " [10].

Beyond the Standard Model, there are a few trends of exploration towards the unification. Some of them are limited to the purpose of reducing the number of parameters by reorganizing particular group terms of Standard Model (*Grand Unification*). The *Supersymmetry* theories explore possibilities of unification by extending group representations of QFT [1]. Other theories like the *string models* investigate a path to a Unified Field Theory (UFT) by introducing *extra-dimensions* (the way originated by T.Kalusa and O.Klein, see [8]) and implementing the *compactification* ideas [11].

Unification of QFT with relativistic theory of gravitation (*Theory of Everything*, ToE) is the most challenging paradigm of the modern theoretical physics. *General Theory of Relativity* (GTR) [12-16], created by A. Einstein on the basis of Special Theory of Relativity (STR) and *equivalence principle*, is interpreting gravity as an effect of the "non-flatness of the space-time geometry" caused by the *macroscopic* material objects. A traditional approach to unification consists of introduction of the gravitational field to the "quantum club" after its "quantization", though quantization of such essentially non-linear classical theory as Relativistic Theory of Gravity (RTG) is a difficult problem itself, if it is solvable at all. By the way, there was discovered a fundamental obstacle to unification: RTG is recognized as a non-renormalizable field theory [17]. Equally, attempts to establish *quantization* of other fields as a regular procedure in a *curved-geometry* space-time manifold have not been successful [18]. On the other hand, such a powerful method of introducing interaction in QFT as the *gauge principle* seems to correspond to the *covariance principle of GTR* [1]. Could this similarity be viewed as an indication of existence of a more general gauge principle as a *universal covariance*?

In our view, there are other questions that should be addressed to the QFT developments in the context of efforts on its unification with GTR. First, what is unclear in QFT is the meaning of the space-time variables $(\vec{r}, t) \equiv \hat{x}$ in the subatomic range. Are they subject to measure – then, by what means? In GTR of Einstein, the space-time intervals are meant as measurable by use of "little" scales and clocks; the fundamental microscopic structure of these instruments is not a subject of GTR. But QFT is specifically a theory of the microstructure of matter and transformations of its "elementary constituents". If so, then what makes the space-time variables so special that they are utilized in the microscopic theory in the same manner as in the macroscopic one?

Other questions connected to the previous one: does it make sense to try to derive a *unified differential law* for *State Vector* as an *irreducible* system of differential equations for SV and related *coefficient functions* (CF)? If nothing prohibits this, then what would be the *geometrical nature* of SV and CFs, in what variables could such equations be formulated, where would there be a place for the space-time variables and what meaning would they have in the structure of the theory?

We would like to note in the context of the reviewing the principles of structuring and mathematical foundations of QFT that it stands as an essentially *differential theory* for *State Vector* object as function of more than 4 variables in general, a theory built in specific methodological way to satisfy the *correspondence* and *superposition* principles established by the founders of the quantum mechanics. At the same time, quite a friable structure of Standard Model certainly does not look close in style to a *unified system* to address the *irreducibility* requirement. In the frame of QFT foundations, there is no principle that would prevent one from introduction of unlimited realm of new items (objects, interactions, symmetry) in order to describe the newly observed phenomena. In our view, when going this path, the fundamental theory of the micro-world has no perspective to attain features of a *self-contained* field theory - as the *unified theory* should be supposed to be so. Approaches to UFT other than those based on the canonical receipts of *quantization* of suitable CFTs and *superposition* principles should be investigated in parallel.

In this paper we present our exploration of the above pointed questions and preliminary results. In Chapters 2 through 9 we disclose the motivations, principles and resulting equations of the *superdimensional dual-covariant* differentia approach to UFT. In our sight, the unified theory should be subordinate to a set of the *irreducibility* principles. The principles should be established based on the logical arguments rather than on the heuristic or esthetic ones. On the other hand, the logic of structuring the unified system cannot be produced "from nothing" but should be navigated by observation of the genetic history of principles and mathematical background of the fundamental theoretical physics. Studies in this work on approaching possible path to unified theory were inspired by the *quantum legacy* of P. Dirac, the *covariance* paradigm of A. Einstein, the *gauge theory* of QFT, and the *irreducibility* demand of W. Pauli, though the meaning and structure of the fundamental objects and equations of the presented approach to UFT distinguish significantly from those of QFT and GTR.

A comparative discussion of the derived equations and open questions is conducted in the conclusive Chapters 10 and 11.

### 1.2. Prolegomena

**Preamble** Notion of *field* is the central one in the modern fundamental theoretical physics. Mathematical theory of atoms, nuclei and "elementary particles" is based on this notion; yet in the methodological aspects the theory of elementary particles is mounted in two levels, the "classical field theory" and the *quantum* one. Here one can observe an interference and sometimes confusing mixture of physical perceptions and mathematical definitions, physical ideas about the elementary objects and processes and mathematical methods of description of the objects and their transformations [1, 2].

Review of the genesis of *quantum theory of fields* (QFT) and explication of essence of QFT as a differential theory for *State Vector* (SV) field as a continuous function of variables of an $N > 4$ dimensions space of variables, together with realization of the constraints and problems of *unification*, prompt a concept of approach to *unified field theory* (UFT) as a *superdimensional* dual-covariant field theory (SFT) which schematically is a kindred of the Maxwell – Dirac electrodynamics as classical field theory but transplanted to a superdimensional space of variables, a *unified manifold* (UM) of a dimensionality $N > 4$, generalized following to the *covariance* requirement and unified under the press of the *irreducibility* demands. In whole, the following insights and arguments have being guiding the approach to UFT under consideration in this paper.

*1. SFT as a self-contained field theory*
The SFT is profiled as a *self-contained* differential theory for State Vector field (in the further texts denoted $\Xi$), subordinate of a *differential law* (DL) as function of $N$ variables of a *unified manifold* (UM). This definition, in accordance to the SV status, implies that, in such theory derivatives of SV on UM variables are connected to SV itself – via *coefficient functions* (CFs) as objects of the theory which are not given (proposed) in advance as explicit functions of UM variables but connected to SV by the related differential equations. Note that, Maxwell-Dirac electrodynamics as a field theory of the *first level* (i.e. a "classical" field theory) is a theory of this type. Let us use notation $\check{\psi}$ to denote a *point* in UM given by $N$ "coordinates" $\check{\psi}^k$:

$$\check{\psi} \equiv \{\check{\psi}^k\}; \quad k = 1, 2, \ldots, N . \tag{1.1}$$

In these notations, the transition from theory of classical fields $\hat{\psi}(\hat{x})$, $\hat{x} = \vec{r}, t$ to a unified theory can be symbolized in the following manner:

$$\hat{\psi}(\hat{x}) \rightarrow \Xi(\check{\psi}); \quad \check{\psi} \equiv (\hat{\psi}, \hat{x}) \tag{1.2}$$

$$\frac{\partial \hat{\psi}(\hat{x})}{\partial \hat{x}} \to \frac{\partial}{\partial \check{\psi}} \Xi(\check{\psi}). \qquad (1.3)$$

In the context of the relation to space of variables of QFT, UM of SFT is envisioned as space of degrees of freedom corresponding (but not being an identity) to the *fermion* fields of QFT, according to the point of view of the author that fermion's degrees of freedom should play a pilot role in a fundamental irreducible theory of the micro-world. In more direct comparison, fermion features of the theory should be addressed to transformation properties of the State Vector field as the global pilot object of SFT. *Boson* objects (fields) can be envisioned to be profiled based on binary combinations of the fermion type of objects. In principle, transformation properties of the *observable objects* should result from the UFT dynamics. In our sight, a background soil for appearance of bosons should be associated with such coefficient functions of equation for SV as *gauge objects* $\check{\mathcal{A}}(\check{\psi})$ introduced for covariant extension of the SV derivatives[1]:

$$\frac{\partial}{\partial \check{\psi}} \Xi \;\to\; \left(\frac{\partial}{\partial \check{\psi}} + \check{\mathcal{A}}\right) \Xi. \qquad (1.4)$$

It may seem at the first glance that SFT cannot lead to *quantum* properties of field dynamics in projection to the 4-dimensional space-time manifold, unless we incorporate postulates of "quantization" in the concept. Remind in this connection that, the non-relativistic wave mechanics of Schrödinger with "non-quantized" wave function leads immediately to quantization of atomic energy levels and uncertainty relations concerning the transition to classical mechanics, as well as the relativistic theory of Dirac immediately explains electron spin and introduces a concept of creation-annihilation of particles before the "secondary quantization". After all, QFT (initiated by P. Dirac) as a mathematical system in essence is a differential theory for *state vector* (SV) of a dynamical system as function of a conglomerate of variables (space-time and variety of *fields* as free variables associated with "elementary particles"), subordinate to a *Schrödinger equation* with *energy operator* ("quantum Hamiltonian") of a certain structure which includes differentials over all variables. In this context, the presented treat is in general correspondence to the QFT establishment.

Description of *matter* in terms of the State Vector field in approach to UFT as a self-contained field theory in space of Unified Manifold might be implied corresponding to quest for a *Universal Wave Function* by S. Hawking [19].

## *2. Unified Manifold as space of free numbers*
C*oordinates* i.e. *variables* of the Unified Manifold of SFT should be regarded as *free numbers* varying continuously. They cannot be referred to "material bodies", "classical objects", etc. Such references are not compatible with the sense of UFT as a fundamental, *irreducible* field theory.

## *3. The dynamical genesis of the physical geometry*
No specific geometrical characterization of UM space should be posed in advance. Definitions of *distance* or *interval* should not be introduced in advance, as well. Geometrical characteristics (metric signature, topology of UM, group properties of SFT objects, etc.) can only be profiled based on the established structure and *solutions* of the SFT differential system.

---

[1] Standard Model of QFT includes *Higgs boson* as a non-gauge field serving for creation of particles' masses. Also, the supersymmetry theories suggest unification of fermions and bosons (including *gravitons*) in one group of independent objects. In the context of the developments toward UFT, we prefer to keep a point of view expressed by the above comment.

## *4. The Unified Manifold− Matter Function homomorphism*

To be irreducible, the fundamental differential law should be associated with a procedure of a background level produced on the UM space, which could be a special *homomorphism* of $N$ variables $\check{\psi}^k$ i.e. existence of $\mu$ functions $\varphi^\alpha(\check{\psi})$ direct by a *differential law* (DL), subject to find out:

$$\check{\psi}^k \rightarrow \varphi^\alpha(\check{\psi}); \quad k = 1,2, \dots, N; \quad \alpha = 1,2, \dots, \mu. \tag{1.5}$$

We will call this homomorphism *matter function* (MF). DL should be associated with derivatives of MF that define connection between the differentials:

$$d\varphi^\alpha = \frac{\partial \varphi^\alpha}{\partial \check{\psi}^k} d\check{\psi}^k ; \tag{1.6}$$

in a symbolic form:

$$d\boldsymbol{\varphi} = F d\check{\boldsymbol{\psi}}; \qquad F \equiv F_k^\alpha \equiv \frac{\partial \varphi^\alpha}{\partial \check{\psi}^k}. \tag{1.7}$$

We will talk about system of derivatives $F_k^\alpha$ as *matrix*, meaning that, generally, it is not quadratic. Note that, the *inverse connection* cannot be formulated similar to treatment in [1[*]], once $\mu > N$. On the other hand, there is still a room for establishing a mutual *isomorphic* correspondence between possible *affine structures* which could be built in two spaces; investigation in this direction can be envisioned as the next step in specifying the SFT concept. The related exploration is lying beyond the scope of this paper.

## *5. Contravariant State Vector as field of directions in MF space*

Object of DL, *state vector* field (SV) to be irreducible representative of *matter*, is supposed to be associated with differentials of MF. An autonomic differential law, however, cannot be derived in terms of functions $\varphi^\alpha$, but it could be derived for an object *collinear* with *differentials* of MF $d\varphi^\alpha$ been regarded as an object of a *finite* magnitude, system of $\mu$ variable numbers (a *column* in matrix terms), denote it $\Psi^\alpha$:

$$\Psi^\alpha \propto d\varphi^\alpha. \tag{1.8}$$

Object $\Psi^\alpha$ can be associated with *tangent vectors* of "world lines" $\varphi^\alpha(\tau)$ in MF space connected to world lines $\check{\psi}^k(\tau)$ in UM space ($\tau$ is a *canonical parameter* of a line [20]):

$$\Psi^\alpha = \frac{d\varphi^\alpha}{d\tau} = \frac{\partial \varphi^\alpha}{\partial \check{\psi}^k} \frac{d\check{\psi}^k}{d\tau} = F_k^\alpha q^k; \quad q^k = \frac{d\check{\psi}^k}{d\tau}; \tag{1.9}$$

in a symbolic manner:

$$\boldsymbol{\Psi} = F \mathbf{q}. \tag{1.10}$$

Tangent vectors of the "world lines" in UM space are transformed as contravariant vectors:

$$\mathfrak{q} \rightarrow \mathfrak{q}' = A\mathfrak{q} ; \tag{1.11}$$

then, according to (1.13) and (1.10), field $\Psi$ is transformed also as a *contravariant vector* but with a matrix $B$ in MF space:

$$\Psi \rightarrow \Psi' = B\Psi. \tag{1.12}$$

Geometrical nature of *state vector* field $\Psi^\alpha$ with respect to the unified manifold i.e. its transformation at transformations of UM variables (in other words, connection $B(A)$) should be established based on possible *inquired invariance* properties of a *covariant differential law* for SV a subject to find out.

## 6. Affine duality of State Vector field
In parallel with field $\Psi$, one can consider a $\mu$-components vector field independent of $\Psi$ but associated with the *inverse* transformation in MF space; denote it $\Phi_\alpha$ or $\Phi$ (*covariant state vector* field, CSV):

$$\Phi' = \Phi B^{-1}; \tag{1.13}$$

If SV can be represented in matrix terms by a column of numbers (functions), then $\Phi$ is represented by a row of numbers. One also may consider an *inverse homomorphism* between covariant vectors in MF and UM space as follows:

$$p_k = F_k^\alpha \Phi_\alpha . \tag{1.14}$$

## 7. *State norm*.
Scalar product of the two introduced vector objects is *invariant* of transformations:

$$\mathbb{N} \equiv (\Phi\Psi) \equiv \Phi_\alpha \Psi^\alpha = \mathbb{N}' = invariant ; \tag{1.15}$$

the identical invariance takes place in the UM space, since:

$$\Phi_\alpha \Psi^\alpha = \Phi_\alpha F_k^\alpha \mathfrak{q}^k = p_k \mathfrak{q}^k. \tag{1.16}$$

We will consider not just one vector field but *dual couple* of *real* vector fields as a *master object* of a *dual-covariant field theory*:

$$\Xi \Longrightarrow (\Psi^\alpha, \Phi_\alpha). \tag{1.17}$$

We call the introduced duality the *affine duality*, and association of these two vector fields the *Dual State Vector* field (DSV). Note that, this duality distinguishes essentially from the conventional *metric duality* usually obtained by lifting up or down of indices of the objects applying *metric tensor* like in GTR: two the above introduced vector fields are considered as two systems of $\mu$ numbers (functions of the UM variables) *independent* of each other. Also note that, field $\Phi_\alpha$ is not associated with gradients of a *scalar field* but is of a more general nature. After all, in our view the scalar objects of irreducible theory should not be introduced to the theory (i.e. postulated) to play role of the basic objects but could only be *composed* as invariant forms based on use of the *affine duality* and (or) *metric tensor* of the theory. Our *affine duality* also distinguishes from the *complex numbers* duality of QFT but could include the last one as a fragment of a more general *matrix structure*.

Introduction of a *covariant state vector* field **Φ** as dual but independent partner to *contravariant state vector* field **Ψ** is a start point for building up the SFT *dynamics*.

## *8. Duality of SV as a presage of UM – MF algebraic structural isomorphism*

Duality of SV can be viewed as a prerequisite for building up the structural forms (i.e. *tensors*) in MF space based on DSV and its derivatives as functions of UM variables. This sight leads to consideration of possible *structural* or *algebraic UM – MF isomorphism*. Investigation of this perspective, however, goes beyond the scope of this paper.

## *9. The homogeneity principle*

Differential system of SFT is considered *homogeneous* in space of UM. This implies that, **t**he differential system should be formulated only as relations between involved basic objects $X^a$ and their derivatives $\partial_k X^a$ and should not include any explicit i.e. *given* functions of UM variables. This requirement is one of those that make SFT a *self-contained* theory.

The sense of this principle consists in the following. Establishing the differential law as relations between SV and its derivatives takes introduction of the *coefficient functions*. When profiling these relations, one should use no assumptions about behavior of these objects as functions of the manifold variables, neither *ad hoc* or with references to "reality". Instead, coefficient functions should be connected to SV by the correspondent differential equations, as above mentioned.

The homogeneity principle may seem a "routine" one at first glance, since it is a basic declaration of QFT as a *quantum* field theory in the 4-dimensional space-time manifold. On the other hand and *in fact*, QFT is a differential field theory for SV as function of variables of $N > 4$ dimensions manifold. However, QFT does not follow the *homogeneity principle* when building up the dynamic law for the SV as the *secondary quantization* function: while considering SV *in fact* as function of fields $Q$, QFT as a mathematical system at the same time utilizes a representation in the style of *Schrödinger equation* in which *Hamiltonian* as *energy operator* is an *explicit given function* (form) of $Q$. This methodology cannot bring QFT to the class of the self-contained theories.

## *10. The uniformity principle*

Equations of SFT should be *uniform* (symmetric, *homogenized*) over all components of the involved objects and UM variables. All the known or expected newly arriving particular *fields* or *particles* should be envisioned to be profiled at possible fragmentation of asymptotic solutions of a *uniform* UFT.

## *11. General covariance principle*

Differential system of SFT should be *generally covariant* (i.e. invariant in *form* of relations between basic objects) relative to arbitrary transformations of UM variables. This property of SFT can be characterized as *extended general relativity* (EGR).

## *12. The dynamical genesis of the transformation properties*

Transformation properties (hence, *"geometrical nature"*) not only of DSV but all SFT objects are envisioned as determined by the SFT dynamic laws.

## *13. Differential irreducibility* (DI) *as principle of the dynamical existence*

Equations for DSV connect first derivative of this object to itself and not include higher order derivatives. Equations on the triadic objects (coefficient functions) are supposed to be formulated in the lowest order derivatives, in the correspondence to the all *irreducibility* demands.

DI can be considered as an expression of an ontological principle of the *dynamical existence*. Namely, DL of DSV is an autonomic system of $2\mu$ equations in first order derivatives for $2\mu$ functions of UM variables. DSV as an object subordinate to such a law does not have points of zero: if there would be one such point, then DSV would be zero everywhere. Search for the correspondence $N \rightarrow \mu$, in our sight, should be directed by principle of an *irreducible structural UM-MF isomorphism* in cooperation with all the posed *irreducibility* demands. Exploration of the dimensionality aspects, however, goes beyond the frame of this publication.

*14. Preview of master equations*

Like in QFT, differential equations for DSV are meant to have a form linear on DSV. The law should connect the *first derivatives* of DSV to DSV itself, in accordance with the *irreducibility* requirements. So it necessarily includes some multi-index *connection objects*, the *coefficient functions* (CF) matrices. Envision of irreducible DL as *covariant relations* between DSV and its derivatives leads to the following primer formulation of equations for DSV:

$$P_\beta^{\alpha k} \mathfrak{D}_k \Psi^\beta = \Psi^\alpha; \qquad \overline{P}_\alpha^{\beta k} \mathfrak{D}_k \Phi_\beta = \Phi_\alpha; \qquad (1.18)$$

here objects $\mathfrak{D}_k \Psi^\beta$ and $\mathfrak{D}_k \Phi_\beta$ are *covariant deri*vatives of DSV:

$$\mathfrak{D}_k \Psi^\beta \equiv \partial_k \Psi^\beta + \mathcal{A}_{\gamma k}^\beta \Psi^\gamma; \qquad \mathfrak{D}_k \Phi_\beta \equiv \partial_k \Phi_\beta - \mathcal{A}_{\beta k}^\gamma \Phi_\gamma. \qquad (1.19)$$

Each of the 3-indices *geometrical objects*: $P_\beta^{\alpha k}$; $\overline{P}_\alpha^{\beta k}$ and $\mathcal{A}_{\beta k}^\alpha$ is association of $N$ matrices (on Greek indices) of rank $\mu$. Introducing notations: $\mathcal{A}_{\beta k}^\alpha \equiv \mathcal{A}_k$; $P_\beta^{\alpha k} \equiv \mathbf{P}^k$; $\overline{P}_\alpha^{\beta k} \equiv \overline{\mathbf{P}}^k$, we can write equations (1.18) in the following symbolic view:

$$\mathbf{P}^k \cdot \mathfrak{D}_k \mathbf{\Psi} + \mathbf{\Psi} = 0; \qquad \mathfrak{D}_k \mathbf{\Phi} \cdot \overline{\mathbf{P}}^k + \mathbf{\Phi} = 0; \qquad (1.20)$$

$$\mathfrak{D}_k \mathbf{\Psi} \equiv (\partial_k + \mathcal{A}_k) \mathbf{\Psi}; \qquad \mathfrak{D}_k \mathbf{\Phi} \equiv \partial_k \mathbf{\Phi} - \mathbf{\Phi} \cdot \mathcal{A}_k; \qquad (1.21)$$

with the following transformation law for $\mathcal{A}$ (here omit Roman index as well):

$$\mathcal{A} \rightarrow \mathcal{A}' = A^{-1} B (\mathcal{A} + \partial) B^{-1}. \qquad (1.22)$$

Transformation law for matrices $\mathcal{A}$ is determined based on requirement of compensation for terms with derivatives of matrix $B$ which arrive in equations (1.20) at transformations of UM variables. Consequently, objects (1.21) transform similar to tensors but with two different matrices, $A$ and $B$, associated with transformations in the UM and MF spaces, respectively:

$$\mathfrak{D}' \mathbf{\Psi}' = A^{-1} B \mathfrak{D} \mathbf{\Psi}; \qquad \mathfrak{D}' \mathbf{\Phi}' = A^{-1} (\mathfrak{D} \mathbf{\Phi}) B^{-1}. \qquad (1.23)$$

Based on equations (1.20), this leads to transformation rule for triadic objects $\mathbf{P}$ and $\overline{\mathbf{P}}$ as follows:

$$\mathbf{P} \rightarrow \mathbf{P}' = AB\mathbf{P}B^{-1}; \qquad \overline{\mathbf{P}} \rightarrow \overline{\mathbf{P}}' = AB\overline{\mathbf{P}}B^{-1}. \qquad (1.24)$$

## 15. Dynamic connection between UM and MF transformations
At transformation of UM variables at a point with matrix $A$, MF differentials and DSV are transformed with some matrix $B$.

## 16. Requirement of the existence of a conservative supercurrent
Existence of a *conservative vector current* $\mathcal{J}^k$ in UM space (*supercurrent*) should be an intrinsic property of the derived equations for DSV as an attribute of the above mentioned principle of the *dynamical existence*. Conservative vector current associated with DSV can be presented in the following form:

$$\mathcal{J}^k = \Lambda^{\alpha k}_\beta \Phi_\alpha \Psi^\beta; \quad \nabla_k \mathcal{J}^k \equiv \frac{1}{\sqrt{w}} \partial_k(\sqrt{w}\mathcal{J}^k) = 0; \quad w = |det w_{kl}| \qquad (1.25)$$

(here $w_{kl}$ is metric tensor of UM, the *Grand Metric*) with undefined h-tensor $\Lambda^{\alpha k}_\beta \equiv \mathbf{\Lambda}^k$. Further, let us represent h-tensors $\mathbf{P}^k$ and $\bar{\mathbf{P}}^k$ in equations (1.25) in the following way, introducing an undefined s-tensor $\mathbf{\lambda} \equiv \lambda^\alpha_\beta$:

$$\mathbf{P}^k = (\mathbf{1} + \mathbf{\lambda})^{-1}\mathbf{\Lambda}^k; \qquad \bar{\mathbf{P}}^k = -\mathbf{\Lambda}^k(\mathbf{1} - \mathbf{\lambda})^{-1}. \qquad (1.26)$$

Drawing then the requirement of a *conservative* current (1.25) and using equations (1.26), we find solution for s-tensor $2\mathbf{\lambda}$ as *covariant divergence* of h-tensor $\mathbf{\Lambda}^k$:

$$2\mathbf{\lambda} \Rightarrow \mathcal{D}_k \mathbf{\Lambda}^k \equiv \frac{1}{\sqrt{w}} \partial_k(\sqrt{w}\mathbf{\Lambda}^k) + [\mathcal{A}_k, \mathbf{\Lambda}^k]; \qquad (1.27)$$

here symbol $[;]$ means commutator of two matrices. Thus, specification of DSV equations for existence of a conservative current (1.30) might result in the following form of these equations:

$$\mathbf{\Lambda}^k \mathcal{D}_k \mathbf{\Psi} + \left(\frac{1}{2}\mathcal{D}_k\mathbf{\Lambda}^k + \mathbf{1}\right)\mathbf{\Psi} = 0; \qquad (\mathcal{D}_k\mathbf{\Phi})\mathbf{\Lambda}^k + \mathbf{\Phi}\left(\frac{1}{2}\mathcal{D}_k\mathbf{\Lambda}^k - \mathbf{1}\right) = 0, \qquad (1.28)$$

with reduction of a couple of triadic h-tensors, matrices $\mathbf{P}^k$ and $\bar{\mathbf{P}}^k$ to a single h-tensor, $N$ matrices $\mathbf{\Lambda}^k$.

## 17. Constraint of CFs – DSV coupling
Due to the *homogeneity* principle, coefficient functions $\mathbf{\Lambda}^k$ and $\mathcal{A}_k$ in equations for DSV (1.28) cannot be *given* i.e. explicated in advance in their structure and as functions of UM variables. They also cannot be viewed as constant matrices, since such foundation would be contrary to the *general covariance* principle. Then, there is the only resolution of this constraint: CFs should be *connected to DSV* by other equations based on some fundamental principle of the differential calculus.

## 18. Abandoning the superposition principle
Once CFs of master equations (1.28) are connected to DSV, differential system of SFT arrives *non-linear in DSV*, thus abandoning basic postulate of QFT− the *superposition principle.*

## 19. Reality of the SFT objects as an attribute of General Covariance
*Differential law* (DL) as an analytical algorithm of UFT should be formulated in all *real* terms. The *imaginary unit "i"*, and *complex* basic objects or variables are not admitted.

This "puritanical" restriction is imposed due to a consideration that the presence of *invariable objects*, like "*i*", is not compatible with the requirement of *general covariance* that implies that all the involved objects should be variable in a *covariant way*. We presume that, the complex analytical structure of the existing "quantum theory" shall be recognized in frame of SFT as a particular asymptotical sector of a more general analytical structure of SFT represented in terms of such background objects as vectors and matrices of the MF space – all real.

## 20. *The hybrid objects, s-tensors, conventional tensors, covariant derivatives and scalar forms*

We call $N$ matrices $\mathcal{A}_k$ *unified gauge field* (UGF) in the context of a general external correspondence to *gauge fields* of QFT. In the context of the correspondence to objects of the conventional differential geometry, they can be characterized as *hybrid affine tensor* or *hybrid Christoffel symbols*. $N$ matrices $\Lambda^k$ can be regarded as a triadic *hybrid tensor*. Treat of a DSV-based field theory requires introduction of covariant derivatives of the hybrid objects including *covariant derivative of UGF* itself; the last one being an h-tensor is recognized as a *hybrid curvature form* (HCF), an extended analog of the Riemann-Christoffel curvature form of differential geometry. Conventional *tensors* (including *vectors*) as objects that are transformed only with matrix *A* can be structured on the introduced basic objects and their *covariant derivatives*. There also can be composed the multi-Greek index objects as transformed only with matrix *B*; we call such ones the *s-tensors*. *Scalar functions* as *invariants of transformations* of UM variables and MF objects all result *in dynamics* from *scalar forms* composed *in presuppositions* of the *Extreme Action* principle. We resort to this principle in order to derive connections of the introduced triadic objects to DSV, together with equations for DSV itself.

## 21. *Extreme Action as principle of a dynamic balance*

The CFs should be connected to DSV by resorting to the variation principle of the *Extreme Action* (EAP):

$$\delta \int \mathcal{L}(X, \partial X) d\Omega = 0; \quad d\Omega = d\psi^1 d\psi^2 \ldots d\psi^N, \tag{1.29}$$

posing, as usual, variations of basic objects $\delta X = 0$ at a (arbitrary) closed surface limiting volume of integration. Lagrangian form $\mathcal{L}$ is structured on basic objects $X$ and their derivatives $\partial X$ as product of *scalar Lagrangian* form $\mathbb{L}$ and *weigh factor* $\sqrt{w}$ :

$$\mathcal{L} = \mathbb{L}\sqrt{w}; \quad w = |det w_{kl}|; \tag{1.30}$$

where $w_{kl}$ is a symmetric non-degenerated tensor, so that $\sqrt{w} d\Omega$ is *invariant differential volume*. EAP generally results in *Euler-Lagrange* (EL) *equations* for system of *basic objects* $\{X^a\}$:

$$\partial_k \frac{\partial \mathcal{L}}{\partial(\partial_k X^a)} - \frac{\partial \mathcal{L}}{\partial X^a} = 0. \tag{1.31}$$

EAP is the unique methodological principle for deriving the fundamental equations of a field theory. It is one of the corner stones of the QFT methodology, though the way it is used therein – building the "quantum Hamiltonian" (energy operator) by a transition from Lagrangian of a "classical" ("non-quantized") field theory – look more like a mnemonic rule or postulated receipt rather than a logically conditioned principle. In approach to SFT as "classical" field theory in a superdimensional manifold, Euler-Lagrange equations (including, of course, EL equations for DSV itself) are immediately derived as a fundamental law of the theory with no resorting to further procedures as "quantization", etc. The superdimensional EAP is viewed as replacing the quantization paradigm of QFT; quantum behavior of the observable material objects could be interpreted as associated with projecting of a superdimensional field dynamics to the *intelligible* 4-dimensional space-time manifold (STM). Dimensionalities $N$ and $\mu$ of SFT are supposed to be determined in the frame of the theory itself as a minimum required for a self-consistent irreducible SFT. An associated "home task" of the theory should be explanation of special STM role as a realm for the intelligible world that is immediately grasped by the senses and apparatus.

## *22. Scale Invariance principle*
Principles of building up the unified theory should eliminate sensitivity of its dynamical properties to introduction of arbitrary real constants as multipliers at scalar items of Lagrangian. We call such property *scale invariance*, considering it as a feature necessary for a field theory to be a candidate in UFT. To be noted that, it can be realized in a logically consistent way only based on the EAP.

## *23. The mini-max principle*
To be in consistence with the irreducibility principles, Lagrangian of UFT should be subordinate to the *mini-max principle*: while under the restrictive press of the exhibited requirements, number of different scalar items in Lagrangian should be *maximum* at *minimum* collection of the basic objects**.**

# 2. Synopsis

## 2.1. Lagrangian

*Scalar Lagrangian and Grand Metric*

Under the press of the above listed *irreducibility* demands, scalar form $\mathbb{L}$ and tensor $w_{kl}$ are composed on basic objects in the following way:

$$\mathbb{L} = \mathbb{M} + \mathbb{G}; \quad \mathbb{M} = \mathbb{N} + \mathbb{D}; \tag{2.1.1}$$

$$\mathbb{N} \equiv \Phi_\alpha \Psi^\alpha; \quad \mathbb{D} \equiv Tr\mathbf{\Lambda}^k \boldsymbol{\mathfrak{D}}_k \equiv \Lambda_\alpha^{\beta k} \mathfrak{D}_{\beta k}^\alpha; \quad \mathfrak{D}_{\beta k}^\alpha \equiv \frac{1}{2}(\Phi_\beta \mathfrak{D}_k \Psi^\alpha - \Psi^\alpha \mathfrak{D}_k \Phi_\beta) \equiv \boldsymbol{\mathfrak{D}}_k; \tag{2.1.2}$$

$$\mathfrak{D}_k \Psi^\beta \equiv \partial_k \Psi^\beta + \mathcal{A}_{\gamma k}^\beta \Psi^\gamma; \quad \mathfrak{D}_k \Phi_\beta \equiv \partial_k \Phi_\beta - \mathcal{A}_{\beta k}^\gamma \Phi_\gamma; \tag{2.1.3}$$

$$\mathbb{G} \equiv \frac{1}{4} \Lambda^{kl} \Lambda^{mn} \mathbb{G}_{km;ln}; \quad \mathbb{G}_{km;ln} \equiv \mathfrak{R}_{\beta km}^\alpha \mathfrak{R}_{\alpha ln}^\beta \equiv Tr(\boldsymbol{\mathfrak{R}}_{km} \boldsymbol{\mathfrak{R}}_{ln}); \tag{2.1.4}$$

$$\mathfrak{R}_{\beta kl}^\alpha \equiv \boldsymbol{\mathfrak{R}}_{kl} \equiv \partial_k \boldsymbol{\mathcal{A}}_l - \partial_l \boldsymbol{\mathcal{A}}_k + [\boldsymbol{\mathcal{A}}_k, \boldsymbol{\mathcal{A}}_l] = -\boldsymbol{\mathfrak{R}}_{lk}; \tag{2.1.5}$$

$$w^{kl} \Rightarrow \Lambda^{kl} \equiv Tr\mathbf{\Lambda}^k \mathbf{\Lambda}^l \equiv \Lambda_\beta^{\alpha k} \Lambda_\alpha^{\beta l} = \Lambda^{lk}; \quad \Lambda^{km} \Lambda_{lm} = \delta_l^k. \tag{2.1.6}$$

Here objects in (2.1.3) are the above introduced covariant derivatives forms of DSV; symbol $[\boldsymbol{\mathcal{A}}_k, \boldsymbol{\mathcal{A}}_l]$ denotes commutator of two gauge matrices. As one can see, Lagrangian is structured on four basic *geometrical objects* $X^a$: DSV, *h-tensor* $\Lambda_\beta^{\alpha k} \equiv \mathbf{\Lambda}^k$, and *affine h-tensor* $\mathcal{A}_{\beta k}^\alpha \equiv \boldsymbol{\mathcal{A}}_k$:

$$\{X^a\} = \Psi^\alpha, \Phi_\alpha; \quad \Lambda_\beta^{\alpha k}; \quad \mathcal{A}_{\beta k}^\alpha. \tag{2.1.7}$$

We call h-tensor $\Lambda_\beta^{\alpha k}$ *split metric* (SM), affine h-tensor $\mathcal{A}_{\beta k}^\alpha$ *unified gauge field* (UGF) or simply *gauge*. Form $\mathfrak{D}_{\beta k}^\alpha$ is named *matter matrices* (MM), and form $\mathfrak{R}_{\beta kl}^\alpha$ *hybrid curvature form* (HCF); the last one is uniquely recognized as *covariant derivative of gauge* $\mathcal{A}_{\beta k}^\alpha$ itself. Tensor forms $\mathbb{G}_{km;ln}$ and $\Lambda^{kl}$ are named *gauge 4-tensor* and *grand metric* (GM), respectively. Scalar forms $\mathbb{N}$, $\mathbb{D}$, $\mathbb{M}$ and $\mathbb{G}$ are named *state norm, kinetic scalar, matter scalar* and *gauge scalar*, respectively.

Note that, all definitions (2.1.2) through (2.1.6) are unambiguous, since contractions between Roman and Greek indices are not legitimate in the differential theory under treat.

*Scale invariance of Lagrangian*

Scalar Lagrangian (2.1.1) as well as the whole Lagrangian (1.30) is *scale-invariant* i.e. it possesses the *immunity* of its form relative introduction of arbitrary real numbers (positive or negative) as multipliers of its scalar items: by a simple proper scaling the DSV and SM magnitudes, whole the scalar

Lagrangian can be returned to the initial form (2.1.1). Same is true relative introduction of arbitrary real multipliers of UGF; in this case the restoring of Lagrangian form is achieved by the correspondent re-scaling of the UM variables. It should be noted, by the way, that scale invariance as an intrinsic property of the constant-less irreducible field theory can be implemented in a logically consistent way only based on EAP as the background dynamical principle. This commitment makes EAP an indispensable, no-alternative receipt of deriving basic equations of a unified covariant differential theory.

*Mini-max principle*

As an aspect of scale invariance in the context of the *irreducibility* demand, the above introduced *mini-max principle* has been applied to structuring the Lagrangian: at a *minimum* (necessary) association of basic objects, the scale-invariant composition of Lagrangian should include *maximum* variety of the related scalar forms. In our case, when weigh factor $\sqrt{\Lambda}$ is structured as shown above (under the press of the irreducibility demand), the mini-max requirement is referred directly to the scalar Lagrangian $\mathbb{L}$. Any addition to $\mathbb{L}$ been built on the same basic objects violate the feature of *scale invariance*.

## 2.2. Euler-Lagrange equations

*Master equations*

Taking into account definition of DSV covariant derivatives (2.1.3), we can write EL equations (1.31) on DSV in the following form:

$$\mathbf{\Lambda}^k \mathfrak{D}_k \mathbf{\Psi} + (\frac{1}{2}\mathfrak{D}_k \mathbf{\Lambda}^k + 1)\mathbf{\Psi} = 0 ; \qquad (2.2.1)$$

$$(\mathfrak{D}_k \mathbf{\Phi})\mathbf{\Lambda}^k + \mathbf{\Phi}(\frac{1}{2}\mathfrak{D}_k \mathbf{\Lambda}^k - 1) = 0 ; \qquad (2.2.2)$$

here

$$\mathfrak{D}_k \mathbf{\Lambda}^k \equiv \frac{1}{\sqrt{\Lambda}} \partial_k (\sqrt{\Lambda} \mathbf{\Lambda}^k) + [\boldsymbol{\mathcal{A}}_k, \mathbf{\Lambda}^k] . \qquad (2.2.3)$$

Note that, object $\mathfrak{D}_k \Lambda_\beta^{\alpha k}$ is an s-tensor, since it can be represented as covariant derivative $\mathfrak{D}_l \Lambda_\beta^{\alpha k}$ of h-tensor $\Lambda_\beta^{\alpha k}$ contracted on Roman indices $l = k$:

$$\mathfrak{D}_l \mathbf{\Lambda}^k \equiv \partial_l \mathbf{\Lambda}^k + \Gamma_{ml}^k \mathbf{\Lambda}^m + [\boldsymbol{\mathcal{A}}_l, \mathbf{\Lambda}^k] ; \qquad (2.2.4)$$

here $\Gamma_{ml}^k$ are the conventional *Christoffel symbols* or *matched connection* form [8,9]:

$$\Gamma_{ml}^k = \frac{1}{2} \Lambda^{kn} (\partial_m \Lambda_{ln} + \partial_l \Lambda_{mn} - \partial_n \Lambda_{lm}) . \qquad (2.2.5)$$

Object $\mathfrak{D}_k \Lambda_\beta^{\alpha k}$ can be characterized as *covariant divergence* of Split Metric $\Lambda_\beta^{\alpha k}$.

Note that, EL equations (2.2.1), (2.2.2) coincide with the previously derived pair of equations (1.34) taking into account specification (2.1.6).

*Gauge equations*

Equations (1.31) on UGF $\mathcal{A}_k$ can be written in the following symbolic covariant view:

$$\mathfrak{D}_l \mathfrak{R}^{kl} = \mathcal{J}^k, \tag{2.2.7}$$

with covariant divergence of hybrid tensor $\mathfrak{R}^{kl}$ on the left-hand side:

$$\mathfrak{D}_l \mathfrak{R}^{kl} \equiv \frac{1}{\sqrt{\Lambda}} \partial_l (\sqrt{\Lambda} \mathfrak{R}^{kl}) + [\mathcal{A}_l, \mathfrak{R}^{kl}] \tag{2.2.8}$$

and *supercurrent* matrix, an h-tensor $\mathcal{J}_\beta^{\alpha k} \equiv \mathcal{J}^k$ on the right-hand side:

$$\mathcal{J}_\beta^{\alpha k} \equiv \frac{1}{2} (\Lambda_\gamma^{\alpha k} \Phi_\beta \Psi^\gamma + \Lambda_\beta^{\gamma k} \Phi_\gamma \Psi^\alpha). \tag{2.2.9}$$

Note that, object $\mathfrak{D}_l \mathfrak{R}^{kl}$ is an h-tensor as well, since it can be represented as covariant derivative of HCF:

$$\mathfrak{D}_m \mathfrak{R}^{kl} \equiv \partial_m \mathfrak{R}^{kl} + \Gamma_{nm}^k \mathfrak{R}^{nl} + \Gamma_{nm}^l \mathfrak{R}^{kn} + [\mathcal{A}_m, \mathfrak{R}^{kl}]$$

after contraction on indices $m = l$ (taking into account the *skew symmetry* of HCF on Roman indices vs the *even symmetry* of the above shown Christoffel symbols $\Gamma_{nm}^k$).

Equations (2.2.7) connect the affine h-tensor, *unified gauge field* matrices $\mathcal{A}_{\beta k}^\alpha$ to DSV and SM.

*Metric equations*

Performing variation derivatives of Lagrangian (1.30) on *split metric* (SM) matrices $\Lambda_\beta^{\alpha k}$ according to general equations (1.31), we obtain the following EL equations:

$$(\mathbb{G}_{kl} - \mathbb{L} \Lambda_{kl}) \Lambda^l = -\mathfrak{D}_k \ ; \tag{2.2.10}$$

here notation $\mathbb{G}_{kl}$ is for *gauge tensor* defined as follows:

$$\mathbb{G}_{kl} \equiv \Lambda^{mn} \mathbb{G}_{km;ln} = \mathbb{G}_{lk}. \tag{2.2.11}$$

Since Lagrangian (1.30) given by equations (2.1.1.) through (2.1.6) does not include derivatives of SM, EL equations on SM result in a system of *algebraic equations* on SM considered as function of DSV and UGF.

*Covariance of EL equations as an attribute of the Extreme Action*

Structural form of EL equations *is not* and *cannot be* thought as connected (or referred in advance) to a certain "frame of coordinates". Raising such question does not have sense with respect to the background differential system of an irreducible field theory. Realization of this circumstance is the essence of the *general covariance* paradigm in the context of the UFT foundations. The only legitimate question in this context is: how the basic objects are transformed at transformations of the UM variables ? At this stage of profiling dynamic properties of SFT, there is no unambiguous answer of this question. Namely, in accordance with the proposed UM−MF homomorphism, one can assume that, at transformation of UM variables (at a point) with matrix $A$, state vector $\Psi^\alpha$ is transformed with some matrix $B$ as a contravariant vector in MF space. Then, connections between basic objects by the derived system of EL equations determine transformations of covariant state vector $\boldsymbol{\Phi}$, unified gauge field $\mathcal{A}$ and split metric $\boldsymbol{\Lambda}$ according to equations (1.13), (1.22) and (1.24), respectively.

As pointed above, matrix $B$ is different from $A$ but connected to the last one via the dynamic law of SFT. Now, connection $B(A)$ can be considered as aspect of the dynamics based on the derived system of EL equations. Exploration of this issue, however, goes beyond the scope of this paper.

## 2.3. Dynamic Identities

In addition to the *grand metric* tensor, a series of objects and equations contracted on Greek indices, and related equations can be extracted from system of objects and equations of SFT. In turn, some of them can be used for reduction of initially defined forms and derived EL equations.

### Extended Faraday-Maxwell equations
*Extended Faraday equations*

By taking trace of HCF form (2.1.8) on Greek indices $\beta = \alpha$ we obtain a skew-symmetric covariant tensor $\mathbb{F}_{kl}$ defined as:

$$\mathbb{F}_{kl} \equiv \frac{1}{N}\mathfrak{R}^{\alpha}_{\alpha kl} = \partial_k \mathcal{A}_l - \partial_l \mathcal{A}_k = -\mathbb{F}_{lk} ; \qquad (2.3.1)$$

here

$$\mathcal{A}_k \equiv \frac{1}{N}\mathcal{A}^{\alpha}_{\alpha k} . \qquad (2.3.2)$$

Note that object (2.3.1) is tensor despite that $\mathcal{A}^{\alpha}_{\alpha k}$ is not a vector. This tensor satisfies the identity equations similar to the *first pair* of Maxwell equations (but now in $N$-dimensional space of UM):

$$\partial_m \mathbb{F}_{kl} + \partial_l \mathbb{F}_{mk} + \partial_k \mathbb{F}_{lm} = 0 . \qquad (2.3.3)$$

*Extended Maxwell equations*

By taking trace of gauge equations (2.2.7) on Greek indices $\beta = \alpha$ we obtain the following $N$ equations (similar to the *second pair* of Maxwell equations):

$$\frac{1}{\sqrt{\Lambda}}\partial_l(\sqrt{\Lambda}\mathbb{F}^{kl}) = \mathcal{J}^k. \qquad (2.3.4)$$

These $N$ equations connect two contravariant objects: a skew-symmetric *contravariant* tensor field, a *metrical image* of the covariant tensor $\mathbb{F}_{kl}$:

$$\mathbb{F}^{kl} \equiv \Lambda^{km}\Lambda^{ln}\mathbb{F}_{mn} = -\mathbb{F}^{lk} \qquad (2.3.5)$$

and a contravariant *vector* field, the *supercurrent*:

$$\mathcal{J}^k \equiv \frac{1}{N}\mathcal{J}^{\alpha k}_{\alpha} = \frac{1}{N}\Lambda^{\beta k}_{\alpha}\Phi_\beta \Psi^\alpha . \qquad (2.3.6)$$

*Scalar dynamic identities of DSV equations*

Considering contraction of DSV equations (2.3.1) in products with DSV itself, we find the following two scalar equations.

*Conservation of the supercurrent* :

$$\frac{1}{\sqrt{\Lambda}} \partial_k (\sqrt{\Lambda} \mathcal{J}^k) = 0 \,. \tag{2.3.7}$$

Note that, this equation is in a direct consistence with equations (2.3.4), since

$$\partial_k \partial_l (\sqrt{\Lambda} \mathbb{F}^{kl}) \equiv 0,$$

as for any skew-symmetric contravariant tensor.

*Nullification of matter scalar in dynamics*

Other important direct consequence of DSV equations is nullification *in dynamics* of form $\mathbb{M}$, *matter scalar*:

$$\mathbb{M} \Rightarrow 0 \tag{2.3.8}$$

i.e. there is a *dynamical identity*:

$$\mathbb{D} = -\mathbb{N}. \tag{2.3.9}$$

*Equation for gauge scalar*

Multiplying equations (2.2.10) by $\Lambda^{\beta k}_\alpha$, producing contraction on all indices and taking into account the dynamic identity (2.3.9), we find the following dynamic relation:

$$(N - 4)\mathbb{G} = \mathbb{D} = -\mathbb{N} \,. \tag{2.3.10}$$

So at $N \neq 4$ we find that gauge scalar $\mathbb{G}$ in dynamics is a proportion to *state norm* $\mathbb{N}$:

$$\mathbb{G} = -\frac{\mathbb{N}}{N - 4} \,. \tag{2.3.11}$$

When considering case $N = 4$ in equation (2.3.10), we have to accept dynamic condition $\mathbb{D} = -\mathbb{N} = 0$, instead of the proportion between $\mathbb{G}$ and $\mathbb{N}$ as at $N \neq 4$. It should be noted, however, that such condition for *mathematical consistence* of the theory as $\mathbb{N} = 0$ at $N = 4$ is not in complete consistence with the foundation of the *autonomic duality* of *state vector* as represented by the two *independent* vector fields in the *matter function* space, contravariant $\Psi^\alpha$ and covariant $\Phi_\alpha$ . Therefore, this peculiarity should be regarded as *standing out* of the frame of the treated superdimensional field theory (meaning case $N = 4$ *inconsistent* with the derived dualistic structure of SFT).

*Dynamic reduction of metric equations*

It follows from dynamic identity (2.3.9) that, scalar Lagrangian $\mathbb{L}$ in *metrics equations* (2.2.10) can be replaced by gauge scalar $\mathbb{G}$:

$$\mathbb{L} \Longrightarrow \mathbb{G} \Longrightarrow -\frac{\mathbb{N}}{N-4}, \qquad N \neq 4. \tag{2.3.12}$$

Using this dynamic reduction, we can write metric equations (2.2.10) in the following view:

$$\mathbb{H}_{km}\Lambda_\beta^{\alpha m} = -\mathfrak{D}_{\beta k}^{\alpha}; \tag{2.3.13}$$

here

$$\mathbb{H}_{km} \equiv \mathbb{G}_{km} - \mathbb{G}\Lambda_{km}. \tag{2.3.14}$$

At $N \neq 4$ scalar $\mathbb{G}$ can be replaced by its *dynamic identity* according to relation (2.3.11).

There are two important outcomes from dynamic reduction (2.3.12) of metrics equations (2.2.10).

### *Solution for Split Metric as function of DSV, UGF and GM*

Equations (2.3.13) can directly be solved relative Split Metric $\Lambda_\beta^{\alpha k}$ considered as function of DSV, UGF and GM:

$$\Lambda_\beta^{\alpha k} = -\breve{\mathbb{H}}^{kl}\mathfrak{D}_{\beta l}^{\alpha}; \tag{2.3.15}$$

here $\breve{\mathbb{H}}^{kl}$ is tensor inverse to $\mathbb{H}_{kl}$, i.e. :

$$\breve{\mathbb{H}}^{kn}\mathbb{H}_{ln} = \delta_l^k. \tag{2.3.16}$$

### *Algebraic equations for Grand Metrics as function of matter tensor and gauge 4-tensor*

Contracting on Greek indices product of two equations (2.3.13) with Roman indices $k$ and $l$, we obtain algebraic equations for Grand Metric $\Lambda^{mn}$ as function of DSV and UGF:

$$\mathbb{H}_{km}\mathbb{H}_{ln}\Lambda^{mn} = \mathbb{D}_{kl}; \tag{2.3.17}$$

here we have introduced notation $\mathbb{D}_{kl}$ for *matter tensor* defined as follows:

$$\mathbb{D}_{kl} \equiv \mathfrak{D}_{\beta k}^{\alpha}\mathfrak{D}_{\alpha l}^{\beta}. \tag{2.3.18}$$

Equations (2.3.17) can also be written in the following view:

$$\mathbb{H}_n^l \mathbb{H}_k^n = \mathbb{D}_k^l; \tag{2.3.19}$$

here

$$\mathbb{H}_k^l = \mathbb{G}_k^l - \mathbb{G}\delta_k^l; \quad \mathbb{G}_k^l = \Lambda^{lm}\mathbb{G}_{km}; \quad \mathbb{D}_k^l = \Lambda^{lm}\mathbb{D}_{km}. \tag{2.3.20}$$

As one can see, equations (2.3.19) are a system of algebraic equations of the fourth power for *grand metric* tensor $\Lambda^{kl}$ as function of *matter tensor* $\mathbb{D}_{kl}$ and *gauge 4-tensor* $\mathbb{G}_{km;ln}$; in other words, as function of DSV and UGF. Once GM arrives from equations (2.3.19) as function of DSV and UGF, so is about SM given by equations (2.3.15).

### *Hamilton-Nöther equation*

Considering the derivatives of Lagrangian $\mathcal{L}$ while taking into account general Euler-Lagrange equations, we find the following generic *dynamic identities* :

$$\check{\partial}_l \mathcal{T}_k^l = 0 , \qquad (2.3.21)$$

where $\mathcal{T}_k^l$ is a mix valence 2 *pseudo-tensor* object:

$$\mathcal{T}_k^l = \frac{1}{2}\Lambda_\beta^{\alpha l}(\Phi_\alpha \partial_k \Psi^\beta - \Psi^\beta \partial_k \Phi_\alpha) + \mathfrak{R}_\beta^{\alpha l m}\partial_k \mathcal{A}_{\alpha m}^\beta + \frac{\mathbb{N}}{N-4}\delta_k^l . \qquad (2.3.22)$$

It should be noted that, though this object is not tensor, its structure (i.e. form as a composition of basic objects and their derivatives) does not change at arbitrary transformations of UM variables.

Discussion of the presented equations is placed in special Chapters 10 following the comprehensive consistent exposition of SFT principles and derivations in Chapters 3 through 9.

## 2.4. Summary and Outlook

*Summary*

This paper presents the results of exploration of an approach to a Unified Field Theory based on the concept of *matter function* (MF) as a real $\mu$-component function $\varphi^\alpha$ of $N$ variables $\check{\psi}^k$ of a *unified manifold* ($k = 1,2,...N$; $\alpha = 1,2,...\mu > N$) subordinate to an *invariant differential law* (DL) $\partial_k \varphi^\alpha \equiv F$, subject to find out. The basic result of the paper is the derived system of covariant Euler-Lagrange equations for a collection of basic objects of DL. The pilot object of the theory is *Dual State Vector* field (DSV) representing *matter* in DL. It consists of a dual couple of $N$-component fields $\Psi^\alpha, \Phi_\alpha$, contra- and co-variant vector, respectively. Vector field $\Psi^\alpha(\{\check{\psi}^k\})$ is introduced as *field of directions* in the MF space collinear with differentials $d\varphi^\alpha$ and transformed with a matrix at transformation of variables $\check{\psi}^k$ with matrix *A*. Co-vector field $\Phi_\alpha$ is introduced independent of $\Psi^\alpha$ but transformed with matrix $B^{-1}$. Connection of two matrices ($BF = FA$) is not established *ad hoc* but is supposed to be found as an attribute of DL itself.

The rest of the basic objects are the triadic coefficient functions transformed with both *A* and *B* differential matrices: *Split Metric* matrices on Greek indices (SM) $\Lambda_\beta^{\alpha k}$ (*hybrid tensor)* and *gauge matrices* $\mathcal{A}_{\beta k}^\alpha$ of the *unified gauge field* (UGF, a *hybrid affine tensor*). Metric tensor $\Lambda_{kl}$ of UM (*Grand Metric*, GM) is introduced for invariant integration of scalar forms; it is built on SM with no assumptions about its signature. This connection can be considered as an extended analog of connection between $\gamma$-matrices and Minkowski metric tensor in Dirac equations; however, it is introduced in a way inverse to that of Dirac theory – due to the logic of the *irreducibility* demands. Gauge matrices, a hybrid analog of the *affine connections* of the differential geometry, are introduced for covariant extension of the DSV derivatives; they can be considered as a *hybrid* non-Abelian extension of vector-potential of Electrodynamics to the superdimensional UM − MF spaces. The *hybrid curvature form*, matrices (on Greek indices) $\mathfrak{R}_{\beta k l}^\alpha = -\mathfrak{R}_{\beta l k}^\alpha$, an analog of Riemann-Christoffel tensor of differential geometry, has been established and introduced to Lagrangian as a *covariant curl-derivative* of the gauge matrices.

A unique *maxim* Lagrangian form has been composed under the press of the *irreducibility* principle specified as a set of requirements suited for UFT in the author sight: *homomorphism, duality, homogeneity, uniformity, reality, covariance, extreme action, existence of a conservative current, differential irreducibility, scale invariance, and the mini-max principle*.

Euler-Lagrange equations on DSV (*master equations*) play in the theory a role corresponding to Dirac equation in the classical field theory of Electrodynamics. EL equations on UGF (*gauge* *equations)* couple UGF to DSV and Split Metric; they play a role corresponding to Maxwell equations in Electrodynamics and, in general, role of equations for gauge fields in QFT. Finally, equations for SM (*metric equations*) connect SM (hence, Grand Metric as well) to DSV and UGF; they correspond to equations for metric tensor in GTR.

### *Outlook*

A preliminary comparison in basic detail of the properties of the exposed approach to UFT and derived equations with the existing classical and quantum theories of elementary particles and gravitation is conducted in Chapter 10. Here we would like to mark several generic insights and open questions of the presented superdimensional concept.

1. *Quantum behavior as projection of a Superdimensional "Classical Field" Dynamics*

The exposed approach to UFT as a homomorphic "classical field theory" in a superdimensional manifold suggests an interpretation of the quantum behavior of the matter fields as *projection* of a totally deterministic $N$-dimensional picture of *supermatter* dynamics onto the intelligible 4d world, with no "quantization" procedure required as a specific principle.

2. *Breaking of the superposition principle*

Setting UFT as a *homogenous* differential law for DSV as a *universal wave function* (UWF) leads directly to *breaking of the superposition principle* of QFT, since the theory arrives naturally and unavoidably non-linear on the *state vector* (note that, a linear *homogeneous* differential field theory is empty of content). In the context of a comparison between the principles and methodology of SFT and QFT, the last one is viewed as a theory that does not match the principle of *homogeneity* in space of its variables as degrees of freedom. In the context of the interpretation of the theory, it should be noted that, non-linearity of the differential system relatively the Universal Wave Function (Dual State Vector field in our case) i.e. breaking of the *superposition principle* may lead to a significant modification of the conceptual texture and methodology of a fundamental field theory of interactions and transformations of "elementary particles" as particular fields recognized in the asymptotic limit. On the other hand, property of the *scale invariance* of a unified theory could be viewed as a replacement for the superposition principle. This property looks necessary for the ability to pick out the fundamental dimensionless constants of QFT as quantities related to the asymptotic substructures of UFT.

3. *Issue of the dimensionality*

This paper leaves open questions about dimensionality of both the *unified manifold* and *matter function* spaces. It seems to be not out of sense that, analysis of connection of two dimensionalities could be performed by consideration of possible irreducible isomorphic correspondence between possible *irreducible* geometrical structures of two spaces. Such *structural isomorphism* might be induced by suitable physical requirements of a specific transformational invariance of the derived differential system. After this, establishing of the absolute dimensionality presumably should be guided by the demand of *reality* of *grand metric* tensor as solution of the algebraic *metric equations* derived in this paper.

4. *Constraint of the reality*

One of the critical questions to the presented approach to UFT is the compatibility of the all-real superdimensional covariant differential system (equations for DSV, first of all) with *complexity* of the existing quantum equations (Dirac equation and other, and QFT master equation for State Vector, overall) associated with the definitions of the *energy* and *momentum operators,* on one hand, and with *charged particles duality*, on the other hand. Namely, in our view, the general covariance of a unified theory is not compatible with use of invariable objects such as the imaginary unit, "*i*". A preliminary answer to this question is that the required correspondence can be envisioned and recognized based on the analysis of the matrix structure of UGF in the equations for DSV. This structure being considered as an aspect of the extended and unified gauge principle may suggest a real representation of the imaginary unit and complex numbers as part of more general group of *variable* objects like the UGF matrices.

5. *Possible interpretation of singularities in QFT*

Presence of singularities in QFT is conventionally considered as a fundamental defect of the theory. Superdimensional Field Theory may suggest interpretation of singularities in QFT as manifestation of the super-dimensional nature of the microworld (hence, whole the true world). Namely, asymptotic solutions for DSV of SFT being finite could concentrate along unbound "world lines" and could be interpreted as "elementary particles". In such sight, infinities of the binary "vacuum averages" ("Green functions") of QFT might be recognized as associated with the squared superdimensional coordinates i.e. unbound variables related to particular "fields" as degrees of freedom, − which all are infinite on definition, hence, have no physical sense (as well as a certain mathematical one) in the context of the dynamics.

6. *Possibility of the space-time − matter transformations*

Perhaps, one of the most intriguing and speculative consequences of the presented SFT sight on UFT is a logical possibility to consider the processes of the *mutual space-time − matter transformations.* Despite of a striking strangeness of such an outcome, after a few meditations it appears look well aligning the *relativity tendency* of the fundamental theoretical physics.

7. *Search for solutions*

Search for solutions of the derived equations may and, likely, should start with investigation of possible *asymptotes* as possessing the simplest structure and behavior but perhaps delivering the results, which may play a key role in approaching analysis and solutions for dynamics of possible clusters, transformation processes and interpretation of the theory. Profiling the asymptotes might be associated with imposing some specific invariance properties on the solutions. In this way, it might not be insignificant to use the right variables, perhaps associated with the geodesics. Another envisioned instrument of building up the asymptotes is *Nöther theorem* [1, 2]. Even before starting a search for solutions and interpretations, it would be necessary to conduct a comparison in detail between presented approach to a unified theory and the existing field theories (Standard Model), and well as other trends of unification.

*8. The gravitation probe of the theory*

As known, the Einstein-Hilbert and Weyl theories of gravitation both lead to *Newton gravitation law* (NGL) in the non-relativistic limit and show the same lowest order *relativistic corrections* to the non-relativistic dynamics in Newtonian gravitational field [8]. The higher order corrections of two theories are different, while to-date there is no experimental data on measuring the higher order effects. This situation stimulates the attempts to explore alternative approaches to the theory of gravitation. Sure, those other theories should show the Newton law and same lowest order relativistic corrections − but yet they must response to the quest of unification with theory of the micro-world i.e. QFT. Ultimate such theories are supposed to serve unification of QFT itself, as well.

Explorations in the *superdimensional dual-covariant field theory* (SFT) towards *unification* presented here are directed, in particular, by a point of view that phenomenon of gravitation should arrive as an intrinsic and imprescriptible attribute of a generally covariant fundamental field theory. The present paper resulted in algebraic equations for metric tensor of SFT (*grand metric*, GM) as function of dual state vector (DSV) and unified gauge field (UGF) (in turn, driven by DSV). Analysis of asymptotic solutions of the derived EL equations should answer a question, can the *Newton gravitation law* and the known lowest order *relativistic corrections* be derived in frame of the explored approach to UFT.

9. *Fragmentation*

There is still, of course, a number of other critical questions to investigate and answer based on analysis of the derived system of equations and their solutions: "geometrical nature" and fragmentation of Dual State Vector, Split Metric and Unified Gauge Field; correspondence to the Fermi-Dirac quantization in QFT; the general dynamic invariants and the asymptotic ones; concept of the observables; profiling spins, charges, masses, symmetries and other group characteristics which could be identified or associated with particular sectors of QFT and beyond; and more – all beyond the scope of this paper.

**Acknowledgements**

During many years of this investigation, the discussions and consultations with colleagues of Jefferson Laboratory and outside were so helpful for the author to keep going. The author is grateful to Yury Orlov, Iakov Azimov, Dmitry Diakonov and Leonid Ponomarev for useful discussions during the initial period of these studies, to Johann Bengtsson for his valuable advice to resort in this investigation to the variation principle, and to Victor Mokeev for his profound consultations on the state and history of the Standard Model. The author expresses his special appreciation to Martha Harrell and Rui Li for the accompanying discussions and help in reviewing of the backgrounds and history of theory of the elementary particles, General Relativity, and the related mathematics. The author also thanks Stanley Brodsky, James Bjorken and Arkady Vainshtein for review of the texts of publication [1[*]] and useful comments. I use the present publication also as opportunity to express my deepest gratitude to Stanislav Podosenov for discussions and useful comments about the aspects of covariance of the General Relativity and structure of the Relativistic Theory of Gravitation.

# 3. Pre-view of Differential Law for DSV

In this Chapter we will profile differential law for DSV as equations corresponding to equation for State Vector in QFT.

## 3.1. Principles of DL for DSV

In accordance with the presuppositions about properties of a fundamental differential law of UFT discussed in Chapter 1 (*Prolegomena*), the following requirements should be claimed to the differential equations for DSV.

*Dual autonomy*: differential equations for DSV should be formulated as relations between DSV and its derivatives – though these relations should include some objects as *coefficient functions* i.e. multipliers at DSV and its derivatives. These functions are supposed to be connected to DSV by other, complementary differential equations.

*Homogeneity:* the equations should not include any explicit i.e. *given* coefficient functions of variables $\breve{\psi}^k$ of the *N*-dimensional manifold.

*Uniformity*: the equations should be uniform over all UM variables as well as over all components of every involved object.

*Differential irreducibility of the master equations*: equations for DSV should be formulated only in terms of the *first order* derivatives of DSV.

*General covariance*: the equations must not change their general structure i.e. *form* at *arbitrary* transformations of variables.

*Reality*: DL as an analytical algorithm should be formulated in all real terms; the *imaginary unit* "*i*", and *complex* objects are not admitted.

*Existence of a conservative current* in the UM space.

## 3.2. Primer Relations and Constraint of Covariance

*DSV as a primary object of a self-contained field theory*

In correspondence to the quest for a differential formulation of the fundamental equations of UFT, objects of the differential equations themselves should be defined in the differential terms. This requirement is due to the demand of certainty of transformation properties of objects and equations relative to transformations of differentials of the *unified manifold* variables. The matter function $\breve{\varphi}^\alpha(\breve{\psi})$ itself does not belong to this class of objects, in contrary to differentials $d\breve{\varphi}^\alpha$ as discussed above. On the other hand, the physical equations cannot be written for differentials but should be written for the finite objects. Resolution of this constraint is that, the objects can be associated with *directions* in space of matter function $\breve{\varphi}$ i.e. they should be represented by or associated with *ratios* between differentials of $\breve{\varphi}$. Dual State Vector field $\Xi = (\Phi_\alpha, \Psi^\alpha)$ above introduced is of this type of an object. It is also worthwhile to underline an irreducible, background nature of this object; all complicate objects can be introduced, composed and connected to $\Xi$ basing on considering it as an irreducible geometrical object. DSV can be viewed a candidate for a pilot role in the differential system of a fundamental field theory. Differential equations for DSV as a pilot object of the system should connect derivatives of DSV to DSV itself:

$$\{\partial \Xi\} \propto \Xi. \tag{3.2.1}$$

We thus come up to the main paradigm of *field theory* of *matter*, the *wave equation*.

In accordance to the *duality* principle, two partner objects, $\Psi^\alpha$ and $\Phi_\alpha$, should not serve each other as a source, so there should be two equations of an autonomic type:

$$\{\partial_k \Psi^\alpha\} \propto \Psi^\alpha;$$
$$\{\partial_k \Phi_\alpha\} \propto \Phi_\alpha. \tag{3.2.2}$$

*Primer form of DSV equations*

Now we are making a step from *relations* principle (3.2.2) to *equations* for DSV. Apparently, connection (3.2.2) should be "utilized" by mean of introduction of the multi-index coefficients. The differential law for DSV in the considered approach to unified field theory then can be viewed as system of differential equations consisting of the following type of equations for DSV:

$$P^{\alpha k}_\beta \partial_k \Psi^\beta + \Pi^\alpha_\beta \Psi^\beta = \Psi^\alpha; \tag{3.2.3}$$

$$\overline{P}^{\beta k}_\alpha \partial_k \Phi_\beta + \overline{\Pi}^\beta_\alpha \Phi_\beta = \Phi_\alpha. \tag{3.2.4}$$

Here we have introduced coefficient functions $P^{\alpha k}_\beta, \overline{P}^{\beta k}_\alpha; \Pi^\alpha_\beta, \overline{\Pi}^\beta_\alpha$ as matrices on Greek indices.

Using the following symbolic notations for the introduced matrices:

$$\mathbf{P}^k \equiv P^{\alpha k}_\beta; \quad \overline{\mathbf{P}}^k \equiv \overline{P}^{\alpha k}_\beta; \tag{3.2.5}$$

$$\mathbf{\Pi} \equiv \Pi^\alpha_\beta; \quad \overline{\mathbf{\Pi}} \equiv \overline{\Pi}^\alpha_\beta; \tag{3.2.6}$$

we can rewrite relations (3.2.3) and (3.2.24) in a symbolic form as follows:

$$\mathbf{P} \partial \mathbf{\Psi} + \mathbf{\Pi} \mathbf{\Psi} = \mathbf{\Psi}; \tag{3.2.7}$$

$$(\partial \mathbf{\Phi}) \overline{\mathbf{P}} + \mathbf{\Phi} \overline{\mathbf{\Pi}} = \mathbf{\Phi}. \tag{3.2.8}$$

We do not assume that matrices in these equations are constants, as well as we do not pose in advance any specific relations between their components.

These relations being taken literally as *equations* should satisfy the *covariance* requirement as stated above.

*Constraint of General Covariance*

Transformation of the derivatives of DSV includes terms proportional to the DSV components in product with the derivatives of the transformation matrix $B$. Applying differentiation on the transformed variables to the transformed DSV:

$$\Psi^{\alpha'} = B_\alpha^{\alpha'} \Psi^\alpha; \qquad \Phi_{\alpha'} = B_{\alpha'}^{\alpha} \Phi_\alpha,$$

and taking into account background relations

$$\partial_{k'} = A_{k'}^{k} \partial_k ; \quad \partial_k = A_k^{k'} \partial_{k'}$$

we obtain the following relations:

$$\partial_{k'} \Psi^{\alpha'} = A_{k'}^{k} B_\alpha^{\alpha'} \partial_k \Psi^\alpha + (\partial_{k'} B_\alpha^{\alpha'}) \Psi^\alpha ; \tag{3.2.9}$$

$$\partial_{k'} \Phi_{\alpha'} = A_{k'}^{k} B_{\alpha'}^{\alpha} \partial_k \Phi_\alpha + (\partial_{k'} B_{\alpha'}^{\alpha}) \Phi_\alpha . \tag{3.2.10}$$

It may be convenient to write these relations in a symbolic view as follows:

$$\boldsymbol{\Psi}' = B\boldsymbol{\Psi} ; \quad \boldsymbol{\Phi}' = \boldsymbol{\Phi} B^{-1};$$

$$\partial' = A^{-1} \partial ; \quad \partial = A \partial'$$

$$\partial' \boldsymbol{\Psi}' = A^{-1} \partial \boldsymbol{\Psi}' = A^{-1}[B \partial \boldsymbol{\Psi} + (\partial B) \boldsymbol{\Psi}];$$

$$\partial' \boldsymbol{\Phi}' = A^{-1} \partial \boldsymbol{\Phi}' = A^{-1}[(\partial \boldsymbol{\Phi}) B + \boldsymbol{\Phi} \partial B .]$$

(3.2.11)

So the left-hand side of equations (3.2.7), (3.2.8) written in the new variables acquire terms with derivatives of transformation matrix $B$ as multipliers at DSV.

Following the principle of general covariance, we have to require that, the form of the fundamental differential law should not change at arbitrary transformations of the UM variables. Can we reach this *covariance*, taking into account that, in distinction to the conventional differential geometry, matrix $B$ is different from matrix $A$? It should be noted that, from a formal point of view, equations (3.2.7), (3.2.8) can be considered as covariant relative of transformations with *constant* matrix $B$, at least. However, assumption of a constant matrix $B$ cannot be made voluntary, in contrary to always existing possibility to consider transformations with a constant matrix $A$ – since matrix $B$ is associated with *matter function* which is unlikely to be represented by a linear function of UM variables.

## 3.3. The Hybrid Affine Tensor as Unified Gauge Field

The structure and transformation properties of the introduced matrices $\Pi_\beta^\alpha$ and $\overline{\Pi}_\alpha^\beta$ are supposed been determined under a requirement that their transformation law should absorb terms with derivatives of transformation matrix $B$. At transformations, terms with derivatives of matrix $B$ arrive as multipliers at DSV in product with objects $P_\beta^{\alpha k}$ and $\overline{P}_\alpha^{\beta k}$ in relations (3.2.7), (3.2.8). Therefore, objects $\Pi_\beta^\alpha$ and $\overline{\Pi}_\beta^\alpha$ can be structured as products of matrices $P_\gamma^{\alpha k}$ and $\overline{P}_\gamma^{\alpha k}$ with some matrices $\mathcal{A}_{\beta k}^\gamma \equiv \mathcal{A}_k \equiv \mathcal{A}$ and $\overline{\mathcal{A}}_{\beta k}^\gamma \equiv \overline{\mathcal{A}}_k \equiv \overline{\mathcal{A}}$, respectively, and contracted on Roman indices $k$ as follows:

$$\Pi_\beta^\alpha \Rightarrow P_\gamma^{\alpha k} \mathcal{A}_{\beta k}^\gamma \ ; \qquad \overline{\Pi}_\beta^\alpha \Rightarrow \overline{P}_\gamma^{\alpha k} \overline{\mathcal{A}}_{\beta k}^\gamma \qquad (3.3.1)$$

or

$$\mathbf{\Pi} \Rightarrow \mathbf{P}^k \mathcal{A}_k \equiv \mathbf{P}\mathcal{A} \ ; \qquad \overline{\mathbf{\Pi}} \Rightarrow \overline{\mathbf{P}}^k \overline{\mathcal{A}}_k \equiv \overline{\mathbf{P}}\overline{\mathcal{A}} \ . \qquad (3.3.2)$$

Equations for DSV then can be written in matrix notations as follows:

$$\mathbf{P}(\partial + \mathcal{A})\mathbf{\Psi} = \mathbf{\Psi} \ ; \qquad (3.3.3)$$

$$[\partial \mathbf{\Phi} + \mathbf{\Phi}\overline{\mathcal{A}}]\overline{\mathbf{P}} = \mathbf{\Phi} \ , \qquad (3.3.4)$$

or, in the explicit form:

$$P_\beta^{\alpha k}(\partial_k \Psi^\beta + \mathcal{A}_{\gamma k}^\beta \Psi^\gamma) = \Psi^\alpha \ ;$$

$$\overline{P}_\alpha^{\beta k}\left(\partial_k \Phi_\beta + \overline{\mathcal{A}}_{\beta k}^\gamma \Phi_\gamma\right) = \Phi_\alpha \ . \qquad (3.3.5)$$

Here we have four coefficient functions, and we have to consider issue of their transformations: $\mathbf{P}, \overline{\mathbf{P}} \rightarrow \mathbf{P}', \overline{\mathbf{P}}'$ and $\mathcal{A}, \overline{\mathcal{A}} \rightarrow \mathcal{A}', \overline{\mathcal{A}}'$ at transformation of UM variables with matrix $A$, while DSV is transformed with matrix $B$. Let us first rewrite equation (3.3.3) in terms of transformed field $\mathbf{\Psi}' = B\mathbf{\Psi}$ and derivatives on the transformed variables of UM:

$$AB\mathbf{P}B^{-1}[\partial' + B(\partial' B^{-1}) + A^{-1}B\mathcal{A}B^{-1}]\mathbf{\Psi}' = \mathbf{\Psi}'. \qquad (3.3.6)$$

Based on a comparison with equation (3.3.3) written in terms of "frame" $\breve{\psi}'$:

$$\mathbf{P}'(\partial' + \mathcal{A}')\mathbf{\Psi}' = \mathbf{\Psi}'$$

we have to assume the following transformation laws for object $\mathcal{A}$:

$$\mathcal{A}' = A^{-1}B(\mathcal{A} + \partial)B^{-1} \qquad (3.3.7)$$

or, in the explicit notations:

$$\mathcal{A}_{\beta' k'}^{\alpha'} = A_{k'}^k B_\alpha^{\alpha'} \left(\mathcal{A}_{\beta k}^\alpha B_{\beta'}^\beta + \partial_k B_{\beta'}^\alpha\right). \qquad (3.3.8)$$

The structural role of matrices $\mathcal{A}$ with their assumed transformation law (3.3.8), thus, is in compensation for terms with derivatives of transformation matrix $B$ of state vector $\boldsymbol{\Psi}$ those arrive at transformation of primer equation (3.3.3). Introducing symbolic notations:

$$\mathfrak{D}\boldsymbol{\Psi} \equiv (\partial + \mathcal{A})\boldsymbol{\Psi} \qquad (3.3.9)$$

or

$$\mathfrak{D}_k \Psi^\alpha \equiv \partial_k \Psi^\alpha + \mathcal{A}^\alpha_{\beta k} \Psi^\beta \qquad (3.3.10)$$

we can write equations for the contravariant state vector field in the following view:

$$\mathbf{P}\mathfrak{D}\boldsymbol{\Psi} = \boldsymbol{\Psi} \qquad (3.3.11)$$

or

$$\mathrm{P}^{\alpha k}_\beta \mathfrak{D}_k \Psi^\beta = \Psi^\alpha , \qquad (3.3.12)$$

or

$$\mathrm{P}^{\alpha k}_\beta (\partial_k \Psi^\beta + \mathcal{A}^\beta_{\gamma k} \Psi^\gamma) = \Psi^\alpha . \qquad (3.3.13)$$

The correspondent derivation relatively equation (3.3.4) for the covariant partner of $\boldsymbol{\Psi}$, field $\boldsymbol{\Phi}$ leads to the following equation (compare (3.3.6)):

$$[(\partial'\boldsymbol{\Phi}') + \boldsymbol{\Phi}'(\partial'B)B^{-1} + \boldsymbol{\Phi}'B\bar{\mathcal{A}}B^{-1}A^{-1}]AB\bar{\mathbf{P}}B^{-1} = \boldsymbol{\Phi}'. \qquad (3.3.14)$$

Based on comparison with equation (3.3.4) written in terms of the transformed frame of the UM variables:

$$[(\partial'\boldsymbol{\Phi}') + \boldsymbol{\Phi}'\bar{\mathcal{A}}']\bar{\mathbf{P}}' = \boldsymbol{\Phi}'$$

one can conclude:

$$\bar{\mathcal{A}}' = B\bar{\mathcal{A}}B^{-1}A^{-1} + (\partial'B)B^{-1} = A^{-1}B(\bar{\mathcal{A}} - \partial)B^{-1}. \qquad (3.3.15)$$

As one can directly observe from comparison with equation (3.3.7), object $\bar{\mathcal{A}}$ can be identified with $-\mathcal{A}$:

$$\bar{\mathcal{A}} \Rightarrow -\mathcal{A}. \qquad (3.3.16)$$

Introducing a notation:

$$\mathfrak{D}\boldsymbol{\Phi} \equiv \partial\boldsymbol{\Phi} - \boldsymbol{\Phi}\mathcal{A} \qquad (3.3.17)$$

or

$$\mathfrak{D}_k \Phi_\alpha \equiv \partial_k \Phi_\alpha - \mathcal{A}^\beta_{\alpha k} \Phi_\beta , \qquad (3.3.18)$$

one can write equation (3.3.4) in the following view:

$$\mathfrak{D}\boldsymbol{\Phi} \cdot \bar{\mathbf{P}} = \boldsymbol{\Phi} \qquad (3.3.19)$$

or

$$\bar{\mathrm{P}}^{\beta k}_\alpha \mathfrak{D}_k \Phi_\beta = \Phi_\alpha , \qquad (3.3.20)$$

or

$$\bar{\mathrm{P}}^{\beta k}_\alpha \left( \partial_k \Phi_\beta - \mathcal{A}^\gamma_{\beta k} \Phi_\gamma \right) = \Phi_\alpha . \qquad (3.3.21)$$

It is important that, only one object, $\mathcal{A}^{\alpha}_{\beta k}$ is actually required for *covariant extension* of derivatives of both DSV partners, $\Psi^{\alpha}$ and $\Phi_{\alpha}$.

*Association property of UGF*

In correspondence to the transformation law for tensors and Christoffel symbols of differential geometry, transformation law for UGF $\mathcal{A}^{\alpha}_{\beta k}$ does possess the *association* property: if a transformation from $\check{\psi}'$ to $\check{\psi}''$ follows a transformation from $\check{\psi}$ to $\check{\psi}'$, then the resulting transformation of the connection coincides with the immediate transformation from $\check{\psi}$ to $\check{\psi}''$.

*UGF as the hybrid Christoffel symbols*

Matrices $\mathcal{A}_k$ of our treatment play a role corresponding to *Christoffel symbols* in differential geometry. They can be considered as *hybrid Christoffel's symbols*, since they are transformed with two distinguish differential matrices, $A$ and $B$; but *derivatives only of matrix B* are involved in the transformation law (3.3.7). Matrices $\mathcal{A}_k$ correspond more to the *general Christoffel symbols* or *affine tensor* $G^m_{kl} \neq G^m_{lk}$ rather than to ordinary Christoffels or "*matched connection*" $\Gamma^m_{kl} = \Gamma^m_{lk}$ [20, 21]. However, the hybrid Christoffels $\mathcal{A}_k$ distinct substantially from object $G^m_{kl}$, since the last one is transformed with the only matrix, $A$, as for transformation of differentials of UM variables, and also vectors and tensors. Being considered from this point of view, collection of $N$ matrices $\mathcal{A}_k$ can be regarded as a *hybrid affine tensor*. We will call this collection *unified gauge field* (UGF).

### 3.4. The Hybrid Tensors

*Covariant derivatives of DSV as Hybrid Tensors*

There is an important question about transformation of objects $\mathfrak{D}_k \Psi^{\alpha}$ and $\mathfrak{D}_k \Phi_{\alpha}$, once object $\mathcal{A}^{\alpha}_{\beta k}$ is assumed to transform according to law (3.3.7). Using the matrix symbolism and equation (3.3.11), one can find:

$$\mathfrak{D}'\Psi' \equiv (\partial' + \mathcal{A}')\Psi' = A^{-1}\partial(B\Psi) + A^{-1}(B\partial B^{-1} + B\mathcal{A}B^{-1})B\Psi. \qquad (3.4.1)$$

After using the identity:
$$B\partial B^{-1} + (\partial B)B^{-1} = \partial(BB^{-1}) \equiv 0 \qquad (3.4.2)$$

equation (3.4.1) results in the following transformation law:

$$\mathfrak{D}'\Psi' = A^{-1}B(\partial + \mathcal{A})\Psi + A^{-1}(\partial B - \partial B)\Psi = A^{-1}B\mathfrak{D}\Psi, \qquad (3.4.3)$$

or
$$\mathfrak{D}_{k'}\Psi^{\alpha'} = A^k_{k'} B^{\alpha'}_{\alpha} \mathfrak{D}_k \Psi^{\alpha}. \qquad (3.4.4)$$

Similar to that as we derived transformation law for object $\mathfrak{D}\Psi$, one can find:

$$\mathfrak{D}'\Phi' \equiv \partial'\Phi' - \Phi'\mathcal{A}' = \partial'(\Phi B^{-1}) - \Phi A^{-1}(\mathcal{A} + \partial)B^{-1} = \qquad (3.4.5)$$

$$= [(\partial \Phi) - \Phi \mathcal{A}]A^{-1}B^{-1} \equiv (\mathfrak{D}\Phi)A^{-1}B^{-1}$$

or

$$\mathfrak{D}_{k'}\Phi_{\alpha'} = A_{k'}^{k} B_{\alpha'}^{\alpha} \mathfrak{D}_{k}\Phi_{\alpha}. \tag{3.4.6}$$

Each of two bjects $\mathfrak{D}\Psi$ and $\mathfrak{D}\Phi$ is transformed with two matrices, $A$ and $B$, in the direct product; the transformations do not include derivatives of both matrices – once it is assumed that there exists object $\mathcal{A} \equiv \mathcal{A}_{\beta k}^{\alpha}$, $N$ matrices of rank $\mu$ transformed according to law (3.3.7). Objects $\mathfrak{D}\Psi$ and $\mathfrak{D}\Phi$ can be regarded as *covariant derivatives* of $\Psi$ and $\Phi$, respectively. We come to a conclusion that, if there does exist object $\mathcal{A}_{\beta k}^{\alpha}$ with transformation properties (3.3.11), then objects $\mathfrak{D}\Psi$ and $\mathfrak{D}\Phi$ are transformed according to equations (3.3.18) and (3.3.29), respectively. Vice versa, if these objects are defined as transformed with two matrices $A$ and $B$ as shown by equations (3.3.28) and (3.3.29), then object $\mathcal{A}_{\beta k}^{\alpha}$ is transformed according to law (3.3.7).

Objects $\mathfrak{D}\Psi$ and $\mathfrak{D}\Phi$ are transformed similar to *tensors* in differential geometry, but in our case with the direct products of *two different matrices*, $A$ and $B$ as shown in equations (3.3.22) and (3.3.33). The objects transformed as products of components ("coordinates") of vectors and s-vectors can be regarded as *hybrid tensors* or *h-tensors*. Apparently, covariant derivatives of DSV, objects $\mathfrak{D}_k\Psi^\alpha$ and $\mathfrak{D}_k\Phi_\alpha$, belong this class of objects. They can be defined as h-tensors of *valence 2*. In fact, they give birth to all the class of the *hybrid tensors*.

Correspondence between covariant derivatives $\mathfrak{D}_k\Psi^\alpha$ and $\mathfrak{D}_k\Phi_\alpha$ can also be derived by considering derivative of scalar form $N = \Phi_\alpha \Psi^\alpha$ introduced above. Derivative of this form is a *covariant vector* (so is the derivative of any *scalar form*):

$$\partial_{k'}N(\breve{\psi}') = \partial_{k'}N(\breve{\psi}) = \frac{\partial N(\breve{\psi})}{\partial \breve{\psi}^k} \frac{\partial \breve{\psi}^k}{\partial \breve{\psi}^{k'}} \equiv A_{k'}^{k} \partial_k N(\breve{\psi}).$$

On the other hand, this derivative can be represented in the following way:

$$\partial_k N \equiv \Phi_\alpha \mathfrak{D}_k \Psi^\alpha + \Psi^\alpha \left( \partial_k \Phi_\alpha - \mathcal{A}_{\alpha k}^{\beta} \Phi_\beta \right) \equiv \Phi_\alpha \mathfrak{D}_k \Psi^\alpha + \Psi^\alpha \mathfrak{D}_k \Phi_\alpha. \tag{3.4.7}$$

Once the first term on the right-hand side of this equation is transformed as vector, then so is the second term; thus, combined object $\mathfrak{D}_k \Phi_\alpha$ transforms as shown in equations (3.4.6).

*The Triadic Hybrid Tensors*

Once the right-hand side of equations (3.3.12) is an s-vector, then so must be the left-hand side. Further, since the covariant derivatives of DSV in these equations are transformed as the hybrid tensors according to equations (3.4.4), (3.4.6), then so must be matrices $P_\beta^{\alpha k}$ and $\overline{P}_\beta^{\alpha k}$:

$$\mathbf{P}' = AB\mathbf{P}B^{-1}; \quad \overline{\mathbf{P}}' = AB\overline{\mathbf{P}}B^{-1}; \tag{3.4.8}$$

or in the explicit view:

$$P^{\alpha'k'}_{\beta'} = A^{k'}_k B^{\alpha'}_\alpha B^{\beta}_{\beta'} P^{\alpha k}_\beta; \qquad \overline{P}^{\alpha'k'}_{\beta'} = A^{k'}_k B^{\alpha'}_\alpha B^{\beta}_{\beta'} \overline{P}^{\alpha k}_\beta. \qquad (3.4.9)$$

Note that, order of matrices $A$ and $B$ in these symbolic forms does not matter, since the Roman and Greek indices do not interfere, in accordance to definition of all the objects and structures of the theory.

Objects of valence 3 transformed according to equations (3.4.8) or (3.4.9) can be called the *triadic hybrid tensors*.

*Variation equivalence of UGF to a triadic h-tensor*

If there exists an object satisfying transformation law (3.3.7), then this object being changed by adding an arbitrary h-tensor of the same valence, satisfies the same law. In the other words, the difference of two the hybrid *affine* tensors, $\Delta \mathcal{A}^\alpha_{k\beta}$, is an *h-tensor* :

$$\Delta \mathcal{A}^{\alpha'}_{\beta'k'} = B^{\alpha'}_\alpha B^{\beta}_{\beta'} A^k_{k'} \Delta \mathcal{A}^\alpha_{\beta k} . \qquad (3.4.10)$$

We call this property of hybrid Christoffels the *variation equivalence*.

*Generally covariant form of DSV equations*

Thus, generally covariant equations for the autonomic dual s-field can be written in the following form:

$$P^{\alpha k}_\beta (\partial_k \Psi^\beta + \mathcal{A}^\beta_{\gamma k} \Psi^\gamma) = \Psi^\alpha,$$

$$\overline{P}^{\beta k}_\alpha (\partial_k \Phi_\beta - \mathcal{A}^\gamma_{\beta k} \Phi_\gamma) = \Phi_\alpha ; \qquad (3.4.11)$$

– *assuming* that object $\mathcal{A}^\alpha_{\beta k}$ is transformed according to the law (3.3.7).

Together with matrices $\mathbf{P}^k$ and $\overline{\mathbf{P}}^k$, matrices $\boldsymbol{\mathcal{A}}_k$ are supposed to be connected to DSV by specific differential equations, subject to find out.

*The associate hybrid tensors*

Based on the above introduced h-tensors, one can compose *associated* h-tensors, for example:

$$\Phi_\beta \mathcal{D}_k \Psi^\alpha ; \qquad \Psi^\alpha \mathcal{D}_k \Phi_\beta ; \qquad (3.4.12)$$

$$P^{\alpha k}_\gamma \Phi_\beta \Psi^\gamma ; \qquad \overline{P}^{\gamma k}_\beta \Phi_\gamma \Psi^\alpha; \qquad (3.4.13)$$

$$P^{\alpha k}_\gamma \overline{P}^{\gamma l}_\beta \pm P^{\alpha l}_\gamma \overline{P}^{\gamma k}_\beta , etc. \qquad (3.4.14)$$

## 3.5. The Associate Vectors and Tensors

Based on the triadic hybrid tensors, one can compose the *associated vector* and *tensor* objects, i.e. objects transformed with matrices $A$ and $A^{-1}$ of transformation of the manifold variables. They generally are obtained by contracting of the hybrid objects and their products on the s-indices.

1. Contravariant vector objects:

$$P^k \equiv P_\alpha^{\alpha k}; \qquad \overline{P}^k \equiv \overline{P}_\alpha^{\alpha k}; \qquad (3.5.1)$$

$$P_\beta^{\alpha k}\Phi_\alpha\Psi^\beta; \qquad \overline{P}_\beta^{\alpha k}\Phi_\alpha\Psi^\beta. \qquad (3.5.2)$$

These objects are transformed as differentials of manifold variables, $d\breve{\psi}^k$, similar to the tangent vectors of lines $q^k$ introduced above.

2. Contravariant tensors.
Based on contracted products of triadic objects, one also can compose *tensors* e.g. objects transformed as products of vector coordinates, for instance:

$$\mathfrak{w}^{kl} \equiv P_\beta^{\alpha k}\overline{P}_\alpha^{\beta l} + P_\beta^{\alpha l}\overline{P}_\alpha^{\beta k} = \mathfrak{w}^{lk}; \qquad \mathfrak{w}^{k'l'} = A_k^{k'}A_l^{l'}\mathfrak{w}^{kl} \qquad (3.5.3)$$

$$\sigma^{kl} \equiv P_\beta^{\alpha k}\overline{P}_\alpha^{\beta l} - P_\beta^{\alpha l}\overline{P}_\alpha^{\beta k} = -\sigma^{lk}; \qquad \sigma^{k'l'} = A_k^{k'}A_l^{l'}\sigma^{kl}, \qquad (3.5.4)$$

and other.
Triadic h-tensors in equations (3.4.11), apparently, could have a common genetic origin: they could be connected to a single h-tensor $\Lambda_\beta^{\alpha k}$. The even-symmetric tensor $\mathfrak{w}^{kl}$ then can be defined as follows:

$$\mathfrak{w}^{kl} \equiv \Lambda_\beta^{\alpha k}\Lambda_\alpha^{\beta l}. \qquad (3.5.5)$$

3. Covariant vector objects.
In our considerations at this point, such objects are connected to derivatives of matter field:

$$\partial_k(\Phi_\alpha\Psi^\alpha); \qquad (3.5.6)$$

$$\Phi_\alpha\mathfrak{D}_k\Psi^\alpha; \qquad \Psi^\alpha\mathfrak{D}_k\Phi_\alpha. \qquad (3.5.7)$$

4. *Associate dual objects* can be composed by contracting vectors and tensors and hybrid tensors with tensor $\mathfrak{w}^{kl}$ or tensor $\mathfrak{w}_{kl}$ inverse to $\mathfrak{w}^{kl}$ (procedures known as "lifting" or "lowering" indices), for instance:

$$P_{\beta k}^{\alpha} \equiv P_\beta^{\alpha l}\mathfrak{w}_{kl}, \text{ etc.} \qquad (3.5.8)$$

Note that, introduction of tensor $\mathfrak{w}_{kl}$ inverse to $\mathfrak{w}^{kl}$:

$$\mathfrak{w}_{km}\mathfrak{w}^{lm} = \delta_k^l$$

implies a condition

$$det\mathfrak{w}^{kl} \neq 0.$$

This necessary property is maintained at arbitrary non-degenerate transformation $|A_k^{k'}| \neq 0$:

$$|\mathfrak{w}^{k'l'}| = |A_k^{k'}||A_l^{l'}||\mathfrak{w}^{kl}| = |\mathfrak{w}^{kl}||A|^2. \tag{3.5.9}$$

By use of tensor $\mathfrak{w}^{kl}$, one can built two particular norms, scalar forms:

$$\mathbb{N}_\mathfrak{p} = \mathfrak{p}_k \mathfrak{p}_l \mathfrak{w}^{kl}; \qquad \mathbb{N}_\mathfrak{q} = \mathfrak{q}^k \mathfrak{q}^l \mathfrak{w}_{kl}. \tag{3.5.10}$$

5. Existence of tensor $\mathfrak{w}^{kl}$ also allows one recognize the following valence 2 tensor $\mathfrak{w}^{\alpha\beta}$ in MF space:

$$\mathfrak{w}^{\alpha\beta} = \frac{\partial \check{\varphi}^\alpha}{\partial \check{\psi}^k} \frac{\partial \check{\varphi}^\beta}{\partial \check{\psi}^l} \mathfrak{w}^{kl} = \mathfrak{w}^{\beta\alpha}; \tag{3.5.11}$$

$$\mathfrak{w}^{\alpha\gamma}\mathfrak{w}_{\beta\gamma} = \delta^\alpha_\beta, \tag{3.5.12}$$

and the correspondent particular norms as the following scalar forms:

$$\mathbb{N}_\Psi \equiv \Psi^\alpha \Psi^\beta \mathfrak{w}_{\alpha\beta}; \qquad \mathbb{N}_\Phi \equiv \Phi_\alpha \Phi_\beta \mathfrak{w}^{\alpha\beta} \tag{3.5.13}$$

Tensor type objects $\mathfrak{w}_{kl}$ and $\mathfrak{w}_{\alpha\beta}$ could be envisioned as playing role of metric tensors in the UM and MF space, respectively. Note that, they have been introduced not a prior but based on the triadic hybrid tensors $P_\beta^{k\alpha}$, $\bar{P}_\beta^{k\alpha}$ those to be determined by a dynamic connection to DSV.

### 3.6. Specification of DSV equations for existence of a Conservative Supercurrent

Existence of a *conservative vector current* $\mathcal{J}^k$ in UM space (*supercurrent*) should be an intrinsic property of the derived equations for DSV as an attribute of the above mentioned principle of the *dynamical existence*. Vector current associated with DSV can be presented in the following form:

$$\mathcal{J}^k = \Lambda_\beta^{\alpha k} \Phi_\alpha \Psi^\beta \tag{3.6.1}$$

with undefined h-tensor $\Lambda_\beta^{\alpha k} \equiv \mathbf{\Lambda}^k$. Further, let us represent h-tensors $\mathbf{P}^k$ and $\bar{\mathbf{P}}^k$ in equations (3.4.11) in the following way, introducing an undefined matrices (s-tensors) $\boldsymbol{\lambda} \equiv \lambda_\beta^\alpha$ and $\bar{\boldsymbol{\lambda}} \equiv \bar{\lambda}_\beta^\alpha$:

$$\mathbf{P}^k = (\mathbf{1} + \boldsymbol{\lambda})^{-1} \mathbf{\Lambda}^k; \qquad \bar{\mathbf{P}}^k = -\mathbf{\Lambda}^k (\mathbf{1} + \bar{\boldsymbol{\lambda}})^{-1}. \tag{3.6.2}$$

Drawing then the requirement of a *conservative* current:

$$\nabla_k \mathcal{J}^k \equiv \frac{1}{\sqrt{w}} \partial_k (\sqrt{w} \mathcal{J}^k) = 0; \quad w = |det w_{kl}| \tag{3.6.3}$$

(here $w_{kl}$ is metric tensor of UM, the *Grand Metric*) and using equations (3.4.11), we find the following equation:

$$\boldsymbol{\lambda} - \bar{\boldsymbol{\lambda}} = \mathfrak{D}_k \boldsymbol{\Lambda}^k; \qquad (3.6.4)$$

here

$$\mathfrak{D}_k \boldsymbol{\Lambda}^k \equiv \frac{1}{\sqrt{w}} \partial_k (\sqrt{w} \boldsymbol{\Lambda}^k) + [\mathcal{A}_k; \boldsymbol{\Lambda}^k], \qquad (3.6.5)$$

symbol [; ] means commutator of two matrices. Applying a symmetry consideration when comparing reduction of $\mathbf{P}^k$ and $\bar{\mathbf{P}}^k$ in (3.6.2), we can consider two possible relations between $\boldsymbol{\lambda}$ and $\bar{\boldsymbol{\lambda}}$ :

$$\bar{\boldsymbol{\lambda}} = \pm \boldsymbol{\lambda}.$$

Choice of $\bar{\mathbf{d}} = -\mathbf{d}$ leads to the following specification of DSV equations:

$$\boldsymbol{\Lambda}^k \mathfrak{D}_k \boldsymbol{\Psi} + \left(\frac{1}{2}\mathfrak{D}_k \boldsymbol{\Lambda}^k + \mathbf{1}\right)\boldsymbol{\Psi} = 0; \qquad (\mathfrak{D}_k \boldsymbol{\Phi})\boldsymbol{\Lambda}^k + \boldsymbol{\Phi}\left(\frac{1}{2}\mathfrak{D}_k \boldsymbol{\Lambda}^k - \mathbf{1}\right) = 0. \qquad (3.6.6)$$

Choice $\bar{\boldsymbol{\lambda}} = \boldsymbol{\lambda}$ would lead to an additional requirement $\mathfrak{D}_k \boldsymbol{\Lambda}^k = 0$. Both options reveal a conservative supercurrent (3.6.1), and in both the couple of initially introduced triadic tensors $\mathbf{P}^k, \bar{\mathbf{P}}^k$ is reduced to a single one, $N$ matrices $\boldsymbol{\Lambda}^k$. However, there is no a substantial motivation for condition $\mathfrak{D}_k \boldsymbol{\Lambda}^k = 0$, at least, before that the equations that connect the triadic objects $\boldsymbol{\Lambda}^k$ and $\mathcal{A}_k$ to DSV are derived and investigated. Option $\bar{\boldsymbol{\lambda}} = -\boldsymbol{\lambda}$ is preferred since it is free of unmotivated assumptions about properties of matrices $\boldsymbol{\Lambda}^k$.

### 3.7. Constraints of the Triads − DSV Coupling

Differential law for dual vector field includes two necessary types of triadic object, h-tensors $P_\beta^{k\alpha}$, $\bar{P}_\beta^{k\alpha}$ and affine h-tensor, *unified gauge field* $\mathcal{A}_{k\beta}^{\alpha}$. According to the *homogeneity* principle, these objects should not be considered as functions of manifold variables given or chosen *ad hoc*. They also cannot be constant, if not to admit trivialization of whole the theory and loss of its logical consistence. In particular, such "simplification" is not in a consistence with the *extended general relativity* principle, since a non-linear transformation of variables would make triadic objects non-constant.

To build a self-contained unified field theory, one has to establish coupling of triadic objects to DSV. The coupling should allow one finally explicate and utilize these objects on basis of s-vector objects and their derivatives. Apparently, the connections should have a form of the algebraic and differential relations. These equations should be formulated in terms of the possibly the lowest order derivatives (in accordance to the principle of *differential irreducibility*), and they preferably should not include any new objects beyond the collection exhibited in equations for DSV. Thus, we are coming to formulation of UFT as a *self-contained system of the lowest order derivative differential equations* for DSV and associated triadic objects, including the introduced affine h-tensor $\mathcal{A}_{k\beta}^{\alpha}$ .

Coupling between the triadic objects and matter field should be formulated as equations that might include or be associated with derivatives of the triads or, more generally, with derivatives of the h-tensors and UGF. As well as equations for the dualistic matter field DSV, these equations should satisfy the *extended general relativity* or *covariance* principle. Therefore, we have to consider *covariant extension of derivatives of h-tensors and the affine h-tensor.*

*Resume of Chapter 3*

In Chapter 3 we have being exploring a path to a Unified Field Theory as a system of real covariant differential equations for Dual State Vector (DSV) $\Psi^\alpha, \Phi_\alpha$ (Greek indices are related to the *µ*-dimensional *Matter Function* space) and related coefficient functions, all as functions of variables of a real *N*-dimensional manifold (Roman indices). Differential equations for DSV have been profiled as relations between DSV and its derivatives based on a set of principles or the requirements logically suited for UFT. Imposing a requirement of the *extended covariance*, we have pointed out the necessity to introduce two different types of the coefficient functions, mixed-valence *hybrid* triadic objects or matrices: a *hybrid triadic tensor* $\Lambda_\beta^{\alpha k}$, an extended analog of Dirac's matrices, and a *hybrid affine tensor* or *unified gauge field* $\mathcal{A}_{\beta k}^\alpha$, an extended analog of Christoffel symbols of Rimannian geometry used in *General Theory of Relativity* and gauge fields of QFT. Transformation laws of the hybrid objects include both differential matrices *A* and *B* of transformations of variables in Unified Manifold and Matter Function spaces, respectively.

In order to attain a consistent unified theory, the hybrid triadic objects as coefficient functions of DSV equations should be connected to DSV by other covariant differential equations. To explore the ways to build this coupling, we have to consider possibilities of building up *covariant derivatives* of the *hybrid objects*.

# 4. Covariant Derivatives of the Hybrid Objects

## 4.1. Covariant Derivatives of the Hybrid Tensors

### 4.1.1. The Affine Connection

*Constraint of covariance for derivatives of h-tensors*

Now, let us consider transformation of derivatives of an h-tensor $t_{\alpha k}$. The transformation equation now contains derivatives of transformation matrix $A_k^{k'}$ of manifold variables:

$$\partial_l t_{\alpha k} = A_l^{l'} A_k^{k'} B_\alpha^{\alpha'} \partial_{l'} t_{\alpha' k'} + A_k^{k'} (\partial_l B_\alpha^{\alpha'}) t_{\alpha' k'} + B_\alpha^{\alpha'} (\partial_l A_k^{k'}) t_{\alpha' k'} . \qquad (4.1.1)$$

Using definition of differential matrix $A$ as

$$A_k^{k'} \equiv \frac{\partial \check{\psi}^{k'}}{\partial \check{\psi}^k},$$

one can replace derivatives of matrix $A_k^{k'}$ in equation (4.1.1) by the second derivatives of transformation function:

$$\partial_l t_{\alpha k} = A_l^{l'} A_k^{k'} B_\alpha^{\alpha'} \partial_{l'} t_{\alpha' k'} + A_k^{k'} (\partial_l B_\alpha^{\alpha'}) t_{\alpha' k'} + B_\alpha^{\alpha'} \frac{\partial^2 \check{\psi}^{k'}}{\partial \check{\psi}^l \partial \check{\psi}^k} t_{\alpha' k'} . \qquad (4.1.2)$$

Acting further similar to the above produced covariant extension for derivative of object $\Phi_\alpha$, let us compliment derivatives $\partial_l t_{\alpha k}$ now by two terms linear on the h-tensor coordinates with two different coefficient functions, $\mathcal{A}_{\alpha l}^{\beta}$ and $G_{kl}^m$ as follows:

$$\partial_l t_{\alpha k} \to \mathfrak{D}_l t_{\alpha k} \equiv \partial_l t_{\alpha k} - \mathcal{A}_{\alpha l}^{\beta} t_{\beta k} - G_{kl}^m t_{\alpha m} , \qquad (4.1.3)$$

and impose the requirement to combined object, $\mathfrak{D}_l t_{\alpha k}$, to be transformed as an h-tensor (e.g. to be an h-tensor):

$$\mathfrak{D}_l t_{\alpha k} = A_l^{l'} A_k^{k'} B_\alpha^{\alpha'} \mathfrak{D}_{l'} t_{\alpha' k'} . \qquad (4.1.4)$$

Requirement (4.1.4) leads immediately to the necessity of the following transformation laws for the introduced objects:

$$B^{\alpha}_{\alpha'}\mathcal{A}^{\alpha'}_{\beta'k'} = B^{\beta}_{\beta'}A^{k}_{k'}\mathcal{A}^{\alpha}_{\beta k} + \partial_{k'}B^{\alpha}_{\beta'} ; \tag{4.1.5}$$

$$A^{l}_{l'}G^{l'}_{m'k'} = A^{m}_{m'}A^{k}_{k'}G^{l}_{mk} + \partial_{k'}A^{l}_{m'} . \tag{4.1.6}$$

Transformation law for object $\mathcal{A}^{\alpha}_{\beta k}$, obviously, is the same as above derived for the affine h-tensor when considering covariant derivatives of DSV. Transformation law (4.1.6) for *affine tensor* $G^{l}_{mk}$, naturally, is similar in form to that of the affine h-tensor $\mathcal{A}^{\beta}_{\alpha k}$, replacing matrix $B$ as for transformation of the s-field by matrix $A$ as for transformation of the manifold variables. An important difference, however, is that, term with derivatives of matrix $A$ in law (4.1.6) can be expressed through second derivatives of transformation functions $\check{\psi}(\check{\psi}')$:

$$A^{l}_{l'}G^{l'}_{m'k'} = A^{k}_{k'}A^{m}_{m'}G^{l}_{mk} + \frac{\partial^{2}\check{\psi}^{l}}{\partial\check{\psi}^{k'}\partial\check{\psi}^{m'}} , \tag{4.1.7}$$

so the transformation law for affine tensor $G^{l}_{mk}$ is symmetric relatively the down indices, both Roman.

Thus, we obtained definition of covariant derivative for a hybrid tensor $t_{\alpha k}$ :

$$\mathfrak{D}_{l}t_{\alpha k} \equiv \partial_{l}t_{\alpha k} - \mathcal{A}^{\beta}_{\alpha l}t_{\beta k} - G^{m}_{kl}t_{\alpha m} . \tag{4.1.8}$$

In similar way, one can find covariant derivatives for hybrid tensors $t^{\alpha}_{k}$, $t^{\alpha k}$ and $t^{k}_{\alpha}$ :

$$\mathfrak{D}_{l}t^{\alpha}_{k} \equiv \partial_{l}t^{\alpha}_{k} + \mathcal{A}^{\alpha}_{\beta l}t^{\beta}_{k} - G^{m}_{kl}t^{\alpha}_{m} ; \tag{4.1.9}$$

$$\mathfrak{D}_{l}t^{\alpha k} \equiv \partial_{l}t^{\alpha k} + \mathcal{A}^{\alpha}_{\beta l}t^{\beta k} + G^{k}_{ml}t^{\alpha m}; \tag{4.1.10}$$

$$\mathfrak{D}_{l}t^{k}_{\alpha} \equiv \partial_{l}t^{k}_{\alpha} - \mathcal{A}^{\beta}_{\alpha l}t^{k}_{\beta} + G^{k}_{ml}t^{m}_{\alpha}. \tag{4.1.11}$$

Indeed, it is straightforward to check that cancellation of terms with derivatives of both transformation matrices $A$ and $B$ takes place at arbitrary transformation of variables, once it is assumed that objects $G^{m}_{lk}$ and $\mathcal{A}^{\alpha}_{\beta l}$ transform in accordance to the law (4.1.6) and (4.1.5), respectively, so that:

$$\mathfrak{D}_{l}t^{\alpha}_{k} = A^{l'}_{l}A^{k'}_{k}B^{\alpha}_{\alpha'}\mathfrak{D}_{l'}t^{\alpha'}_{k'} ;$$

$$\mathfrak{D}_{l'}t^{\alpha'k'} = A^{l}_{l'}A^{k'}_{k}B^{\alpha'}_{\alpha}\mathfrak{D}_{l}t^{\alpha k} ; \qquad \mathfrak{D}_{l'}t^{k'}_{\alpha'} = A^{l}_{l'}A^{k'}_{k}B^{\alpha}_{\alpha'}\mathfrak{D}_{l}t^{k}_{\alpha} .$$

Formulas (4.1.3) and (4.1.8) – (4.1.11) determine covariant derivatives of h-tensors of the lowest valence. Covariant extension of the derivatives of higher valence h-tensors is obvious:

$$\mathfrak{D}_{l}t^{\alpha k}_{\beta} \equiv \partial_{l}t^{\alpha k}_{\beta} + \mathcal{A}^{\alpha}_{\gamma l}t^{\gamma k}_{\beta} - \mathcal{A}^{\gamma}_{\beta l}t^{\alpha k}_{\gamma} + G^{k}_{ml}t^{\alpha m}_{\beta} ;$$

$$\mathfrak{D}_{l}t^{\alpha}_{\beta k} \equiv \partial_{l}t^{\alpha}_{\beta k} + \mathcal{A}^{\alpha}_{\gamma l}t^{\gamma}_{\beta k} - \mathcal{A}^{\gamma}_{\beta l}t^{\alpha}_{\gamma k} - G^{m}_{kl}t^{\alpha}_{\beta m} ; \tag{4.1.12}$$

$$\mathfrak{D}_l t_\beta^{\alpha km} \equiv \partial_l t_\beta^{\alpha km} + \mathcal{A}_{\gamma l}^\alpha t_\beta^{\gamma km} - \mathcal{A}_{\beta l}^\gamma t_\gamma^{\alpha km} + G_{nl}^k t_\beta^{\alpha nm} + G_{nl}^m t_\beta^{\alpha kn}, \quad \text{etc.} \tag{4.1.13}$$

*Covariant derivatives of vectors and tensors*

Apparently, covariant derivatives of vector and tensors can be defined in the following way:

$$\mathcal{D}_l U^k \equiv \partial_l U^k + G_{ml}^k U^m; \tag{4.1.14}$$

$$\mathcal{D}_l V_k \equiv \partial_l V_k - G_{kl}^m V_m; \tag{4.1.15}$$

$$\mathcal{D}_k t^{lm} \equiv \partial_k t^{lm} + G_{nk}^l t^{nm} + G_{nk}^m t^{ln}, \tag{4.1.16}$$

$$\mathcal{D}_k t_{lm} \equiv \partial_k t_{lm} - G_{lk}^n t_{nm} - G_{mk}^n t_{ln}, \tag{4.1.17}$$

$$\mathcal{D}_k t_l^m \equiv \partial_k t_l^m + G_{nk}^m t_l^n - G_{lk}^n t_n^m, \text{ etc.} \tag{4.1.18}$$

Note that, these formulas can also be obtained by contracting on Greek indices of covariant derivatives of the correspondent h-tensors.

Formulas for covariant derivatives of a vector and co-vector, in particular, can also be applied to the above introduced tangent vector of a line $q^k$ and its dual partner $\mathfrak{p}_k$ (see *Prolegomena*).

*Covariant derivatives of the products*

Apparently, the following general formulas are valid for both kinds of covariant derivatives of products of tensors, h-tensors, etc., same as for the ordinary derivatives:

$$\mathcal{D}_k(t_1 t_2) = t_1 \mathcal{D}_k t_2 + t_2 \mathcal{D}_k t_1; \quad \mathfrak{D}_k(t_1 t_2) = t_1 \mathfrak{D}_k t_2 + t_2 \mathfrak{D}_k t_1, \text{ etc.} \tag{4.1.19}$$

Triadic object $G_{kl}^m$ subordinated to the transformation law (5.1.7) is generally known in the differential geometry as *affine connection* [20, 21]. Since the word "connection" can be used in a field theory in a more general context (including "interaction", etc.), we call this object an "affine tensor" (AT), following the terminology used by W. Pauli in the context of field theory [8]. This characterization reflects property of this object to behave as tensor only at linear transformations of the manifold variables.

In the differential geometry object $G_{kl}^m$ is conventionally introduced to a manifold based on the recourse to the pictorial paradigm of *parallel displacement (translation)* of vectors [20, 21]. In our approach to covariant field theory we do not resort to the notion of a translation, considering it as a heritage of a pictorial methodology. As one can see from the treatment above and below, the covariance can be acquired and affine tensor $G_{mk}^n$, as well as affine h-tensor $\mathcal{A}_{\alpha k}^\beta$, can be found within the frame of a pure differential methodology based on the requirements of consistence of the *fundamental differential law* relative to an arbitrary transformations of variables $\breve{\psi}$ i.e. requirement of *general covariance*.

A principal distinction of AT from a *hybrid affine tensor* (HAT) is that all three indices of AT are associated with matrix $A$ of transformation of UM variables $\breve{\psi}^k$, while in case of HAT there is only one index of this nature; two others are associate with transformation matrices *B* of the Dual State Vector. This "uniformity" of AT allows one to immediately explicate it, i.e. realize its general structure based on the requirement of covariance itself, not resorting to any other conditions or notions. To demonstrate this, we first have to consider general properties of AT that can be found based on the required transformation law (4.1.6).

*General properties of the affine tensor*

1. Association property.

In correspondence to the transformation law for tensors, the transformation law for AT is an *association* law: if the transformation from $\check{\psi}'$ to $\check{\psi}''$ follows the transformation from $\check{\psi}$ to $\check{\psi}'$, then the resulting transformation of the connection coincides with the immediate transformation from $\check{\psi}$ to $\check{\psi}''$ [20].

2. The local tensor equivalence.

If there exists an object satisfying transformation law (4.1.6), then this object changed by adding an arbitrary *tensor* of the same valence, satisfies the same law. In other words, the difference of two AT, $\Delta G_{kl}^m$, is tensor :

$$\Delta G_{k'l'}^{m'} = A_{k'}^k A_{l'}^l A_m^{m'} \Delta G_{kl}^m.$$

3. The skew-symmetric part of AT:

$$S_{kl}^m \equiv \frac{1}{2}(G_{kl}^m - G_{lk}^m) = -S_{lk}^m \qquad (4.1.20)$$

is tensor, as this immediately follows from transformation equation (4.1.6):

$$S_{k'l'}^{m'} = A_{k'}^k A_{l'}^l A_m^{m'} S_{kl}^m ;$$

tensor $2S_{kl}^m$ is known in the differential geometry as *torsion* [20, 21].

4. The even-symmetric part of an affine tensor:

$$G_{\{kl\}}^m \equiv \frac{1}{2}(G_{kl}^m + G_{lk}^m) \qquad (4.1.21)$$

satisfies transformation law (4.1.6), i.e. it is an AT, as well.

5. General decomposition of affine tensor can be produced based on properties (4.1.19) and (4.1.21): if there exists a particular symmetric form

$$\acute{G}_{kl}^m = \acute{G}_{lk}^m ,$$

which satisfies transformation law (4.1.6 ), then one can define a general AT form as

$$G_{kl}^m = \acute{G}_{kl}^m + \acute{T}_{kl}^m,$$

where $\acute{T}_{kl}^m$ is an arbitrary, unspecified *tensor*.

6. Local equivalence to *torsion*.

It is proven in the differential geometry, that an even-symmetric AT can be turned to zero at a point (and even along a line) by a specific non-linear transformation of variables. The proof is based just on transformation law (4.1.7) [20]. Since the even-symmetric part of AT, $\acute{G}_{kl}^m$, satisfies transformation law (4.1.7) regardless to presence of torsion, $\acute{G}_{kl}^m$ possesses this property in general case. This *matter-fact* of the differential geometry made this discipline the mathematical background of relativistic theory of the gravitational field built by A. Einstein based on his *equivalence principle* [8].

### *4.1.2. Matched Connection*

Based on the forms of the covariant derivatives of valence 2 tensors, one can immediately explicate an affine tensor i.e. derive a form which is transformed *in fact* according to equation (4.1.7).

Consider some tensor $u_{mkl}$ symmetric in indices $k, l$ but arbitrary in the rest:

$$u_{mkl} = u_{mlk}. \qquad (4.1.22)$$

Let us try to represent this tensor as covariant derivative of an even-symmetric tensor $\mathfrak{w}_{kl} = \mathfrak{w}_{lk}$, with symmetric affine connection $\acute{G}_{kl}^n = \acute{G}_{lk}^n$:

$$u_{mkl} = \partial_m \mathfrak{w}_{kl} - \acute{G}_{km}^n \mathfrak{w}_{nl} - \acute{G}_{lm}^n \mathfrak{w}_{kn}. \qquad (4.1.23)$$

Tensor $\mathfrak{w}_{kl}$ can be assumed to be the inverse to the even-symmetric tensor (4.4.7):

$$\mathfrak{w}^{kl} \equiv P_\beta^{\alpha k} \overline{P}_\alpha^{\beta l} + P_\beta^{\alpha l} \overline{P}_\alpha^{\beta k}, \qquad (4.1.24)$$

assuming $|\mathfrak{w}^{kl}| \neq 0$.

Relations (4.1.23) can be considered as algebraic equations for object $\acute{G}_{km}^n$:

$$\acute{G}_{km}^n \mathfrak{w}_{ln} + \acute{G}_{lm}^n \mathfrak{w}_{kn} = \partial_m \mathfrak{w}_{kl} - u_{mkl} \equiv V_{mkl}. \qquad (4.1.25)$$

The solution is straightforward. Let us rewrite these equations as:

$$V_{mkl} = \acute{G}_{km}^n \mathfrak{w}_{ln} + \acute{G}_{lm}^n \mathfrak{w}_{kn}.$$

By performing a cyclic substitution of indices $k, l, m$, we obtain two more similar systems of equations:

$$V_{lmk} = \acute{G}_{ml}^n \mathfrak{w}_{kn} + \acute{G}_{kl}^n \mathfrak{w}_{mn};$$

$$V_{klm} = \acute{G}_{lk}^n \mathfrak{w}_{mn} + \acute{G}_{mk}^n \mathfrak{w}_{ln}.$$

Performing a summation of two of these three systems of equations and subtracting the third one, we obtain:
$$2\acute{G}^n_{kl}\mathfrak{w}_{mn} = V_{kml} + V_{lmk} - V_{mlk}.$$

So solution for $\acute{G}^k_{ml}$ is as follows:

$$\acute{G}^k_{ml} = \frac{1}{2}\mathfrak{w}^{kn}(V_{mnl} + V_{lnm} - V_{nlm}).$$

Now we pick term $\partial_m \mathfrak{w}_{kl}$ in $V_{mkl}$ and denote the corresponding solution part as $\Gamma^k_{ml}$, and the rest as $\acute{T}^k_{ml}$:

$$\acute{G}^k_{ml} = \Gamma^k_{ml} + \acute{T}^k_{ml};$$

$$\Gamma^n_{mk}\mathfrak{w}_{nl} + \Gamma^n_{ml}\mathfrak{w}_{kn} = \partial_m \mathfrak{w}_{kl}; \qquad (4.1.26)$$

$$\acute{T}^n_{mk}\mathfrak{w}_{nl} + \acute{T}^n_{ml}\mathfrak{w}_{kn} = u_{klm};$$

then we find:

$$\Gamma^k_{ml} = \frac{1}{2}\mathfrak{w}^{kn}(\partial_l \mathfrak{w}_{nm} + \partial_m \mathfrak{w}_{nl} - \partial_n \mathfrak{w}_{ml}) \qquad (4.1.27)$$

$$\acute{T}^k_{ml} = \frac{1}{2}\mathfrak{w}^{kn}(u_{mnl} + u_{lnm} - u_{lmn}) = \acute{T}^k_{lm}.$$

Form $\acute{T}^k_{ml}$ is tensor, while form $\Gamma^k_{ml}$ should be an affine tensor. Indeed, form $\Gamma^k_{ml}$ satisfies *in fact* transformation law (4.1.7), so it is an affine tensor. One then can introduce explicit tensor forms containing derivatives of vectors and tensor $\mathfrak{w}_{kl}$, *short covariant derivatives* of vector objects as:

$$\nabla_l U^k \equiv \partial_l U^k + \Gamma^k_{ml} U^m,$$

$$\nabla_l V_k \equiv \partial_l V_k - \Gamma^m_{kl} V_m,$$

as well as the short covariant derivatives of tensors, replacing symbol $\mathcal{D}_l$ by $\nabla_l$ in equations (4.1.14) – (4.1.18). Note that,

$$\nabla_m \mathfrak{w}_{kl} \equiv \partial_m \mathfrak{w}_{kl} - \Gamma^n_{mk}\mathfrak{w}_{nl} - \Gamma^n_{ml}\mathfrak{w}_{kn} \equiv 0, \qquad (4.1.28)$$

since object $\Gamma^n_{mk}$ is determined as a solution of equation (4.1.26). Note that, contraction of equation (4.1.27) on indices $m = k$ leads to the following relation:

$$2\Gamma^k_{lk} = \mathfrak{w}^{km}\partial_l \mathfrak{w}_{km} = \frac{\partial_l \mathfrak{w}}{\mathfrak{w}}; \quad \mathfrak{w} \equiv |\mathfrak{w}_{km}|; \qquad (4.1.29)$$

here we used a background algebraic relation valid for any non-degenerated tensor $w_{kl}$:

$$\partial_l w = w w^{km} \partial_l w_{km}; \quad w \equiv |w_{kl}| \neq 0. \tag{4.1.30}$$

Also, considering tensor $\mathbfit{w}_{kl}$ as even-symmetric part of tensor $w_{kl}$, one can derive the following relation:

$$\frac{\partial_l w}{2w} = \Gamma^k_{kl} + w_l; \quad w_l \equiv \frac{1}{2} w^{[km]} \nabla_l w_{[km]} \tag{4.1.31}$$

here $w_{[km]}$ and $w^{[km]}$ is the odd-symmetric part of tensor $w_{kl}$ and inverse tensor $w^{km}$, respective.

Form (4.1.27) is known in the differential geometry as *matched connection*, once symmetric tensor $v_{kl}$ is assumed to play the role of a *metric tensor* to define the *interval* or *length* and the *scalar product* of vectors (under a requirement that the scalar product is conserved at *parallel displacement*, etc.) [20]. In fact, none of these notions but simply the existence of a non-degenerated symmetric tensor $v_{kl}$ is actually inquired in order to derive form (4.1.27) as an affine tensor.

*General tensor-based explicit form of the affine tensor*

In accordance to property (4.1.19), the general affine tensor form can be represented as a sum of a matched connection (4.1.27) and an arbitrary tensor $\mathrm{T}^k_{ml}$:

$$\mathrm{G}^k_{ml} = \Gamma^k_{ml} + \mathrm{T}^k_{ml}. \tag{4.1.32}$$

Note that, the difference of matched connection forms built on different (non-degenerate) types of tensor $v_{kl}$ is a tensor. This follows from general property (4.1.19) of affine tensor.

### 4.1.3. Versified covariant derivatives of tensors and h-tensors

In accordance to general definition of affine tensor (4.1.32), one also may consider the *versified* covariant derivatives of tensors and h-tensors in which one or more terms with affine tensor $\mathrm{G}^n_{kl}$ are replaced by the correspondent terms with matched connection $\Gamma^n_{kl}$, for instance:

$$\breve{\mathcal{D}}_l t^m_k \equiv \partial_l t^m_k + \Gamma^m_{nl} t^n_k - \mathrm{G}^n_{kl} t^m_n, \tag{4.1.33}$$

$$\breve{\mathcal{D}}_l t^{km} \equiv \partial_l t^{km} + \Gamma^m_{nl} t^{kn} + \mathrm{G}^k_{nl} t^{nm}, \tag{4.1.34}$$

$$\breve{\mathcal{D}}_l t_{km} \equiv \partial_l t_{km} - \Gamma^n_{kl} t_{nm} - \mathrm{G}^n_{ml} t_{nk}, \quad \text{etc.} \tag{4.1.35}$$

$$\breve{\mathcal{D}}_l t^{\alpha k} \equiv \partial_l t^{\alpha k} + \Gamma^k_{ml} t^{\alpha m} + \mathcal{A}^\alpha_{\beta l} t^{\beta k}; \tag{4.1.36}$$

$$\breve{\mathcal{D}}_l t^\alpha_k \equiv \partial_l t^\alpha_k - \Gamma^m_{kl} t^\alpha_m + \mathcal{A}^\alpha_{\beta l} t^\beta_k; \tag{4.1.37}$$

$$\breve{\mathcal{D}}_l t^{\alpha k}_\beta \equiv \partial_l t^{\alpha k}_\beta + \Gamma^k_{ml} t^{\alpha m}_\beta + \mathcal{A}^\alpha_{\gamma l} t^{\gamma k}_\beta - \mathcal{A}^\gamma_{\beta l} t^{\alpha k}_\gamma; \tag{4.1.38}$$

$$\breve{\mathcal{D}}_l t^{\alpha km}_\beta \equiv \partial_l t^{\alpha km}_\beta + \Gamma^k_{nl} t^{\alpha nm}_\beta + \Gamma^m_{nl} t^{\alpha kn}_\beta + \mathcal{A}^\alpha_{\gamma l} t^{\gamma km}_\beta - \mathcal{A}^\gamma_{\beta l} t^{\alpha km}_\gamma, \quad \text{etc.} \tag{4.1.39}$$

Such versification is always possible, since difference of two affine tensors ($G_{kl}^n$ and $\Gamma_{kl}^n$ in this case) is tensor.

*Covariant divergences of h-tensors*

By contraction of index $l$ in equations (4.1.38), (4.1.39) with one of the upper Roman indices one can build *covariant divergences* of h-tensors:

$$\breve{\mathfrak{D}}_k t_\beta^{\alpha k} \equiv \partial_k t_\beta^{\alpha k} + \Gamma_{mk}^k t_\beta^{\alpha m} + \mathcal{A}_{\gamma k}^\alpha t_\beta^{\gamma k} - \mathcal{A}_{\beta k}^\gamma t_\gamma^{\alpha k} \ ; \tag{4.1.40}$$

$$\breve{\mathfrak{D}}_l t_\beta^{\alpha k l} \equiv \partial_l t_\beta^{\alpha k l} + \Gamma_{nl}^k t_\beta^{\alpha n l} + \Gamma_{nl}^l t_\beta^{\alpha k n} + \mathcal{A}_{\gamma l}^\alpha t_\beta^{\gamma k l} - \mathcal{A}_{\beta l}^\gamma t_\gamma^{\alpha k l} \ ; \text{ etc.} \tag{4.1.41}$$

Note that, in particular, form (4.1.40) is a mixed valence 2 s-tensor, while form (4.1.41) is a triadic h-tensor. In case of the odd symmetry of h-tensor $t_\beta^{\alpha k l}$ (i.e. $t_\beta^{\alpha k l} = -t_\beta^{\alpha l k}$), the second term in form (4.1.41) turns to zero, then we obtain the following result:

$$\breve{\mathfrak{D}}_l t_\beta^{\alpha k l} \Rightarrow \partial_l t_\beta^{\alpha k l} + \Gamma_{nl}^l t_\beta^{\alpha k n} + \mathcal{A}_{\gamma l}^\alpha t_\beta^{\gamma k l} - \mathcal{A}_{\beta l}^\gamma t_\gamma^{\alpha k l} \ . \tag{4.1.42}$$

*Necessity and possibility to exist for an even-symmetric tenor $\mathfrak{w}_{kl}$*

We can conclude our treatment of the affine tensors with an important statement.
Definition of covariant derivatives of the *vectors, tensors,* and *hybrid tensors* requires introduction of affine tensor $G_{mk}^l$ in addition to the hybrid affine tensor, $\mathcal{A}_{\beta k}^\alpha$. On the other hand, explication of the symmetric part of affine tensor, $\Gamma_{mk}^l = \Gamma_{km}^l$, responsible for covariant extension of these derivatives, requires the existence of a non-degenerated symmetric tensor $\mathfrak{w}_{kl} = \mathfrak{w}_{lk}$ as a necessary and sufficient condition. Presumably, this tensor can be realized as structured on the triadic hybrid tensors as shown by equations (4.5.3) or (4.5.5). We may conclude that, in a field theory where DSV plays a pilot role as a matter field, finding a coupling of this tensor to DSV may not arrive as a constraint additional to establishing coupling of the triadic h-tensors to DSV.

*General notations for covariant derivatives*

In this paper (section 5.5) we eventually will use generalized notations for covariant derivatives of vector and tensor type objects as follows:

$$\widehat{\nabla}_l t_\mathfrak{b}^a \equiv \partial_l t_\mathfrak{b}^a + \widehat{\Gamma}_{cl}^a t_\mathfrak{b}^c - \widehat{\Gamma}_{\mathfrak{b}l}^c t_c^\alpha \ ; \tag{4.1.44}$$

$$\breve{\nabla}_l t_\mathfrak{b}^{a k} \equiv \partial_l t_\mathfrak{b}^{a k} + \Gamma_{ml}^k t_\mathfrak{b}^{a m} + \widehat{\Gamma}_{cl}^a t_\mathfrak{b}^{c k} - \widehat{\Gamma}_{\mathfrak{b}l}^c t_c^{a k} \ ; \tag{4.1.45}$$

$$\breve{\nabla}_l t_\mathfrak{b}^{a k m} \equiv \partial_l t_\mathfrak{b}^{a k m} + \Gamma_{nl}^k t_\mathfrak{b}^{a n m} + \Gamma_{nl}^m t_\mathfrak{b}^{a k n} + \widehat{\Gamma}_{cl}^a t_\mathfrak{b}^{c k m} - \widehat{\Gamma}_{\mathfrak{b}l}^c t_c^{a k m} \ , \text{etc.} \tag{4.1.46}$$

Here notations $t_\mathfrak{b}^a$, $t_\mathfrak{b}^{ak}$, $t_\mathfrak{b}^{akm}$ are generally for vector, s-vector, tensor and h-tensor objects; notations $\widehat{\Gamma}_{\mathfrak{b}l}^c$ for connection objects (either for affine tensor or affine h-tensor); $\Gamma_{ml}^k$ for *Matched Connection* (4.1.27). The Roman indices $k, l, m, ...$ always are associated with symbols of manifold variables (in connection to derivatives). The script symbols $a, b, c, ...$ indicate the objects' "coordinates" (i.e. components); they belong either to Greek or Roman group of indices.

## 4.2. Covariant Derivatives of the Unified Gauge Field

### 4.2.1. Constraints of covariant extension for derivatives of UGF

Affine tensor $G_{kl}^m$ has an immediate intrinsic explication (4.1.32) based on tensor forms, namely, its essential part, *Matched Connection* $\Gamma_{kl}^m = \Gamma_{lk}^m$ is expressed through derivatives of an even-symmetric valence 2 non-degenerate tensor $\mathfrak{w}_{kl} = \mathfrak{w}_{lk}$. In this context, there is no necessity-in-principle in finding out an additional specific connection of this object to the basic tensor type objects of a differential system of a UFT; instead, it might be enough to establish structure of tensor $\mathfrak{w}_{kl}$ basing on these objects (for instance, as shown by equation (4.1.24).

Such disposition, however, does not take place with respect to the affine h-tensor $\mathcal{A}_{\beta k}^{\alpha}$. This object is introduced due to the necessity of covariant extension for derivatives of the Dual State Vector field $\Psi^\alpha, \Phi_\alpha$ which presumably is transformed with a matrix $B$ other than matrix $A$ as for transformation of differentials of the Unified Manifold variables. Correspondingly, indices $\alpha$ and $\beta$ of the triadic object $\mathcal{A}_{\beta k}^{\alpha}$ are related to transformation matrix $B$, while index $k$ is associated with matrix $A$. Therefore, transposition procedure $\beta \rightleftarrows k$ together with notion of symmetry (either even or skew) on down indices is not applicable to this object (as well as with respect to h-tensors $P_\alpha^{\beta k}$ and $\bar{P}_\alpha^{\beta k}$ concerning the upper indices). So we have to find out other way to connect the affine h-tensor $\mathcal{A}_{\beta k}^{\alpha}$ to the dual state vector and h-tensors. A step to finding of this connection might consist of establishing a *covariant derivative* of object $\mathcal{A}_{\beta k}^{\alpha}$ itself:

$$\partial_l \mathcal{A}_{\beta k}^{\alpha} \longrightarrow \widehat{\nabla}_l \mathcal{A}_{\beta k}^{\alpha}.$$

Covariant extension of derivative $\partial_l \mathcal{A}_{\beta k}^{\alpha}$ should turn it to a hybrid tensor form, possibly avoiding introduction of any new objects in its structure. This form could be associated with the h-tensor forms already profiled.

Such covariant extension occurs uniquely derivable based on analogy to structure of a fundamental object established in the differential geometry, the *Riemann-Christoffel curvature form* (RCF).

### 4.2.2. Riemann-Christoffel Form as covariant derivative of the Affine Tensor

The Riemann-Christoffel form (RCF) is structured on the affine tensor $G_{mk}^n$ and its derivatives [8, 9, 20, 21]:

$$\mathcal{R}_{mkl}^n \equiv \partial_k G_{ml}^n + G_{pk}^n G_{ml}^p - \partial_l G_{mk}^n - G_{pl}^n G_{mk}^p = -\mathcal{R}_{mlk}^n. \qquad (4.2.1)$$

This form is transformed as a tensor, once object $G_{mk}^n$ is considered as an *affine tensor* i.e. transformed according to transformation law (5.1.7) [20, 21]. Vice versa, if form $\mathcal{R}_{mkl}^n$ is considered a tensor, then object $G_{mk}^n$ transforms according to law (4.1.7).

RCF is usually introduced based on paradigm of "*parallel displacement*" ("translation") of vectors and tensors [20]. On the other hand, as known, this form can be immediately and simply recognized by considering the second covariant derivatives of vector functions $U^m$ or $V_m$ [21]:

$$\mathcal{D}_{kl}U^m \equiv \mathcal{D}_k(\mathcal{D}_l U^m) = \partial_k(\mathcal{D}_l U^m) + G_{nk}^m \mathcal{D}_l U^n - G_{lk}^n \mathcal{D}_n U^m,$$

$$\mathcal{D}_{kl}V_m \equiv \mathcal{D}_k(\mathcal{D}_l V_m) = \partial_k(\mathcal{D}_l V_m) - G_{mk}^n \mathcal{D}_l V_n - G_{lk}^n \mathcal{D}_n V_m.$$

Once $\mathcal{D}_l U^m$ and $\mathcal{D}_l V_m$ are tensors, so are these two forms. Next, one can calculate the *alternated* second covariant derivatives:

$$\mathcal{D}_{[kl]}U^m \equiv \mathcal{D}_k(\mathcal{D}_l U^m) - \mathcal{D}_l(\mathcal{D}_k U^m\ ;$$

$$\mathcal{D}_{[kl]}V_k \equiv \mathcal{D}_k(\mathcal{D}_l V_m) - \mathcal{D}_l(\mathcal{D}_k V_m).$$

The calculations result in the following formulas:

$$\mathcal{D}_{[kl]}U^m = \mathcal{R}_{nkl}^m U^n + G_{[kl]}^n \mathcal{D}_n U^m, \qquad (4.2.2)$$

$$\mathcal{D}_{[kl]}V_m = -\mathcal{R}_{mkl}^n V_n + G_{[kl]}^n \mathcal{D}_n V_m\ ; \qquad (4.2.3)$$

here $G_{[kl]}^n \equiv G_{kl}^n - G_{lk}^n$ is the above mentioned *torsion tensor,* while $\mathcal{R}_{mkl}^n$ is a notation for form (4.2.1). It is important that the same, *unique* RCF tensor results in a bundle with co-and contra-vector components in the expression for the second covariant derivatives.

An intrinsic property of RCF is that it is odd-symmetric on indices $k, l$:

$$\mathcal{R}_{mkl}^n = -\mathcal{R}_{mlk}^n\ .$$

Reflecting this property of RCF, it is convenient to represent it as resulting from a form

$$\check{\mathcal{R}}_{mkl}^n \equiv \partial_k G_{ml}^n + G_{pk}^n G_{ml}^p$$

by alternating this form on indices $k, l$:

$$\mathcal{R}_{mkl}^n = \check{\mathcal{R}}_{m[kl]}^n \equiv \check{\mathcal{R}}_{mkl}^n - \check{\mathcal{R}}_{mlk}^n\ .$$

Note that, form $\check{\mathcal{R}}_{mkl}^n$, together with $G_{mk}^n$, belongs the class of the *affine tensors*, since it does not transform as tensor at the non-linear transformations of UM variables. In contrary to this, form $\mathcal{R}_{mkl}^n$ is tensor, once object $G_{mk}^n$ is transformed according to law (4.1.7). Form $\mathcal{R}_{mkl}^n$ is structured on the affine tensor $G_{mk}^n$ being linear on its derivatives; so it can be considered as *covariant (tensor) derivative of the affine tensor* $G_{mk}^n$:

$$\partial_l G_{mk}^n \to \widehat{\nabla}_l G_{mk}^n \Rightarrow \equiv \mathcal{R}_{mkl}^n = -\mathcal{R}_{mlk}^n\ .$$

Thus, covariant extension of AT derivatives can be defined only with respect to the *alternated* derivatives:

$$\partial_k G_{ml}^n \rightarrow \partial_k G_{ml}^n - \partial_l G_{mk}^n \rightarrow \check{\mathcal{R}}_{m[kl]}^n \equiv \mathcal{R}_{mkl}^n.$$

*General explication of RCF*

With explication of affine tensor according to equation (4.1.32), form (4.2.1) can be split in two forms as follows:

$$\mathcal{R}_{mkl}^n = R_{mkl}^n + \mathcal{T}_{mkl}^n ; \qquad (4.2.4)$$

here $R_{mkl}^n$ is a particular $\mathcal{R}_{mkl}^n$ form which is structured on *Matched Connection* (4.1.27):

$$R_{mkl}^n = \check{R}_{m[kl]}^n = \check{R}_{mkl}^n - \check{R}_{mlk}^n ; \qquad \check{R}_{mkl}^n \equiv \partial_k \Gamma_{ml}^n + \Gamma_{pk}^n \Gamma_{ml}^p , \qquad (4.2.5)$$

while other part in (4.2.4) is notation for the following tensor:

$$\mathcal{T}_{mkl}^n = \check{\mathcal{T}}_{m[kl]}^n = \check{\mathcal{T}}_{mkl}^n - \check{\mathcal{T}}_{mlk}^n ; \qquad \check{\mathcal{T}}_{mkl}^n \equiv \nabla_k T_{ml}^n + T_{pk}^n T_{ml}^p ; \qquad (4.2.6)$$

$$\nabla_k T_{ml}^n \equiv \partial_k T_{ml}^n + \Gamma_{pk}^n T_{ml}^p - \Gamma_{mk}^p T_{pl}^n - \Gamma_{lk}^p T_{mp}^n .$$

Form $R_{mkl}^n$ is known in differential geometry as *Riemann-Christoffel* or *curvature tensor* (RCT) and traditionally used in GTR [8].

It can be convenient to use a matrix representation for RCF and RCT. Introducing notations:

$$\mathbf{G}_k \equiv G_{mk}^n ; \quad \mathbf{\Gamma}_k \equiv \Gamma_{mk}^n ; \quad \mathbf{T}_k \equiv T_{mk}^n \qquad (4.2.7)$$

as for matrices on indices $m, n$, we can consider $\mathcal{R}_{mkl}^n$ and $R_{mkl}^n$ also as matrices $\boldsymbol{\mathcal{R}}_{kl}$ and $\boldsymbol{R}_{kl}$ structured on $\mathbf{G}_k$ and $\mathbf{\Gamma}_k$ and their derivatives as follows:

$$\mathcal{R}_{mkl}^n \equiv \boldsymbol{\mathcal{R}}_{kl} = \check{\boldsymbol{\mathcal{R}}}_{kl} - \check{\boldsymbol{\mathcal{R}}}_{lk} = \partial_k \mathbf{G}_l - \partial_l \mathbf{G}_k + [\mathbf{G}_k; \mathbf{G}_l] ; \qquad \check{\boldsymbol{\mathcal{R}}}_{kl} \equiv \partial_k \mathbf{G}_l + \mathbf{G}_k \mathbf{G}_l \qquad (4.2.8)$$

$$R_{mkl}^n \equiv \boldsymbol{R}_{kl} = \check{\boldsymbol{R}}_{kl} - \check{\boldsymbol{R}}_{lk} = \partial_k \mathbf{\Gamma}_l - \partial_l \mathbf{\Gamma}_k + [\mathbf{\Gamma}_k; \mathbf{\Gamma}_l] ; \qquad \check{\boldsymbol{R}}_{kl} \equiv \partial_k \mathbf{\Gamma}_l + \mathbf{\Gamma}_k \mathbf{\Gamma}_l \qquad (4.2.9)$$

here symbol [ ; ] means commutator of two matrices.

*Properties of RCF*

1. Zero trace of matrix $\boldsymbol{R}_{kl}$.
   Tensor (4.2.9) possesses a fundamental property [8, 9, 15, 20, 21]:

$$R_{mkl}^m \equiv Tr \boldsymbol{R}_{kl} = \partial_k \Gamma_{ml}^m - \partial_l \Gamma_{mk}^m \equiv 0 , \qquad (4.2.10)$$

which can be easily proved by use of the explicit $\Gamma_{mk}^n$ form (4.1.27) and taking into account the background relation (4.1.29).

2. Consequently:

$$\mathcal{R}_{nkl}^n \equiv Tr\boldsymbol{\mathcal{R}}_{kl} = Tr(\partial_k \mathbf{T}_l - \partial_l \mathbf{T}_k) = \partial_k \mathrm{T}_l - \partial_l \mathrm{T}_k \equiv \mathcal{R}_{kl} = -\mathcal{R}_{lk}; \qquad (4.2.11)$$

here

$$\mathrm{T}_k \equiv Tr\mathbf{T}_k = \mathrm{T}_{nk}^n.$$

3. Ricci tensor.

Based on RCT form (4.2.5), one can introduce a unique valence 2 even-symmetric tensor, *Ricci tensor*, by contracting the upper index of RCT with one of two down indices on which this tensor is odd-symmetric [20]:

$$R_{mk} \equiv R_{mkl}^l = \partial_k \Gamma_{ml}^l - \partial_l \Gamma_{mk}^l + \Gamma_{pk}^l \Gamma_{ml}^p - \Gamma_{pl}^l \Gamma_{mk}^p = R_{km}. \qquad (4.2.12)$$

Even symmetry of the first term in this tensor form is easy to recognize using relations (4.1.29); the even symmetry of the rest is obvious.

### *4.2.3. Hybrid Curvature Form as covariant derivative of UGF*

The recognition of curvature form (4.2.1) as covariant derivative of the affine tensor $G_{kl}^m$ prompts a way to establish a covariant extension for derivatives of affine h-tensor $\mathcal{A}_{\beta k}^\alpha$. Namely, similar to that how RCF is detached from expression of second covariant derivative of a contra- and co-variant vector field, covariant extension for derivatives of the affine h-tensor $\mathcal{A}_{\beta k}^\alpha$ can be detached from structure of the covariant derivative of specific h-tensors. Such h-tensors are covariant derivatives of DSV introduced above.

Let us consider the second covariant derivatives of s-fields $\Psi^\alpha$ and $\Phi_\alpha$. Considering $\mathfrak{D}_k \Psi^\alpha$ and $\mathfrak{D}_k \Phi_\alpha$ as h-tensors, we can write:

$$\mathfrak{D}_k \mathfrak{D}_l \Psi^\alpha = \partial_k \mathfrak{D}_l \Psi^\alpha + \mathcal{A}_{\beta k}^\alpha \mathfrak{D}_l \Psi^\beta - G_{lk}^m \mathfrak{D}_m \Psi^\alpha \,;$$

$$\mathfrak{D}_k \mathfrak{D}_l \Phi_\alpha = \partial_k \mathfrak{D}_l \Phi_\alpha - \mathcal{A}_{\alpha k}^\beta \mathfrak{D}_l \Phi_\beta - G_{lk}^m \mathfrak{D}_m \Phi_\alpha \,.$$

Calculating then the alternated second covariant derivatives, we find:

$$(\mathfrak{D}_k \mathfrak{D}_l - \mathfrak{D}_l \mathfrak{D}_k)\Psi^\alpha =$$

$$= \partial_k \mathfrak{D}_l \Psi^\alpha - \partial_l \mathfrak{D}_k \Psi^\alpha + \mathcal{A}_{\beta k}^\alpha \mathfrak{D}_l \Psi^\beta - \mathcal{A}_{\beta l}^\alpha \mathfrak{D}_k \Psi^\beta + (G_{kl}^m - G_{lk}^m) \mathfrak{D}_m \Psi^\alpha =$$

$$= \left(\partial_k \mathcal{A}_{\beta l}^\alpha - \partial_l \mathcal{A}_{\beta k}^\alpha + \mathcal{A}_{\gamma k}^\alpha \mathcal{A}_{\beta l}^\gamma - \mathcal{A}_{\gamma l}^\alpha \mathcal{A}_{\beta k}^\gamma\right) \Psi^\beta + S_{kl}^m \mathfrak{D}_m \Psi^\alpha \equiv \mathfrak{R}_{\beta kl}^\alpha \Psi^\beta + S_{kl}^m \mathfrak{D}_m \Psi^\alpha \,.$$

Thus, we have the following result:

$$(\mathfrak{D}_k \mathfrak{D}_l - \mathfrak{D}_l \mathfrak{D}_k)\Psi^\alpha = \mathfrak{R}_{\beta kl}^\alpha \Psi^\beta + S_{kl}^m \mathfrak{D}_m \Psi^\alpha \,. \qquad (4.2.13)$$

Similar:
$$(\mathfrak{D}_k\mathfrak{D}_l - \mathfrak{D}_l\mathfrak{D}_k)\Phi_\alpha = -\mathfrak{R}^\beta_{\alpha kl}\Phi_\beta + S^m_{kl}\mathfrak{D}_m\Phi_\alpha.$$

Here $\mathfrak{R}^\alpha_{\beta kl}$ is notation for the following form composed on the affine h-tensor $\mathcal{A}^\alpha_{\beta k}$:

$$\mathfrak{R}^\alpha_{\beta kl} \equiv \partial_k \mathcal{A}^\alpha_{\beta l} - \partial_l \mathcal{A}^\alpha_{\beta k} + \mathcal{A}^\alpha_{\gamma k}\mathcal{A}^\gamma_{\beta l} - \mathcal{A}^\alpha_{\gamma l}\mathcal{A}^\gamma_{\beta k}, \tag{4.2.14}$$

while object
$$S^m_{kl} \equiv G^m_{kl} - G^m_{lk} = -S^m_{lk},$$

again, is an odd-symmetric part of the affine tensor, *torsion*, same tensor that arrives in expressions (4.2.2) and (4.2.3) for the second covariant derivative of vectors in the term with the first covariant derivative. So we conclude that, second term on the right hand side of equation (4.2.13) is an h-tensor, then so is the first one; hence, form (4.2.14) is also an h-tensor. More correct, object $\mathfrak{R}^\alpha_{\beta kl}$ is ascertain to transform as an h-tensor:

$$\mathfrak{R}^{\alpha'}_{\beta' k' l'} = B^{\alpha'}_\alpha B^\beta_{\beta'} A^k_{k'} A^l_{l'} \mathfrak{R}^\alpha_{\beta kl}, \tag{4.2.15}$$

once object $\mathcal{A}^\alpha_{\beta k}$ is transformed according to law (3.3.13). Vice versa, if form $\mathfrak{R}^\alpha_{\beta kl}$ is considered as an h-tensor, then object $\mathcal{A}^\alpha_{\beta k}$ has to be subordinate to transformation law (3.3.13).

It may be convenient to represent form (4.2.14) as resulting from a form

$$\breve{\mathfrak{R}}^\alpha_{\beta kl} \equiv \partial_k \mathcal{A}^\alpha_{\beta l} + \mathcal{A}^\alpha_{\gamma k}\mathcal{A}^\gamma_{\beta l} \tag{4.2.16}$$

by alternating it on indices $k, l$:

$$\mathfrak{R}^\alpha_{\beta kl} = \breve{\mathfrak{R}}^\alpha_{\beta[kl]} \equiv \breve{\mathfrak{R}}^\alpha_{\beta kl} - \breve{\mathfrak{R}}^\alpha_{\beta lk}. \tag{4.2.17}$$

Thus, covariant extension for derivatives of UGF can be defined only with respect to the *alternated* (on Roman indices) derivatives:

$$\partial_k \mathcal{A}^\alpha_{\beta l} \longrightarrow \partial_k \mathcal{A}^\alpha_{\beta l} - \partial_l \mathcal{A}^\alpha_{\beta k} \longrightarrow \mathfrak{R}^\alpha_{\beta kl}.$$

Similar to the case of affine tensor $G^n_{mk}$, this extension does not require inclusion of any other objects but is structured on affine h-tensor $\mathcal{A}^\alpha_{\beta k}$ itself. A mnemonic rule of this extension is that the priming derivatives are replaced by the extension terms only relatively to the upper index of the affine h-tensor.

To reflect the structural similarity of two forms (4.2.1) and (4.2.14) but also underline difference in their genesis, we call the introduced form (4.2.14) the *hybrid curvature form*, HCF. Eventually, we also may talk in general about *curvature forms*, meaning in common both the RCF and HCF.

*Matrix representation of HCF*

Based on the above introduced representation of the hybrid affine tensor as matrix:

$$\mathcal{A}^{\alpha}_{\beta k} \equiv \mathcal{A}_k, \qquad (4.2.18)$$

it also is convenient to represent HCF in matrix form:

$$\mathfrak{R}^{\alpha}_{\beta kl} \equiv \mathfrak{R}_{kl} = -\mathfrak{R}_{lk}. \qquad (4.2.19)$$

Matrix $\mathfrak{R}_{kl}$ can be represented as resulting from alternating the following a simpler matrix:

$$\breve{\mathfrak{R}}_{kl} \equiv \partial_k \mathcal{A}_l + \mathcal{A}_k \mathcal{A}_l = (\partial_k + \mathcal{A}_k)\mathcal{A}_l; \qquad (4.2.20)$$

then

$$\mathfrak{R}_{kl} = \breve{\mathfrak{R}}_{kl} - \breve{\mathfrak{R}}_{lk} = \partial_k \mathcal{A}_l - \partial_l \mathcal{A}_k + [\mathcal{A}_k, \mathcal{A}_l]. \qquad (4.2.21)$$

*Properties of HCF*

There are general properties of HCF structure compared to RCF and RCT presented by forms (4.2.5) and (4.2.8), respectively.

1. Similar to RCF, the HCF structure is skew-symmetric on indices $k, l$:

$$\mathfrak{R}^{\alpha}_{\beta kl} = - \mathfrak{R}^{\alpha}_{\beta lk}.$$

2. In distinction to RCT, contraction on Greek indices does not result in zero but in a skew-symmetric covariant tensor, we denote it $\mathfrak{R}_{kl}$:

$$\mathfrak{R}^{\alpha}_{\alpha kl} = \partial_k \mathcal{A}^{\alpha}_{\alpha l} - \partial_l \mathcal{A}^{\alpha}_{\alpha k} \equiv \mathfrak{R}_{kl} = -\mathfrak{R}_{lk}. \qquad (4.2.22)$$

This result also follows immediately from matrix representation (4.2.21):

$$Tr\mathfrak{R}_{kl} = \partial_k \mathcal{A}_l - \partial_l \mathcal{A}_k \equiv \mathfrak{R}_{kl} = -\mathfrak{R}_{lk}, \qquad (4.2.23)$$

since $Tr[\mathcal{A}_k; \mathcal{A}_l] \equiv 0$; here we have introduced a notation:

$$\mathcal{A}_k \equiv Tr\mathcal{A}_k = \mathcal{A}^{\alpha}_{\alpha k}. \qquad (4.2.24)$$

Definition of tensor $\mathfrak{R}_{kl}$ reminds tensor of the electromagnetic field in STM defined as curl of a covariant 4-vector-potential $A_k$:

$$F_{kl} \equiv \partial_k A_l - \partial_l A_k. \qquad (4.2.25)$$

It should be noted, however, that, $\mathcal{A}_k$ is not a vector: as one can see from equation (3.3.8), $\mathcal{A}_k$ could be transformed as a vector only at transformations with a constant matrix $B$:

$$\mathcal{A}^{\alpha'}_{\alpha'k'} = A^{k}_{k'}\mathcal{A}^{\alpha}_{\alpha k} + B^{\alpha'}_{\alpha}\partial_{k'}B^{\alpha}_{\alpha'} ; \qquad (4.2.26)$$

however, choice of a constant matrix $B$ is not an attribute of the presented theory, in contrary to always existing possibility of choice of a constant matrix $A$. Despite of this, object (4.2.22) is tensor relative arbitrary (linear and non-linear) transformations of the UM variables (once HCF is supposed to be or *explicated in fact as* h-tensor). Both identities (4.2.27) and (4.2.29) are generally covariant.

3. Global fragmentation of UGF and HCF.

Let us consider a decomposition of UGF $\mathcal{A}^{\alpha}_{\beta k}$ in two parts:

$$\mathcal{A}^{\alpha}_{\beta k} = \delta^{\alpha}_{\beta}\mathcal{A}_{k} + \mathfrak{U}^{\alpha}_{\beta k} , \qquad (4.2.30)$$

with a condition

$$Tr\mathfrak{U}^{\alpha}_{\beta k} \equiv \mathfrak{U}^{\alpha}_{\alpha k} = 0 \qquad (4.2.31)$$

imposed on the *specific* UGF (SUGF), $\mathfrak{U}^{\alpha}_{\beta k}$. This condition is equivalent to definition of object $\mathcal{A}_{k}$ as:

$$\mathcal{A}_{k} = \frac{1}{N}Tr\mathcal{A}^{\alpha}_{\beta k} = \frac{1}{N}\mathcal{A}^{\alpha}_{\alpha k} . \qquad (4.2.32)$$

This decomposition results in the corresponding decomposition of HCF (4.2.14):

$$\mathfrak{R}^{\alpha}_{\beta kl} = \frac{1}{N}\delta^{\alpha}_{\beta}\mathfrak{R}_{kl} + \mathfrak{S}^{\alpha}_{\beta kl} , \qquad (4.2.33)$$

where object $\mathfrak{R}_{kl}$ is an odd-symmetric tensor, the *curl* of field $\mathcal{A}_{k}$:

$$\mathfrak{R}_{kl} = \partial_{k}\mathcal{A}_{l} - \partial_{l}\mathcal{A}_{k} , \qquad (4.2.34)$$

while object $\mathfrak{S}^{\alpha}_{\beta kl}$ (SHCF, *specific* HCF) is a form structured on object $\mathfrak{U}^{\alpha}_{\beta k}$ similar to that as $\mathfrak{R}^{\alpha}_{\beta kl}$ is structured on $\mathcal{A}^{\alpha}_{\beta k}$:

$$\mathfrak{S}^{\alpha}_{\beta kl} \equiv \partial_{k}\mathfrak{U}^{\alpha}_{\beta l} - \partial_{l}\mathfrak{U}^{\alpha}_{\beta k} + \mathfrak{U}^{\alpha}_{\gamma k}\mathfrak{U}^{\gamma}_{\beta l} - \mathfrak{U}^{\alpha}_{\gamma l}\mathfrak{U}^{\gamma}_{\beta k} . \qquad (4.2.35)$$

Form $\mathfrak{S}^{\alpha}_{\beta kl}$ is specified by a condition

$$Tr\mathfrak{S}^{\alpha}_{\beta kl} = \mathfrak{S}^{\alpha}_{\alpha kl} \equiv \partial_{k}\mathfrak{U}^{\alpha}_{\alpha l} - \partial_{l}\mathfrak{U}^{\alpha}_{\alpha k} = 0 . \qquad (4.2.36)$$

Then

$$Tr\,\mathfrak{R}^{\alpha}_{\beta kl} = \mathfrak{R}^{\alpha}_{\alpha kl} \equiv \mathfrak{R}_{kl} . \qquad (4.2.37)$$

4. It should be underlined that, the existence, unambiguous definition, tensor nature and skew symmetry of the contracted hybrid curvature form, $\mathfrak{R}_{kl}$, together with identities (4.2.28) all are due to the hybrid nature and background skew symmetry of the *hybrid curvature form* $\mathfrak{R}^{\alpha}_{\beta kl}$, HCF, as *covariant derivative* of *unified gauge field* $\mathcal{A}^{\alpha}_{\beta k}$. To be reminded, however, that transformation properties of UGF (3.3.7) so far are only *assumed as necessary prerequisite* of general covariance of the theory under treat.

### *4.2.4. The HCF to DSV coupling constraints*

HCF is the unique form structured on UGF $\mathcal{A}^{\alpha}_{\beta k}$ and its derivatives which is recognized as an h-tensor once $\mathcal{A}^{\alpha}_{\beta k}$ is considered as satisfying transformation law (3.3.7). Vice versa, if form (4.2.14) is validated as an h-tensor, then UGF shows transformation property (3.3.7). To validate HCF as an h-tensor means to connect it to the triadic h-tensors and finally, to DSV that represents *matter*. So object HCF is uniquely viewed to play a key role in the resolution of the covariance constraint in the considered approach to UFT.

Problem of connection of HCF to DSV corresponds to the issue of *RCT – matter* connection in *General Theory of Relativity* as the *relativistic theory of gravitation* (RTG) in the 4-dimensional space-time manifold. Equations that have been found by A. Einstein [12] and D. Hilbert [13] connect the *geometry tensors* to a symmetric *energy-momentum tensor of matter* $T_{kl} = T_{lk}$ (EMT):

$$R_{kl} - \frac{1}{2} R g_{kl} = \kappa T_{kl}, \qquad (4.2.38)$$

where $g_{kl} = g_{lk}$ is the *metric tensor*, $R_{kl}$ is Ricci tensor (4.2.12), $R \equiv R_{mn} g^{mn}$ is the *scalar curvature* [8, 9, 21, 22], and $\kappa$ is the I. Newton's *gravitational constant*. A. Einstein's way was based on the considerations of a structural correspondence between $T_{kl}$ and tensors of *geometry*. Hilbert's method was based on the *extreme action principle* (EAP); he was relying on this principle in an attempt to develop a covariant unified relativistic theory of the electromagnetic and gravitational fields [8]. Einstein resorted to EAP in the latest period of his search for UFT [8].

Considering a possibility of following the Einstein – Hilbert theory of the gravitation field or, generally, incorporating their approaches to a covariant field theory in the system of UFT, we recognize the following constraints. First of all and in general, GTR has been built based on observations of gravity as a *macroscopic* phenomenon produced by macro-clusters of a *neutral matter*; therefore, there is no a direct reason to consider that the EH equations should be immediately put in foundations of UFT (even in the concept of an extended or Unified Manifold). It seems more suitable to envision that, these equations will result from UFT equations at an asymptotic derivation, despite they perhaps could not be immediately seen in the system of the basic equations.

Further, affine tensor $G^m_{kl}$ is immediately explicated in form of *matched connection* (4.1.7) based on derivatives of *metric tensor*, − as a matter-fact of the conventional differential geometry or, more exact, of the *covariant differential calculus of vectors and tensors.* After that, Einstein – Hilbert equations of GTR establish a connection between *metric tensor*, an attribute of *geometry*, and energy-momentum tensor of *matter* (EMT) as an attribute of *matter*.

In the frame of our approach to UFT based on introduction of DSV as a master object of the theory, a correspondent matter-fact does not exist with respect to the connection object, $\mathcal{A}^{\alpha}_{\beta k}$, because of the extended, *hybrid nature* of this object together with hybrid tensors $P^{\alpha k}_{\beta}$ and $\bar{P}^{\alpha k}_{\beta}$. Its connection to hybrid tensors and DSV cannot be found as a background mathematical relation between objects in frame of the conventional differential geometry. Consequently, the hybrid nature of connection $\mathcal{A}^{\alpha}_{\beta k}$ does not allow for contraction of h-tensor $\mathfrak{R}^{\alpha}_{\beta k l}$ to an even-symmetric tensor type of Ricci tensor (now in $N$ dimensions manifold), since indices $\alpha, \beta$ do not mix i.e. cannot be switched and (or) contracted with indices $k, l$ (remind that, contraction of $\mathfrak{R}^\alpha_{\beta kl}$ on Greek indices results in the *skew-symmetric* tensor $\mathfrak{R}_{kl} \equiv \mathfrak{R}^\alpha_{\alpha kl} = -\mathfrak{R}_{lk}$).

We thus come to a conclusion that, in the DSV-based approach to UFT the equations connecting *"geometry"* to *matter* should be modified compared to that of GTR. It should be stressed that, the above described DSV concept of *matter field* was introduced not eventually but referring to the Dirac's quantum legacy and observation of genesis of theory of the "elementary particles", QFT.

On the other hand, one may think that explication of $\mathcal{A}^\alpha_{\beta k}$ as hybrid affine tensor i.e. an object that transforms according to equation (3.3.13) can be realized once coupling of $P^{\alpha k}_\beta$, $\overline{P}^{\alpha k}_\beta$ and HCF to DSV is found based on a fundamental dynamical principle of an irreducible mathematical background. As it was mentioned above, D. Hilbert derived equation (4.2.30) applying Extreme Action principle. Our studies have led to the persuasion that EAP should be regarded as one the logical principles of the differential approach to UFT. Corresponding to this tendency, we resort to this principle in search for coupling of the hybrid triadic objects to DSV. Yet equations for DSV itself can be derived in this way, as well.

## *Resume of Chapter 4*

Considering a possibility to apply the covariant differential method to the issue of connection between the hybrid triadic objects and DSV, we came to the necessity to introduce covariant derivatives of the *hybrid tensors* (HT) and the *hybrid affine tensor* (HAT) or *unified gauge field* (UGF) $\mathcal{A}^\alpha_{\beta k}$. Covariant derivatives of HT include HAT in their structure as well as the ordinary affine tensor i.e. Christoffel symbols. The Riemann-Christoffel curvature form (RCF) $\mathcal{R}^n_{mkl} = -\mathcal{R}^n_{mlk}$, built on affine tensor $G^n_{mk}$, is recognized as covariant derivative of $G^n_{mk}$. The *hybrid curvature form* (HCF), h-tensor $\mathfrak{R}^\alpha_{\beta kl} = -\mathfrak{R}^\alpha_{\beta lk}$, built in similar way on $\mathcal{A}^\alpha_{\beta k}$, is recognized as *covariant derivative of* $\mathcal{A}^\alpha_{\beta k}$.

Aggregate of objects consisting of DSV, HT, HAT and their covariant derivatives is envisioned to be sufficient in order to compose a closed system of differential equations in which the hybrid tensors and UGF would be coupled to DSV.

A prerogative of deriving the differential system of UFT is committed to method of the *Extreme Action* or *Lagrange formalism* viewed as principle of a *dynamic balance.*

# 5. Extreme Action Principle for a Covariant Field Theory

## 5.1. Differential Law and Invariance

*Differential Law as expression of a dynamic invariance*

The differential law(s) in the field theories i.e. the relations between functions and their derivatives at points of a space of variables expresses a connection between the field values in a region of the space (manifold). In some sense, it gives a possibility to realize the transition from a local consideration to a regional one, though from the logical point of view, an operating with the derivatives of functions immediately goes beyond the "local consideration".

On the other hand, the differential law can be viewed as expressing some fundamental *dynamical* invariance. From this point of view, the differential laws should be established based on consideration of possible *irreducible invariants*. The invariants should not be postulated but recognized as the immediate consequences of the supposed transformation properties of the involved basic objects. Such invariants are *scalar forms* (those include the objects and their first derivatives) and *regional invariants*. The scalar forms are the *local invariants*. The regional invariants are obtained by the *invariant integration* of the scalar forms over a volume of the manifold.

Consideration of the regional invariants then becomes a necessary step to establishing relations between the objects and their derivatives, i.e. to finding the differential law. Namely, one may consider a superposition of the irreducible scalar forms, integrate it over a volume of the manifold, and demand a special feature of this combined regional invariant. The demand should not depend in the choice of the manifold variables and the surface limiting the volume of integration. Such demand is generally known in theory of differential equations as the *variation principle* and in the field theory as the *extreme action principle* (EAP). It results in a system of equations connecting the basic objects (functions) with their derivatives, the *Euler-Lagrange equations*.

The Extreme Action Principle (EAP) was introduced to the fundamental theoretical physics initially as a universal mathematical method to derive equations of motion of a dynamical system in the Newtonian mechanics. Since appearance of the Maxwell – Lorentz electrodynamics, EAP was extended to the area of the *field theory*. The Maxwell-Lorentz electrodynamics ("classical electrodynamics") arrived as a hybrid relativistic theory with the correspondent action integral, which considers the charged micro-objects (constituents of *matter*) as point-like particles interacting with the electromagnetic *field* and radiating the EM waves when accelerated. Since the time of appearance of Schrödinger and then Dirac equation for electrons, EAP is used in theory of *matter* i.e. "elementary particles" as the *field theory*, as well. It is one of the corner-stones of the modern QFT culture, from Quantum Electrodynamics to Standard Model. Growing viability of EAP in the field theory should be viewed as manifestation of its meaning as a *universal invariant balance* principle driving a fundamental *Differential Law* of *matter*.

On the other hand, despite of a very successful practice of use of the Euler-Lagrange-Nöther (ELN) method in QFT, there still exist doubts about the validity of this method and its background meaning in the efforts to reach Unified Field Theory. Eventually, it is regarded as "just a canonical formulation" of equations that could be found based on some "immediate" physical principles or even intuition. However, a reason due to which EAP plays so universal role in the fundamental physical theory cannot be random; in contrary, EAP in our sight is doomed to be discovered as one of the imprescriptible, indispensable principles of a unified field theory.

There are several important issues in the context of comparison between "utilization" of EAP in the frame of presented approach to UFT and the ways how EAP is traditionally used in the developments of QFT. We will touch some of them in the process of the derivations and discuss more afterwards.

In the following section we will introduce a collection of scalar forms that could be engaged in structure of Lagrangian of the superdimensional dual-covariant field theory.

### 5.2. Scalar forms of a DSV based theory

The simplest scalar form of the DSV based theory is the above introduced *state norm*:

$$\mathbb{N} = \Phi_\alpha \Psi^\alpha. \tag{5.2.2}$$

Other scalar forms also binary on the DSV components but including its covariant derivative $\mathfrak{D}_k \Psi^\alpha$ and $\mathfrak{D}_k \Phi_\alpha$ can be composed by use of the triadic h-tensors:

$$\mathfrak{D} \equiv \Lambda^{\alpha k}_\beta \Phi_\alpha \mathfrak{D}_k \Psi^\beta = \Lambda^{\alpha k}_\beta \Phi_\alpha (\partial_k \Psi^\beta + \mathcal{A}^\beta_{\gamma k} \Psi^\gamma);$$

$$\overline{\mathfrak{D}} \equiv \overline{\Lambda}^{\alpha k}_\beta \Psi^\beta \mathfrak{D}_k \Phi_\alpha = \overline{\Lambda}^{\alpha k}_\beta \Psi^\beta (\partial_k \Phi_\alpha - \mathcal{A}^\gamma_{\alpha k} \Phi_\gamma); \tag{5.2.3}$$

here h-tensors $\Lambda^{\alpha k}_\beta$ and $\overline{\Lambda}^{\alpha k}_\beta$ correspond to h-tensors $P^{\alpha k}_\beta$ and $\overline{P}^{\alpha k}_\beta$ in equations (4.3.35) but may distinguish from those.

Other invariant form can be composed based on the hybrid curvature form $\mathfrak{R}^\beta_{\alpha k l}$ (HCF) given by equation (4.2.14). For this we have to introduce an h-tensor $\mathbb{L}^{\alpha k l}_\beta$:

$$\mathbb{L}_G \equiv \mathbb{L}^{\alpha k l}_\beta \mathfrak{R}^\beta_{\alpha k l}. \tag{5.2.4}$$

To avoid unnecessary extension of collection of basic objects, h-tensor $\mathbb{L}^{\alpha k l}_\beta$ should be structured based on the already introduced basic objects. One of the possibilities of this kind is structuring $\mathbb{L}^{\alpha k l}_\beta$ on h-tensors $\Lambda^{\alpha k}_\beta$ and $\overline{\Lambda}^{\alpha k}_\beta$ as follows:

$$\mathbb{L}^{\alpha k l}_\beta \Rightarrow \mathrm{L}^{\alpha k l}_\beta \equiv \Lambda^{\alpha k}_\gamma \overline{\Lambda}^{\gamma l}_\beta + \overline{\Lambda}^{\alpha l}_\gamma \Lambda^{\gamma k}_\beta. \tag{5.2.5}$$

The odd symmetry of HCF on indices $k, l$ reduces binary structure (5.2.5) to a form alternated on indices $k, l$:

$$\mathrm{L}^{\alpha k l}_\beta \rightarrow \Sigma^{\alpha k l}_\beta = \frac{1}{2}\left(\mathrm{L}^{\alpha k l}_\beta - \mathrm{L}^{\alpha l k}_\beta\right) \equiv -\Sigma^{\alpha l k}_\beta; \tag{5.2.6}$$

$$\mathbb{L}_G \Rightarrow \Sigma^{\alpha k l}_\beta \mathfrak{R}^\beta_{\alpha k l}. \tag{5.2.7}$$

In matrix notations:

$$\mathrm{L}^{\alpha k l}_\beta \equiv \mathbf{L}^{kl} = \mathbf{\Lambda}^k \overline{\mathbf{\Lambda}}^l + \overline{\mathbf{\Lambda}}^l \mathbf{\Lambda}^k \equiv \{\mathbf{\Lambda}^k; \overline{\mathbf{\Lambda}}^l\}; \tag{5.2.8}$$

$$\Sigma_\beta^{\alpha kl} \equiv \mathbf{\Sigma}^{kl} \equiv \frac{1}{2}(\mathbf{L}^{kl} - \mathbf{L}^{lk}) ; \tag{5.2.9}$$

$$\mathbb{L}_G = Tr(\mathbf{\Sigma}^{kl} \mathfrak{R}_{kl}) . \tag{5.2.10}$$

*Weyl's-like scalar*

The shown particular collection of scalar forms is limited by the simplest items that can be built based on terms of zero and first power on covariant derivatives of the objects which presence as necessary in the master equations for DSV (3.4.11). The presented examples, obviously, do not exhaust the variety of possible scalar forms that can be built based on the profiled family of objects denoted $\{X\}$ as in (5.2.1) and their covariant derivatives. For instance, h-tensor (5.2.9) in scalar form (5.2.10) can be replaced by dual image of HCF, h-tensor $\mathfrak{R}_\beta^{\alpha kl}$ obtained by lifting indices $k, l$ using symmetric tensor $\Lambda^{kl}$ built on h-tensors $\Lambda_\beta^{\alpha k}$, $\overline{\Lambda}_\beta^{\alpha k}$ :

$$\Sigma_\beta^{\alpha kl} \Longrightarrow \mathfrak{R}_\beta^{\alpha kl} \equiv \mathfrak{R}_{\beta mn}^{\alpha} \Lambda^{km} \Lambda^{ln}$$

$$\mathbb{L}_G \Longrightarrow \mathfrak{R}_{\alpha kl}^{\beta} \mathfrak{R}_\beta^{\alpha kl} \equiv \Lambda^{km} \Lambda^{ln} \mathbb{G}_{kl;mn} \tag{5.2.11}$$

$$\Lambda^{kl} = \frac{1}{2}(\mathrm{L}_\alpha^{\alpha kl} + \mathrm{L}_\alpha^{\alpha lk}) = \Lambda^{lk}. \tag{5.2.12}$$

Here we introduced notation $\mathbb{G}_{kl;mn}$ for a 4-covariant tensor built on HCF:

$$\mathbb{G}_{kl;mn} = \mathfrak{R}_{\alpha kl}^{\beta} \mathfrak{R}_{\beta mn}^{\alpha} . \tag{5.2.13}$$

Scalar form (5.2.11) reminds geometry scalar $\mathbb{W}$ (in our notation) of H. Weyl built on Riemann-Christoffel tensor $R_{nkl}^m$ [8]:

$$\mathbb{W} = R_{nkl}^m R_m^{nkl}; \qquad R_n^{mkl} = g^{kp} g^{lq} R_{npq}^m .$$

Variety and structure of the shown forms can be partially reduced at possible cutback of collection of basic objects from 5 items to 4:

$$\overline{\Lambda}_\beta^{\alpha k} \Longrightarrow \pm \Lambda_\beta^{\alpha k}; \tag{5.2.14}$$

then:

$$\Lambda^{kl} \Longrightarrow \Lambda_\beta^{\alpha k} \Lambda_\alpha^{\beta l}; \qquad \Sigma_\beta^{\alpha kl} \Longrightarrow \Lambda_\gamma^{\alpha k} \Lambda_\beta^{\gamma l} - \Lambda_\gamma^{\alpha l} \Lambda_\beta^{\gamma k}. \tag{5.2.15}$$

*Need more forms?*

Obviously, there exists unlimited variety of more complicate high order tensor and scalar forms that can be built on family of 5 or 4 basic objects. On the other hand, possible collection of items that one could use when sculpturing the SFT differential system apparently will be limited under the press of the *irreducibility* demands claimed to a unified theory.

*No scalars "on definition"*

Due to the principle of the *dynamical invariance*, there is no position for *scalars on definition* in association of basic objects in the presented approach to the irreducible field theory. Scalar functions as *invariants of transformations* of UM variables all are supposed to result *in dynamics* of the *Extreme Action* principle from *scalar forms* composed *in presupposition*. This principle will be applied to derive connections of the introduced triadic objects to DSV, together with equations for DSV itself.

### 5.3. Regional Invariants and Grand Metric Tensor

Next we have to introduce invariant integral forms, the *regional invariants*. Based on scalar forms as local invariants, we also may consider regional invariants of transformations of coordinates. If there is a local invariant, a scalar form $I(\check{\psi}) = I(\check{\psi}')$, then the regional invariant can be composed on the basis of a superposition of infinitesimal elements $I(\check{\psi})d\Omega$, here $d\Omega$ is a *differential volume*:

$$d\Omega \equiv d\check{\psi}^1 d\check{\psi}^2 \ldots d\check{\psi}^n. \tag{5.3.1}$$

However, the infinitesimal element $I(\check{\psi})d\Omega$ is not an invariant of the transformations, since the differential volume $d\Omega$ is not an invariant but is transformed as follows [8, 20, 21]:

$$d\Omega' = J(\check{\psi}'|\check{\psi})d\Omega, \tag{5.3.2}$$

where $J(\check{\psi}'|\check{\psi})$ is the determinant of the transformation matrix:

$$J(\check{\psi}'|\check{\psi}) = |A_k^{k'}| \equiv \left|\frac{\partial \psi^{k'}}{\partial \psi^k}\right|. \tag{5.3.3}$$

So the regional invariants cannot be composed by an immediate integration of infinitesimal elements $I(\check{\psi})d\Omega$. This constraint is solved by mean of introduction of an *invariant differential volume*, referring an arbitrary frame $\check{\psi}'$ to some "priming frame" $\check{\psi}$:

$$d\Omega' \rightarrow \frac{d\Omega'}{J(\check{\psi}'|\check{\psi})} = d\Omega = J(\check{\psi}|\check{\psi}')d\Omega'; \tag{5.3.4}$$

then one can introduce a regional invariant as the following integral form:

$$\mathcal{I}(\sigma) \equiv \int I(\check{\psi})d\Omega = \int I(\check{\psi}')J(\check{\psi}|\check{\psi}')d\Omega' = \mathcal{I}'; \tag{5.3.5}$$

here symbol $\sigma$ denotes integration over a volume limited by a $(N-1)$ dimensional hyper-surface

$$\sigma(\check{\psi}^1, \check{\psi}^2, \ldots, \check{\psi}^N) = const. \tag{5.3.6}$$

Product $IJ(\check{\psi}|\check{\psi}')$ is a *relative invariant*, a notion used in the differential geometry [8, 20, 21].

Though the definition of a regional invariant (5.3.5) is consistent from the point of view of the invariance requirement, it is not satisfactory for use in a field theory, since objects $I \cdot J(\check{\psi}|\check{\psi}')$ are not associated with the *structural* forms i.e. forms structured on *functions as objects* in the manifold. To build this type of relative invariants, let us consider transformation of the magnitude of determinant of a valence 2 tensor $w_{kl}$ inverse to tensor $w^{kl}$ introduced above:

$$w \equiv |det w_{kl}|;$$

$$w' \equiv |det w_{k'l'}| = (detA)^2 |det w_{kl}| = wJ^2(\check{\psi}|\check{\psi}'). \tag{5.3.7}$$

Comparing this transformation with the transformation of the differential volume (5.3.4), we find that the product $\sqrt{w}d\Omega$ is an invariant of the transformations:

$$\sqrt{\frac{w'}{w}} = J(\check{\psi}|\check{\psi}') = \frac{d\Omega}{d\Omega'}; \quad \rightarrow \quad \sqrt{w'}d\Omega' = \sqrt{w}d\Omega = \text{inv}. \tag{5.3.8}$$

Thus, regional invariants can be defined as

$$\mathcal{I} = \int I\sqrt{|det w_{kl}|}d\Omega = \int \hat{I}d\Omega; \tag{5.3.9}$$

We will call scalar form $I$ in product with *weigh factor* $\sqrt{w}$ the *relative scalar* $\hat{I}$:

$$\hat{I} = I\sqrt{|det w_{kl}|}. \tag{5.3.10}$$

Not that, $\sqrt{w}$ itself is a relative scalar.

Since we consider only real functions in a real manifold, both numbers $w$ and $\sqrt{w}$ always can be defined as the positive ones (functions of a point in UM), so the square root procedure does not introduce an ambiguity to the definition of the regional invariant forms.

We call tensor $w_{kl}$ or $w^{kl}$ *Grand Metric* (GM), so far meaning use of it for establishing the invariant differential volume of Unified Manifold.

*Possible structuring of Grand Metric*

Tensor inverse to tensor $\Lambda^{kl}$ shown in equation (5.2.12) could play role of GM tensor:

$$w_{kl} \Rightarrow \Lambda_{kl}; \tag{5.3.11}$$

$$\Lambda_{kl}^{-1} = \Lambda^{kl} \equiv \frac{1}{2}(L_\alpha^{\alpha kl} + L_\alpha^{\alpha lk}). \tag{5.3.12}$$

Other options of structuring tensor $w_{kl}$ will be considered later and denied based on a set of the *irreducibility demands* claimed to the DSV-based UFT.

## 5.4. General formulation of EAP

*Action Integral and Lagrangian*

The EAP is based on the consideration of a composition of regional invariants, an *action integral*

$$S(\sigma) \equiv \int \mathcal{L}(X, \partial X) d\Omega . \qquad (5.4.1)$$

where $\mathcal{L}$ is *Lagrangian*, a *form* composed on the *basic* objects $X(\check{\psi})$ and their *first derivatives*, $\partial_k X$. For simplicity sake, we use for the derivatives a simple general notation $\partial X$ instead of $\partial_k X$. We also will use notation $Y$ for the tensor and h-tensor type objects (including vectors and s-vectors) and notation $\Gamma$ for the *connection* objects:

$$(X, \partial X) = (Y, \partial Y; \Gamma, \partial \Gamma)$$

$$\mathcal{L}(X, \partial X) \Rightarrow \mathcal{L}(Y, \partial Y; \Gamma, \partial \Gamma) .$$

Notations $Y^a$ and $X^a$ will also be used as for a particular type of an object. In our case of SFT in Unified Manifold, the collection of basic objects presumably includes tensor type objects as dual vector field $\Psi^\alpha, \Phi_\alpha$, triadic h-tensors of type $\Lambda_\beta^{\alpha k}, \overline{\Lambda}_\beta^{\alpha k}$ and affine h-tensor $\mathcal{A}_{\beta k}^\alpha$ .

*EAP and general Euler-Lagrange equations*

We remind the formulation of the extreme action principle and related general derivations. Let all objects $X^a$ have certain values (real numbers) at points of a hyper-surface $\sigma$. Then functions $X^a(\check{\psi})$ behave in the volume inside the surface in a way that integral (5.4.1) has an extreme relative to an arbitrary variation of functions $X^a(\check{\psi})$, i.e.:

$$\delta S = \int \delta \mathcal{L} d\Omega = 0, \qquad (5.4.2)$$

here

$$\delta \mathcal{L} = \frac{\partial \mathcal{L}}{\partial X^a} \delta X^a + \frac{\partial \mathcal{L}}{\partial (\partial_k X^a)} \delta \partial_k X^a \qquad (5.4.3)$$

is the variation of the Lagrangian. The two variation terms are not independent in the variation of the action integral. The condition of functions $X^a$ having certain (fixed) values on the surface $\sigma$ means that variations $\delta X^a$ are considered to be equal to zero on the surface:

$$\delta X^a \big|_\sigma = 0. \qquad (5.4.4)$$

Using identity $\delta \partial_k X^a = \partial_k \delta X^a$, one can rewrite variation form (5.4.3) as follows:

$$\delta \mathcal{L} = \left[\frac{\partial \mathcal{L}}{\partial X^a} - \partial_k \frac{\partial \mathcal{L}}{\partial(\partial_k X^a)}\right]\delta X^a + \partial_k \left[\frac{\partial \mathcal{L}}{\partial(\partial_k X^a)}\delta X^a\right]. \tag{5.4.5}$$

Next, applying *Gauss theorem* [8, 21], we transform the integral with complete derivatives to an integral over the boundary hyper-surface:

$$\int \partial_k [\frac{\partial \mathcal{L}}{\partial(\partial_k X^a)}\delta X^a] d\Omega = \oint d\sigma_k \frac{\partial \mathcal{L}}{\partial(\partial_k X^a)}\delta X^a, \tag{5.4.6}$$

here $d\sigma_k$ is a differential element of the boundary hyper-surface [21]. Taking into account the boundary condition (5.4.4), variation of the action integral can be written as

$$\delta S = \int [\partial_k \frac{\partial \mathcal{L}}{\partial(\partial_k X^a)} - \frac{\partial \mathcal{L}}{\partial X^a}]\delta X^a d\Omega. \tag{5.4.7}$$

Since variations $\delta X^a$ are arbitrary over the volume of integration, the extreme principle (5.4.2) results in the requirements:

$$\partial_k \frac{\partial \mathcal{L}}{\partial(\partial_k X^a)} - \frac{\partial \mathcal{L}}{\partial X^a} = 0, \tag{5.4.8}$$

the *Euler-Lagrange equations*.

*Rule of the first-only derivatives for Lagrangian form*

The restriction that the integrand form of the variation principle (Lagrangian) should not include derivatives of objects $X^a$ of order higher than the first one may seem as an assumption *ad hoc*. We consider this rule as coming from the *irreducibility* principle. Namely, one can always represent a Lagrangian form that includes higher order derivatives to a *first-only derivatives* form structured on the *extended* collection of objects, by recourse to the usual simple substitutions.

*Lagrangian as relative invariant*

While the action integral (5.4.1) is supposed to be a (regional) invariant, the Lagrangian $\mathcal{L}$ appears to be a *relative invariant*. Namely, the differential volume $d\Omega$ is a relative invariant, i.e. it is transformed, as known, according to the following equation:

$$d\Omega' = \left|A_k^{k'}\right| d\Omega \equiv |A|d\Omega = Jd\Omega,$$

where $J$ is the determinant of the transformation matrix $A_k^{k'}$. Consequently, the Lagrangian should be transformed in an inverse way:

$$\mathcal{L}' = \mathcal{L}\left|A_{k'}^k\right| = \mathcal{L}/detA.$$

*Scalar Lagrangian and Grand Metric*

To satisfy the invariance requirement, one can generally define Lagrangian $\mathcal{L}$ as a relative invariant as follows:

$$\mathcal{L} = L\sqrt{w} \equiv L\sqrt{|det w_{kl}|} , \tag{5.4.9}$$

here $L$ is an invariant composed form, a *scalar Lagrangian*, $L' = L$, while $w$ is the determinant of a valence 2 covariant tensor $w_{kl}$ introduced above.

Note that, here we do not assume symmetry of tensor $w_{kl}$, i.e. it is considered so far asymmetric, $w_{kl} \neq w_{lk}$. Also note that, tensor $w_{kl}$ is introduced in order to build invariant integral forms, the *regional invariants*. At this stage of our considerations, tensor $w_{kl}$ has no immediate link to the notion of an *interval*, though, based on this tensor, one can consider a second order differential invariant $w_{kl} d\breve{\psi}^k d\breve{\psi}^l$. In our view, the notions of interval, distance, etc. should be *derived* at the latest stages of profiling a fundamental theory i.e. at the *deduction*: we consider these categories as being associated with the properties of the *solutions* of the differential system of SFT.

Variation of action (5.4.7) can be written in the form of an integral with an invariant differential volume:

$$\delta S = \int (\widehat{\nabla}_a \mathcal{L}) \delta X^a \sqrt{w} d\Omega ,$$

here we use notation $\widehat{\nabla}_a \mathcal{L}$ for the integrand forms:

$$\widehat{\nabla}_a \mathcal{L} \equiv \frac{1}{\sqrt{w}} \left[ \partial_k \frac{\partial(L\sqrt{w})}{\partial(\partial_k X^a)} - \frac{\partial(L\sqrt{w})}{\partial X^a} \right] . \tag{5.4.10}$$

These forms should be considered as objects dual to the differentials of objects $X^a$, which (differentials) are vectors or s-vectors and tensors or s-tensors and hybrid tensors. This follows from the supposed invariance of action variation $\delta S$, invariance of the normalized differential volume $\sqrt{w} d\Omega$ and tensor properties of variation $\delta X$ (variation of an affine connection, either $G^k_{lm}$ or $\mathcal{A}^\alpha_{\beta m}$, can be considered as a difference of two affine connections, which is a tensor or h-tensor, as pointed above). Euler-Lagrange equations in an explicit covariant form can be written as $\widehat{\nabla}_a \mathcal{L} = 0$, i.e. :

$$\frac{1}{\sqrt{w}} \left[ \partial_k \frac{\partial(L\sqrt{w})}{\partial(\partial_k X^a)} - \frac{\partial(L\sqrt{w})}{\partial X^a} \right] = 0 .$$

Since $w \neq 0$, these equations also can be written in the usual form of the Euler-Lagrange equations (5.4.8), now with Lagrangian specified as *relative scalar* form:

$$\partial_k \frac{\partial(L\sqrt{w})}{\partial(\partial_k X^a)} - \frac{\partial(L\sqrt{w})}{\partial X^a} = 0 . \tag{5.4.11}$$

*EL equations in the explicit analytical form*

The Lagrangian being represented according to the definition (5.4.9), includes a non-analytical factor $\sqrt{w}$; so do the EL equations written in the general compact form (5.4.11) in terms of the scalar Lagrangian. The related uncertainty, however, goes away from the EL equations after disclosing the variation and usual derivatives of $\sqrt{w}$ in equations as

$$\partial\sqrt{w} = \frac{\partial w}{2\sqrt{w}}\ ; \qquad \partial_k\sqrt{w} = \frac{\partial_k w}{2\sqrt{w}}\ . \tag{5.4.12}$$

After dividing equations (5.4.11) by $\sqrt{w}$, they contain only integer powers of $w$ and its derivatives:

$$(\partial_k + \frac{\partial_k w}{2w})\left[\frac{\partial L}{\partial(\partial_k X^a)} + \frac{L}{2w}\frac{\partial w}{\partial(\partial_k X^a)}\right] - \left(\frac{\partial L}{\partial X^a} + \frac{L}{2w}\frac{\partial w}{\partial X^a}\right) = 0\ . \tag{5.4.13}$$

After introduction of *invariant variation* of Lagrangian $\mathcal{L}$ as

$$\hat{\partial}\mathcal{L} \equiv \frac{1}{\sqrt{w}}\partial\mathcal{L} = \frac{1}{\sqrt{w}}\partial(\sqrt{w}L) = \partial L + \frac{\partial w}{2w}L \tag{5.4.14}$$

and *covariant differential* $\hat{\partial}_k$:

$$\hat{\partial}_k f \equiv \frac{1}{\sqrt{w}}\partial_k(\sqrt{w}f) = \partial_k f + \frac{\partial_k w}{2w}f, \tag{5.4.15}$$

covariant EL equations (5.4.12) can be written in a more compact (but symbolic) analytical form:

$$\hat{\partial}_k \frac{\partial \mathcal{L}}{\partial(\partial_k X^a)} - \frac{\partial \mathcal{L}}{\partial X^a} = 0\ . \tag{5.4.16}$$

Invariance of $\hat{\partial}\mathcal{L}$ is following from the elementary relation [8, 9, 21]:

$$\frac{\partial w}{w} = w^{kl}\partial w_{kl} = -w_{kl}\partial w^{kl}\ . \tag{5.4.17}$$

### 5.5. General covariant reduction of EAP

*5.5.1. Lagrangian as presumably relative invariant form*

We will treat EAP for a superdimensional differential field theory in which the primary role belongs to the Dual State Vector field, DSV. Correspondingly, system of basic objects could be reduced to DSV and triadic objects type of hybrid tensors $P^\alpha_{\beta k}$ and $\bar{P}^\alpha_{\beta k}$ and $\mathcal{A}^\alpha_{\beta k}$ above introduced. We call this kind of a field theory *the s-vector based theory* or *s-theory*. Separate, a theory can be considered in which all objects are supposed to be transformed with a matrix same as for transformation of differentials of the manifold variables; examples of such theories are the Maxwell – Lorentz electromagnetic theory and the *General Theory of Relativity* (GTR) of Einstein or *Relativistic Theories of Gravitation* (RTG) of Einstein – Hilbert – Weyl [8]. We call this type of a theory *the vector-based* *theory* or *v-theory*. Ultimately, one may formally treat a field theory with a Lagrangian which includes vectors and s-vectors and also tensors and h-tensors as the independent basic objects ("mixed theory"; one may observe a relevance of Standard Model of QFT to such situation). One may raise questions concerning the logical consistence of the *v-theories* or *mixed theories* (as well as s-theories) in the context of Pauli's paradigm of the *irreducibility* [8]. Issues of consistence of different types of theories can be discussed after the completing of the derivations presented in this Chapter and Chapters 6 through 9.

The requirement to integrand function $L(X, \partial X)$ to be invariant relatively of an arbitrary transformation of variables $\check{\psi}$ of the unified manifold, as well as the requirement to object $w_{kl}$ to be a tensor leads to the restriction that they should be structured on basic tensor type forms. Those basic tensor type of forms are: usual tensor type objects $Y$ (dual state vector and hybrid tensors), their covariant derivatives $D_k Y^a$, and *covariant derivatives of the connections* $G_{lk}^m$ and $\mathcal{A}_{\alpha k}^\beta$ i.e. the Riemann-Christoffel curvature form (RCF), $\mathcal{R}$, given by formula (4.2.1), and the *hybrid curvature form* (HCF), $\mathfrak{R}$, given by (4.2.14). We will use general notation $\mathfrak{G}_{ak}^\ell$ or $\mathfrak{G}_k$ for connections $G_{lk}^m$ and $\mathcal{A}_{\alpha k}^\beta$ in common, and notation $D_k Y$ for covariant derivative of objects $Y$ as follows:

$$D_k Y = \partial_k Y + [\![\mathfrak{G}_k ; Y]\!] ; \qquad (5.5.1)$$

here symbol $[\![\mathfrak{G}_k ; Y]\!]$ is *gauge commutator* introduced in (4.1.33). Symbols $\mathcal{K}_{akl}^\ell$, $\widehat{\mathcal{K}}_{kl}$ or $\mathcal{K}$ will be used as general notation for curvature forms:

$$\mathcal{K}_{akl}^\ell = \check{\mathcal{K}}_{akl}^\ell - \check{\mathcal{K}}_{alk}^\ell ; \qquad \check{\mathcal{K}}_{akl}^\ell \equiv \partial_k \mathfrak{G}_{al}^\ell + \mathfrak{G}_{ck}^\ell \mathfrak{G}_{al}^c , \qquad (5.5.2)$$

$$\mathfrak{G}_k \equiv \mathfrak{G}_{ak}^\ell ; \qquad \check{\mathcal{K}}_{akl}^\ell \equiv \check{\mathcal{K}}_{kl} = \partial_k \mathfrak{G}_l + \mathfrak{G}_k \mathfrak{G}_l \qquad (5.5.3)$$

$$\mathcal{K}_{akl}^\ell \equiv \mathcal{K}_{kl} = \check{\mathcal{K}}_{kl} - \check{\mathcal{K}}_{lk} = \partial_k \mathfrak{G}_l - \partial_l \mathfrak{G}_k + [\mathfrak{G}_k; \mathfrak{G}_l] \qquad (5.5.4)$$

It should be underlined that, connection forms $\mathfrak{G}_{ak}^\ell$ cannot be included separate in addition in the structure of both, scalar $L$ and tensor $w_{kl}$, since objects $\mathfrak{G}_{ak}^\ell$ are not of the tensor type: separate inclusion of these objects in contraction with themselves and (or) with tensor type objects cannot result in scalars. We thus can and have to consider $L$, $w_{kl}$ and whole Lagrangian $\mathcal{L}$ as forms structured only on $Y, DY$ and $\mathcal{K}_{kl}$ or $\mathcal{K}$:

$$L(X, \partial X) \Longrightarrow \mathbb{L}(Y, DY; \mathcal{K}) ; \qquad w_{kl}(X, \partial X) \Longrightarrow w_{kl}(Y, DY; \mathcal{K}) ; \qquad (5.5.5)$$

$$\mathcal{L}(X, \partial X) \Longrightarrow \mathbb{L}(Y, DY; \mathcal{K}) \sqrt{w} \equiv \mathfrak{L}(Y, DY; \mathcal{K}); \qquad w = |\det w_{kl}| \qquad (5.5.6)$$

Note that, forms $DY$ can be meant as structured with the same connection objects, $G_{nk}^m$ or $\mathcal{A}_{\beta k}^\alpha$, as forms $\mathcal{R}$ or $\mathfrak{R}$, respectively, since the difference of the two connection objects (same type) is tensor or h-tensor, therefore it can be referred to the collection of tensor type objects $Y$. On the other hand side, introduction of RCF $\mathcal{R}_{nkl}^m$ as covariant derivatives of affine tensor $G_{nk}^m$ can be logically justified only if Lagrangian structure includes (covariant) derivatives of h-tensors. In our approach to irreducible field theory, the leader role in structuring the Lagrangian belongs to DSV and its covariant derivatives which do not include affine tensor $G_{nk}^m$. Therefore, necessity of inclusion of $G_{nk}^m$ and, consequently, $\mathcal{R}_{nkl}^m$ in Lagrangian of the DSV-based covariant theory cannot be motivated in a prior. However, in this Chapter we will conduct the Lagrange formalism derivations in general representation of Lagrangian shown in equations (5.5.6).

Now we have to rewrite EL equations (5.4.14) in terms of generally covariant Lagrangian form $\mathfrak{L}(Y, DY; \mathcal{K})$. We have to separate equations on tensor type objects $Y$:

$$\hat{\partial}_k \frac{\hat{\partial} L}{\partial(\partial_k Y)} - \frac{\hat{\partial} L}{\partial Y} = 0, \quad (5.5.7)$$

and equations on *covariant connections* $\mathfrak{G}_k$:

$$\hat{\partial}_l \frac{\hat{\partial} L}{\partial(\partial_l \mathfrak{G}_k)} - \frac{\hat{\partial} L}{\partial \mathfrak{G}_k} = 0. \quad (5.5.8)$$

Our purpose in this section is to show that both systems of EL equations are generally covariant.

### 5.5.2. EL equations on the tensor type objects

Let us first transform equation (5.5.7) to an explicit covariant view. Applying equation (5.5.7) to general covariant Lagrangian form (5.5.6), we can rewrite this equation in the following view:

$$\hat{\partial}_k \mathbb{L}^k - \frac{\partial(D_k Y)}{\partial Y} \mathbb{L}^k = \frac{\hat{\partial} \mathbb{L}}{\partial Y}. \quad (5.5.9)$$

Here we have used relations:

$$\frac{\hat{\partial} L}{\partial Y} = \frac{\hat{\partial} \mathbb{L}}{\partial Y} + \frac{\hat{\partial} \mathbb{L}}{\partial(D_k Y)} \frac{\partial(D_k Y)}{\partial Y} \equiv \frac{\hat{\partial} \mathbb{L}}{\partial Y} + \mathbb{L}^k \frac{\partial(D_k Y)}{\partial Y} \quad (5.5.10)$$

and introduced the following notation:

$$\mathbb{L}^k \equiv \frac{\hat{\partial} \mathbb{L}}{\partial(D_k Y)} = \frac{\hat{\partial} L}{\partial(\partial_k Y)}. \quad (5.5.11)$$

Taking into account general structure of covariant derivatives of objects $Y$ (5.5.1), one can find that equation (5.5.9) can be written in the following view:

$$\hat{\partial}_k \mathbb{L}^k + [\![\mathfrak{G}_k * \mathbb{L}^k]\!] = \frac{\hat{\partial} \mathbb{L}}{\partial Y}; \quad (5.5.12)$$

here symbol $(*)$ denotes an *incomplete* gauge commutator (GC) which is a GC form with exclusion of one term where contravariant index $k$ is transferred from $\mathfrak{L}_a^k$ to $\mathfrak{G}_k$. In rest, reduction of equation (5.5.7) consists of use of the following relation:

$$\hat{\partial}_k = \partial_k + \Gamma^l_{lk} + w_k ; \tag{5.5.13}$$

here object $\Gamma^k_{ml}$ is *matched connection* built on even-symmetric part $\mathfrak{w}_{kl}$ of tensor $w_{kl}$ according to equation (4.1.27), object $w_k$ is vector form (4.1.31). This relation can be derived using relations (5.4.17) and (4.1.29) – (4.1.32). In result, equations (5.5.9) are reduced to an explicit covariant view as follows:

$$\breve{D}_k \mathbb{L}^k + w_k \mathbb{L}^k = \frac{\hat{\partial} \mathbb{L}}{\partial Y} , \tag{5.5.14}$$

here:

$$\breve{D}_k \mathbb{L}^k \equiv \partial_k \mathbb{L}^k + \Gamma^k_{mk} \mathbb{L}^m + [\![\mathfrak{G}_k * \mathbb{L}^k]\!] \tag{5.5.15}$$

is *versified* covariant derivative (VCD) of object $\mathbb{L}^k$ :

$$\breve{D}_l \mathbb{L}^k \equiv \partial_l \mathbb{L}^k + \Gamma^k_{ml} \mathbb{L}^m + [\![\mathfrak{G}_l * \mathbb{L}^k]\!] \tag{5.5.16}$$

contracted on indices $l = k$. The VCD (5.5.15) is a covariant derivative, since it distinguishes from general covariant derivative

$$D_l \mathbb{L}^k \equiv \partial_l \mathbb{L}^k + [\![\mathfrak{G}_l; \mathbb{L}^k]\!] = \partial_l \mathbb{L}^k + G^k_{ml} \mathbb{L}^m + [\![\mathfrak{G}_l * \mathbb{L}^k]\!]$$

only by replacement $G^k_{ml} \to \Gamma^k_{ml}$ in term $G^k_{ml} \mathbb{L}^m$ ; the difference is a tensor type object.

### *5.5.3. EL equations on gauge fields*

Next, we produce a correspondent reduction of equation (5.5.8). Let us introduce the following notations:

$$\Sigma^{alk}_{\mathscr{E}} \equiv \mathbf{\Sigma}^{lk} \equiv \frac{1}{2} \frac{\hat{\partial} \mathbb{L}}{\partial \mathcal{K}_{lk}} = \frac{1}{2} \frac{\hat{\partial} L}{\partial (\partial_l \mathfrak{G}_k)} = -\mathbf{\Sigma}^{kl} ; \tag{5.5.17}$$

$$\boldsymbol{\mathcal{J}}^k \equiv \mathcal{J}^{ak}_{\mathscr{E}} \equiv \frac{1}{2} \mathbb{L}^l \frac{\partial (D_l Y)}{\partial \mathfrak{G}^{\mathscr{E}}_{ak}} . \tag{5.5.18}$$

Note that, matrices (on script indices) $\boldsymbol{\mathcal{J}}^k$ and $\mathbf{\Sigma}^{lk}$ are tensor type objects. We will call object $\mathbf{\Sigma}^{lk}$ *dual curvature form* (DCF), and object $\boldsymbol{\mathcal{J}}^k$ *supercurrent matrix* (CM). Taking into account that,

$$\mathbf{\Sigma}^{mn} \frac{\partial \mathcal{K}_{mn}}{\partial \mathfrak{G}_k} = [\mathfrak{G}_l; \mathbf{\Sigma}^{lk}] , \tag{5.5.19}$$

where $[\mathfrak{G}_l; \mathbf{\Sigma}^{lk}]$ is commutator of matrices $\mathfrak{G}_l$ and $\mathbf{\Sigma}^{lk}$ with contraction on index $l$, we can write equation (5.5.8) in the following view:

$$\hat{\partial}_l \mathbf{\Sigma}^{lk} + [\mathfrak{G}_l; \mathbf{\Sigma}^{lk}] = \boldsymbol{\mathcal{J}}^k , \tag{5.5.20}$$

or

$$\frac{1}{\sqrt{w}}\partial_l(\sqrt{w}\Sigma^{lk}) + [\mathfrak{G}_l; \Sigma^{lk}] = \mathcal{J}^k. \tag{5.5.21}$$

Note that, in this case:

$$[\mathfrak{G}_l; \Sigma^{lk}] = [\![\mathfrak{G}_l ** \Sigma^{lk}]\!]; \tag{5.5.22}$$

here the right hand side of this equation is gauge commutator $[\![\mathfrak{G}_l; \Sigma^{lk}]\!]$ with exemption of two terms where indices $l, k$ of object $\Sigma_{\mathfrak{b}}^{alk}$ are transfer to connection $G_{ml}^n$:

$$[\![\mathfrak{G}_l ** \Sigma^{lk}]\!] \equiv [\![\mathfrak{G}_l; \Sigma^{lk}]\!] - G_{ml}^l \Sigma^{mk} - G_{ml}^k \Sigma^{lm}. \tag{5.5.23}$$

Applying again relation (5.5.13) and taking into account equations (5.5.22), (5.5.23) and odd symmetry of object $\Sigma^{lk}$ as defined by equation (5.5.17), one can represent equation (5.5.20) in the following explicit covariant form:

$$(\breve{D}_l + w_l)\Sigma^{lk} = \mathcal{J}^k; \tag{5.5.24}$$

here

$$\breve{D}_l \Sigma^{lk} = \partial_l \Sigma^{lk} + \Gamma_{nl}^l \Sigma^{nk} + [\mathfrak{G}_l; \Sigma^{lk}] \tag{5.5.25}$$

is *versified* covariant derivative of $\Sigma^{lk}$:

$$\breve{D}_m \Sigma^{lk} \equiv \partial_m \Sigma^{lk} + \Gamma_{nm}^l \Sigma^{nk} + \Gamma_{nm}^k \Sigma^{ln} + [\mathfrak{G}_m; \Sigma^{lk}] \tag{5.5.26}$$

after contraction on indices $m = l$.

*Reduction of EL equations at an even-symmetric Grand Metric tensor*

As discussed above, tensor $w_{kl}$ presumably should have a certain symmetry on its indices: $w_{kl} = \pm w_{lk}$. Choice of the skew symmetry would make impossible receiving covariant EL equations as derived above. Therefore tensor $w_{kl}$ has to be assumed to be the even- symmetric,

$$w_{kl} = w_{lk}. \tag{5.5.27}$$

Then

$$w_l = 0, \tag{5.5.28}$$

and EL equations acquire their final general covariant form as follows:

$$\breve{D}_k \frac{\hat{\partial}\mathbb{L}}{\partial(D_k Y)} = \frac{\hat{\partial}\mathbb{L}}{\partial Y}; \tag{5.5.29}$$

$$\breve{D}_l \frac{\hat{\partial}\mathbb{L}}{\partial \mathcal{K}_{lk}} = \mathcal{J}^k. \tag{5.5.30}$$

## 5.6. EL equations of a DSV-based theory

Let us consider now EL equations in more specific detail referring to approach to UFT exposed in Chapters 3 and 4. According to equations for DSV (3.4.11) profiled in Chapter 3, there are four types of objects on which one may compose differential system of a DSV-based field theory: DSV field $\Psi^\alpha$, $\Phi_\alpha$; UGF $\mathcal{A}_{\alpha k}^\beta$; h-tensors $P_\alpha^{\beta k}$, $\overline{P}_\alpha^{\beta k}$ (below denoted as $\Lambda_\alpha^{\beta k}$); and affine tensor $G_{mk}^n$. Derivations based on general equations (5.5.29), (5.5.30) result in the following EL equations for these objects.

*Equations on DSV*

$$(\Delta_\alpha^\beta \hat{\partial}_k - \mathcal{A}_{\alpha k}^\beta) \frac{\partial L}{\partial(\partial_k \Psi^\beta)} = \frac{\partial \mathbb{L}}{\partial \Psi^\alpha} \ ; \tag{5.6.1}$$

$$(\Delta_\beta^\alpha \hat{\partial}_k + \mathcal{A}_{\beta k}^\alpha) \frac{\partial L}{\partial(\partial_k \Phi_\beta)} = \frac{\partial \mathbb{L}}{\partial \Phi_\alpha} \ . \tag{5.6.2}$$

*Equations on triadic h-tensors*

$$(\Delta_k^n \hat{\partial}_l - G_{kl}^n) \frac{\partial L}{\partial(\partial_l \Lambda_\alpha^{\beta n})} + (\Delta_\beta^\gamma \mathcal{A}_{\varepsilon l}^\alpha - \Delta_\varepsilon^\alpha \mathcal{A}_{\beta l}^\gamma) \frac{\partial L}{\partial(\partial_l \Lambda_\varepsilon^{\gamma k})} = \frac{\partial \mathbb{L}}{\partial \Lambda_\alpha^{\beta k}} \ ; \tag{5.6.3}$$

*Equations on UGF*

$$\frac{1}{2} \mathcal{D}_l \frac{\partial L}{\partial \mathcal{R}_{lk}} = \boldsymbol{J}^k; \tag{5.6.4}$$

here:

$$\boldsymbol{J}^k \equiv \mathcal{J}_\alpha^{\beta k} = \frac{1}{2} \frac{\partial \mathbb{L}}{\partial(\mathcal{D}_l Y)} \frac{\partial(\mathcal{D}_l Y)}{\partial \mathcal{A}_{ak}^\ell} = \frac{1}{2} \left[ \frac{\partial L}{\partial(\partial_k \Psi^\alpha)} \Psi^\beta - \frac{\partial L}{\partial(\partial_k \Phi_\beta)} \Phi_\alpha \right] +$$
$$+ \frac{1}{2} \left[ \frac{\partial L}{\partial(\partial_k \Lambda_\gamma^{\alpha l})} \Lambda_\gamma^{\beta l} - \frac{\partial L}{\partial(\partial_k \Lambda_\beta^{\gamma l})} \Lambda_\alpha^{\gamma l} \right]. \tag{5.6.5}$$

*Equations on affine tensor*

$$\frac{1}{2} \mathcal{D}_l \frac{\partial L}{\partial \mathcal{R}_{lk}} = \boldsymbol{I}^k \ ; \tag{5.6.6}$$

here:

$$\boldsymbol{I}^k \equiv I_n^{mk} = \frac{1}{2} \frac{\partial \mathbb{L}}{\partial(\mathcal{D}_l \Lambda_\beta^{\alpha p})} \frac{\partial(\mathcal{D}_l \Lambda_\beta^{\alpha p})}{\partial G_{mk}^n} = \frac{1}{2} \frac{\partial L}{\partial(\partial_k \Lambda_\beta^{\alpha n})} \Lambda_\beta^{\alpha m} \ . \tag{5.6.7}$$

*A notice on Lagrangian dependence in derivatives of the objects*

Here we have to note the following. DSV and its derivatives are the necessary "elementary" objects of the theory on which, according to consideration in Chapter 3, one can point out a necessity of inclusion of (triadic) h-tensors in Lagrangian structure. On the other hand, inclusion of derivatives of the h-tensors in Lagrangian, in principle, is not necessary in the logical context, since connection of h-tensors to DSV and its (covariant) derivatives could be established without resorting to (covariant) derivatives (CDs) of the h-tensors. In absence of those CDs in Lagrangian, the correspondent EL equations are reduced to algebraic equations relative the h-tensors (assuming a polynomial dependence of Lagrangian on the basic objects as the most corresponding to the *irreducibility* demands):

$$\frac{\hat{\partial}\mathbb{L}}{\partial \Lambda_\alpha^{\beta k}} \Rightarrow \frac{\partial L}{\partial \Lambda_\alpha^{\beta k}} - \frac{1}{2} L w_{mn} \frac{\partial w^{mn}}{\partial \Lambda_\alpha^{\beta k}} = 0 \,. \tag{5.6.8}$$

Expression for the supercurrent matrix (5.6.5) also is reduced to the following:

$$\mathcal{J}^k \equiv \mathcal{J}_\alpha^{\beta k} = \frac{1}{2}\left[\frac{\hat{\partial}L}{\partial(\partial_k \Psi^\alpha)} \Psi^\beta - \frac{\hat{\partial}L}{\partial(\partial_k \Phi_\beta)} \Phi_\alpha\right]. \tag{5.6.9}$$

Consequently, absence of derivatives of the h-tensors in Lagrangian makes inclusion of Riemann-Christoffel form (4.2.1) in Lagrangian also logically not necessary – in distinct to the necessary inclusion of the *hybrid curvature form* (HCF) (4.2.14) required in order to connect *unified gauge field*, hybrid affine tensor $\mathcal{A}_{\beta k}^\alpha$ to dual state vector field $\Psi^\alpha, \Phi_\alpha$.

*Variation derivatives of Lagrangian as structural forms*

Variation derivatives of Lagrangian should be considered not as new objects but as forms structured on basic objects and their covariant derivatives under press of the *irreducibility* demands.

*EAP and covariance of a DSV based theory*

*Extreme Action* method as described reveals a self-contained system of covariant differential equations for a family of basic objects of a field theory. The supposed transformation properties of all the multi-index objects included in Lagrangian are confirmed by (i.e. they are *following from*) the generally covariant form of the derived EL equations (5.5.29) and (5.5.30).

Discussion of covariant connections between objects induced by EAP would be of a significant interest, perhaps not only from a tutorial point of view. Such discussion in full scale, however, will make more sense and clearness when conducted after that the Lagrangian form is specified under press of the rest principles of UFT exhibited in section 3.1. : *differential irreducibility, scale invariance,* and *mini-max.* Here we point the following insights.

1. Before that the EL equations are derived and accepted as *Differential Law*, geometrical nature (i.e. transformation properties) of the multi-index basic objects (triadic h-tensors $P_\beta^{\alpha k}$, $\overline{P}_\beta^{\alpha k}$ and UGF $\mathcal{A}_{\beta k}^\alpha$) in Lagrangian form is only a *supposition*. In fact and in essence, it is explicated not in other way but namely via their coupling to DSV and its derivatives in EL equations.

2. There are in the obtained EL equations the derivatives of the h-tensor forms that result in the left-hand side of equations (5.5.14) and (5.5.24). They automatically acquire a covariant extension due to presence of *Matched Connection* originated by the necessary *weigh factor* $\sqrt{w}$ in generally defined Lagrangian form (5.4.9). This mechanism of covariant extension of the EL differential system is one of the imprescriptible features of the described formulation of EAP. It should be noted that, this extension necessarily presumes the existence of a symmetric non-degenerated tensor $w_{kl}$, regardless to its structural genesis. By the way, this tensor can be viewed not as one of basic objects of EAP but structured on the triadic h-tensors as basic objects.

3. Transformation properties of DSV are not established yet but found to be different from that of vectors of the unified manifold based on *invariance of the Differential Law as necessary logical presumption of a background fundamental law*. However, DL can be formulated only in terms of DSV itself based on the derived EL equations. Therefore, transformation properties of DSV being of an essential relevance to DL can be finally explicated only after establishing the Lagrangian form, deriving EL equations of the considered approach to irreducible field theory and implementing the requirement of *transformational invariance* of DL for DSV (*invariant differential law*, IDL).

*EAP as path to establishing dimensionality of SFT*

Statement of *invariant differential law* (IDL) was generally formulated (see section 3.1.) on basis of the *unified manifold − matter function differential homomorphism* as invariant connection between directions of the "world lines" of two spaces. This disposition, however, is not sufficient for establishing of an unambiguous correspondence between dimensionalities of two spaces. Finding of this correspondence and dimensionality itself could be envisioned to be attained based on consideration of possible (yet irreducible) *local structural isomorphism* of two spaces – in frame of the EAP-based covariant DL complemented with the IDL principle.

Deriving Lagrangian and EL equations is treated in Chapters 6 through 9. Before start this, we will consider *general contracted equations* i.e. system of simplified relations (*dynamic identities*) those follow from the derived general EL equations of a DSV-based field theory.

### 5.7. Contracted Equations

We call *dynamics identity* any particular exact relations between objects or their components and derivatives which are derived based on EL equations and result after contraction on script (matrix) indices. There is quite a sizable collection of such relations.

#### 5.7.1. The extended Faraday equations

First we mark equations that follow directly from the background skew symmetry of the introduced *hybrid curvature form* (HCF) (4.2.14). It should be underlined that, together with other forms, transformation properties of this object as h-tensor are now explicated based on the derived system of EL equations.

Contraction of $\mathfrak{R}^{\alpha}_{\beta kl}$ on Greek indices result in a skew-symmetric covariant tensor $\mathfrak{R}_{kl}$:

$$\mathfrak{R}^{\alpha}_{\alpha kl} = \partial_k \mathcal{A}^{\alpha}_{\alpha l} - \partial_l \mathcal{A}^{\alpha}_{\alpha k} \equiv \mathfrak{R}_{kl} = -\mathfrak{R}_{lk} \,. \tag{5.7.1}$$

This result also follows immediately from matrix representation (4.2.21):

$$Tr\mathfrak{R}_{kl} = \partial_k \mathcal{A}_l - \partial_l \mathcal{A}_k \equiv \mathfrak{R}_{kl} = -\mathfrak{R}_{lk} \,, \tag{5.7.2}$$

since $Tr[\mathcal{A}_k; \mathcal{A}_l] \equiv 0$; here we have introduced a notation:

$$\mathcal{A}_k \equiv Tr\mathcal{A}_k = \mathcal{A}^{\alpha}_{\alpha k} \,. \tag{5.7.3}$$

Definition of tensor $\mathfrak{R}_{kl}$ reminds tensor of the electromagnetic field in STM defined as curl of a covariant 4-vector-potential $A_k$:

$$F_{kl} \equiv \partial_k A_l - \partial_l A_k \,. \tag{5.7.4}$$

As known, the *first pair of Maxwell equations* follows directly from this definition [9]:

$$\partial_m F_{kl} + \partial_l F_{mk} + \partial_k F_{lm} = 0 \,. \tag{5.7.5}$$

These equations could be called *Faraday equations*.

As for any skew-symmetric covariant tensor, due to the even symmetry of *Matched Connection* $\Gamma^m_{kl} = \Gamma^m_{lk}$, there takes place a reduction of the alternated covariant derivative of tensor $\mathfrak{R}_{kl}$ (regardless to the dimensionality of the unified manifold):

$$\nabla_{[k} \mathfrak{R}_{lm]} \Rightarrow \partial_k \mathfrak{R}_{lm} + \partial_m \mathfrak{R}_{kl} + \partial_l \mathfrak{R}_{mk} = \partial_{[k} \mathfrak{R}_{lm]} \,. \tag{5.7.6}$$

Further, definition of tensor $\mathfrak{R}_{kl}$ as in (4.2.25) results in the following covariant equations:

$$\partial_k \mathfrak{R}_{lm} + \partial_m \mathfrak{R}_{kl} + \partial_l \mathfrak{R}_{mk} = 0 \,, \tag{5.7.7}$$

hence,

$$\nabla_k \mathfrak{R}_{lm} + \nabla_m \mathfrak{R}_{kl} + \nabla_l \mathfrak{R}_{mk} = 0 \,.$$

Both identities (5.7.5) and (5.7.7) are generally covariant. We call equations (5.7.7) suerdimensional *Faraday equations* (SFE).

### *5.7.2. Contracted EL equations*

The contracted EL equations are obtained by contraction of EL equations on the matrix indices (i.e. script ones in the previous Chapter).

*Contracted gauge equations*

By contraction of equation (5.6.4) over indices $\beta = \alpha$ we obtain the following vector equation:

$$\breve{\partial}_l Tr \frac{\hat{\partial} L}{\partial \mathfrak{R}_{lk}} \equiv \frac{1}{\sqrt{w}} \partial_l (\sqrt{w} \Sigma^{kl}) = J^k \,, \tag{5.7.8}$$

with vector of *supercurrent*

$$\mathcal{J}^k \equiv \mathcal{J}_\alpha^{\alpha k} = \frac{1}{2}\left[\frac{\partial L}{\partial(\partial_k \Psi^\alpha)}\Psi^\alpha - \frac{\partial L}{\partial(\partial_k \Phi_\alpha)}\Phi_\alpha\right], \tag{5.7.9}$$

where *current matrices* $\mathcal{J}_\beta^{\alpha k}$ are given by equations (5.6.5). Tensor $\Sigma^{kl}$ is a dual party to tensor $\mathfrak{R}_{kl}$, due to that object $\Sigma_\beta^{\alpha kl}$ is a dual party to *hybrid curvature form* $\mathfrak{R}_{\alpha kl}^\beta$ as determined by EL equation (5.5.8).

Equation (5.6.4) has an external similarity to Maxwell's equation in Electrodynamics:

$$\nabla_l F^{kl} = j^k. \tag{5.7.10}$$

We call equations (5.7.8) *superdimensional Maxwell equations* (SME). Equations (5.7.7) and (5.7.8) considered together could be viewed as a presage of a *superdimensional Faraday−Maxwell system* (SFM).

In case when Lagrangian includes derivatives of hybrid triads $\Lambda_\beta^{\alpha k}$, contraction of equations (5.6.6) on script indices results in the following relations:

$$\breve{\partial}_l Tr \frac{\partial L}{\partial \boldsymbol{\mathcal{R}}_{lk}} \equiv \frac{1}{\sqrt{w}} \partial_l \left(\sqrt{w} Tr \frac{\partial L}{\partial \boldsymbol{\mathcal{R}}_{lk}}\right) = Tr \boldsymbol{I}^k \equiv I^k. \tag{5.7.11}$$

*Conservation of the superdimensional current*

The skew symmetry of tensor $\Sigma^{kl}$ leads to *conservation of generalized current* $\mathcal{J}^k$ i.e.:

$$\nabla_k \mathcal{J}^k = 0 \tag{5.7.12}$$

as a direct consequence of contracted equations (5.7.8). This follows from a background property:

$$\nabla_k \nabla_l \Sigma^{kl} \equiv 0, \tag{5.7.13}$$

as for any skew symmetric valence 2 contravariant tensor:

$$\nabla_k \nabla_l \Sigma^{kl} = \frac{1}{\sqrt{w}} \partial_k \left(\sqrt{w}\nabla_l \Sigma^{kl}\right) = \frac{1}{\sqrt{w}} \partial_k \partial_l \left(\sqrt{w}\Sigma^{kl}\right) \equiv 0. \tag{5.7.14}$$

It should be underlined that, in turn, skew symmetry of tensor $\Sigma^{kl}$ is directly due to the background skew symmetry of *curvature form* $\mathcal{K}_{akl}^\ell$ on indices $k, l$ as covariant derivative of the covariant connection $\mathcal{A}_{ak}^\ell$. Our approach to UFT recognizes the generalized curvature form $\mathcal{K}_{akl}^\ell$ as genetic basis for valence 2 skew-symmetric co-and contra-variant tensors of rank $N$ in field theory. It would be premature, however, to consider object $\Sigma^{kl}$ as direct analog of tensor $F^{kl}$ in the 4d Electrodynamics (though extended to $N$ dimensions). Vector object $\mathcal{J}^k$ in SFT is likely to be of a more general meaning than directly ("simply") associated with the electric currents.

Thus, if the supercurrent does arrive in a covariant theory, it arrives *conservative*.

*Conditions of existence of the supercurrent*

Existence of the supercurrent should be considered as *necessary background property* of a unified field theory. Apparently, its existence implies the following requirements to Lagrangian structure:

$$Tr\boldsymbol{\mathcal{J}}^k \equiv \delta_a^{\theta} \mathcal{J}_{\theta}^{ak} \not\equiv 0 \qquad (5.7.15)$$

and

$$Tr\boldsymbol{\Sigma}^{kl} \equiv \delta_a^{\theta} \Sigma_{\theta}^{akl} \not\equiv 0, \qquad (5.7.16)$$

where matrices (on script indices) $\boldsymbol{\mathcal{J}}^k$ and $\boldsymbol{\Sigma}^{kl}$ are given by general formulas (5.5.18) and (5.5.17). Here we omit a treat of contributions to objects $\Sigma^{kl}$ and $\mathcal{J}^k$ from possible various basic objects of EAP, though it should be noted that, contributions to the supercurrent from triadic h-tensors $\Lambda_{\beta}^{\alpha k}$ are nullified as it follows from equation (5.6.5).

*A note about the conservatism of the supercurrent*

It should be noted that, at this stage of profiling Lagrangian structure, equation (5.7.12) may be considered not as requirement to match an intrinsic property (5.7.13) but, more generally, as one of the direct *dynamic consequence*s revealed by the total system of EL equations including equations on *covariant connections* (5.5.30). Indeed, system of EL equations (5.5.29), (5.5.30) connect the involved forms as functions of the manifold variables considered as solution of the EL differential system overall the region of definition of the functions. Therefore, once property (5.6.13) is valid (as a background structural identity) for the form on the left hand side of equations (5.6.1), it also is valid for function of variables on the right hand side considered as *solution* of the derived differential system. However, at further specification of Lagarangian towards the *irreducible* form, property (5.7.12) might arrive as a *dynamical identity* resulting already from EL equations on the tensor type basic objects $Y$ (5.5.29) and associated with the dynamic genesis of these objects in a dual-covariant theory.

*Other contracted EL equations*

In general formalism, other EL equations that can be contracted on the script indices, would be equations on the hybrid tensor type objects $\Lambda_{\beta}^{\alpha k}$ (matrices on Greek indices). We leave such cases to consideration after deriving EL equations with Lagrangian structured under the press of the *irreducibility* demands.

### 5.7.3. The Composite Dynamic Identities

*General formulation*

Euler-Lagrange equations of a covariant field theory arrive as system of covariant relations between tensor type forms structured on basic objects and their covariant derivatives. Tensor type of these relations allows one to recognize a variety of simplified relations which can be obtained by contraction on the MF and (or) UM indices not only of EL equations themselves but also in product with basic objects or their covariant derivatives. Let there is an (irreducible) system of tensor type forms $\mathbb{Z}^a$, then the contracted equations can be written in a symbolic form as follows:

$$\langle \mathbb{Z}^b \hat{\nabla}_a \mathcal{L} \rangle \equiv \langle \mathbb{Z}^b \frac{1}{\sqrt{w}} [\partial_k \frac{\partial(L\sqrt{w})}{\partial(\partial_k X^a)} - \frac{\partial(L\sqrt{w})}{\partial X^a}] \rangle = 0 . \tag{5.7.17}$$

Here angle brackets are for contraction of such products on Greek or Roman indices or both of them.

In particular cases when Lagrangian does not include derivatives of a (tensor type) object $X^a$, the related contracted equations are reduced to the following ones:

$$\langle \mathbb{Z}^b \frac{\partial(L\sqrt{w})}{\partial X^a} \rangle = 0 . \tag{5.7.18}$$

### 5.7.4. Hamilton – Nöther equation

As generally known [1, 2], when considering *complete* derivatives $\partial_k \mathcal{L}$ of Lagrangian $\mathcal{L}$, one finds the following generic *dynamic identities*:

$$\partial_k \mathcal{L} = \partial_l [\frac{\partial \mathcal{L}}{\partial(\partial_l X^a)} \partial_k X^a] ; \tag{5.7.19}$$

or

$$\partial_l (\sqrt{w} \mathcal{H}_k^l) = 0 ; \tag{5.7.20}$$

here we introduced a mixed valence 2 *pseudo-tensor* object:

$$\mathcal{H}_k^l \equiv \frac{1}{\sqrt{w}} \frac{\partial(L\sqrt{w})}{\partial(\partial_l X^a)} \partial_k X^a - \delta_k^l L . \tag{5.7.21}$$

In our case:

$$\mathcal{H}_k^l \equiv \mathbb{L}_a^l \partial_k Y^a + 2\Sigma_{\ell}^{alm} \partial_k \mathfrak{G}_{am}^{\ell} - \delta_k^l \mathbb{L} . \tag{5.7.22}$$

We call object $\mathcal{H}_k^l$ *Hamilton – Nöther form*. Meaning and use of this object and equation (5.7.20) should be considered after that the Lagrangian form and EL equations are specified under press of principles exhibited above in Chapter 1 (*Prolegomena*).

### *Resume of Chapter 5*

Formulation of the Extreme Action method for generally covariant "classical" field theory in a real $N$ dimensions manifold has been conducted. As a step to forming *Lagrangian* and action integral of SFT, we have built the basic scalar forms and introduced the *Grand Metric* tensor, all presumably structured on DSV and the introduced triadic objects. Two generic symmetry principles are implied at use of the EAP method: *homogeneity* and *uniformity* of the differential system. Lagrangian form of the action integral is supposed to be structured on vector, tensor and (or) s-vector and hybrid tensor objects (generally denoted $Y$), their covariant derivatives (CD), those include the connection objects (generally noted $\mathfrak{G}^a_{bk}$): *unified gauge field* (UGF) (or *hybrid Christoffel symbols*) and *affine tensor* (AT) of the differential geometry; and covariant derivatives of the connections (*Riemann-Christoffel form, RCF* for AT and *hybrid curvature form, HCF* for UGF). All objects are implied *real* analytical functions of the manifold variables. Utilization of general covariance in this approach could be considered as unification and extension of the *gauge principle* of QFT and covariance features of GTR.

The generally covariant Euler-Lagrange (EL) equations on the introduced basic objects have been derived. There is an intrinsic generic outcome of equations on the *connection objects*((unified gauge field and conventional Christoffel symbols): existence of contravariant vector fields (generalized currents) and valence 2 the skew-symmetric contravariant tensor fields. These objects are coupled by the divergence equation for the skew-symmetric tensor field (an external analog to Maxwell equations); due to this coupling, vector fields are conservative.
Other possible contracted forms are generally profiled based on the derived EL equations.

Next, we will specify Lagrangian form of the DSV-based differential system by posing the *irreducibility* demands suited for unified field theory in our sight.

# 6. Structuring of the SFT Lagrangian

## 6.1 Principles of Irreducible Superdimensional Field Theory

The differential system of a unified theory should be subordinated to a set of principles. Such principles should be based on possible logical arguments rather than on heuristic or esthetic ones. On the other hand, the logic of structuring the system cannot be produced "from nothing", but should be guided by observation of the genetic history of principles and mathematical background of the fundamental theoretical physics. Studies in this work on approaching the differential system of a unified theory were inspired by the *quantum legacy* of P. Dirac, the *covariance* paradigm of A. Einstein, *gauge theory* of QFT, and the *irreducibility* demand of W. Pauli. The considerations have led to the following set of principles.

1. **Homomorphism:** the fundamental law is considered as system of $\mu$ homomorphic functions $\varphi^\alpha$ (*matter function*; $\alpha = 1.2, \ldots \mu$) of variables of $N$-dimensional *unified manifold* $\check{\psi}^k$ (UM; $k = 1,2,\ldots N$), subordinate to an autonomic differential algorithm (*differential law*, DL).

2. **Duality:** DL can be formulated not directly for MF but for *Dual State Vector* field (DSV) $\Psi^\alpha, \Phi_\alpha$ associated with derivatives of the UM $\rightarrow$ MF homomorphism.
   This direction is due to a consideration that, not MF itself but its derivatives can possess certain transformation properties based on transformation of MF differentials, hence to be subordinate to a differential law.

3. **Homogeneity:** the differential system should be formulated only as relations between the involved basic objects $X^a$ and their derivatives and should not include any explicit functions of variables $\check{\psi}^k$ of the $N$-dimensions manifold.
   The sense of this principle consists of that, when pursuing formulation of a fundamental law, one should not resort to assumptions neither *ad hoc* nor with references to "reality" about behavior of the involved objects as function of the manifold variables.
   This principle may seem a "routine" one at the first glance, since it is a basic declaration of QFT as a field theory in the 4-dimensional space-time manifold. Note, however, that, QFT does not follow this principle when treating the dynamic law for the *state vector* (SV) i.e. the secondary quantization function: while considering SV *in fact* as function of fields $Q$, QFT at the same time utilizes representation of the *differential law* for *state vector* as *Schrödinger equation* in which *Hamiltonian* as *energy operator* is an *explicit function* (form) of $Q$ (and differentials $\frac{\partial}{\partial Q}$).

4. **Uniformity:** the equations should be symmetric (i.e. *homogenized*) over all variables and components of every involved object.

5. **Reality:** DL as an analytical algorithm should be formulated in all real terms; the *imaginary unit* "$i$", and *complex* objects are not admitted.
   This critical restriction is imposed due to a consideration that the presence of *invariable objects*, like "$i$", is not compatible with the requirement of *general covariance* of the theory.

6. *General covariance*: the equations must not change their tensor structural form at arbitrary transformations of variables.

Principles 1 to 6 have already been applied for deriving equations for DSV in Chapter 3. Now, when targeting the deriving of the complete system of equations for whole collection of objects including the triadic coefficient functions, we add the following principles and requirements based on the treat exposed in Chapters 4 and 5.

7. *Extreme Action* principle (EAP): equations for DSV and the related triadic objects should be derived based on the variation method of the *extreme action* as a universal principle of the *dynamic balance*.

8. *Differential irreducibility*: Euler-Lagrange equations on DSV should connect first derivative of this object to itself and not include higher order derivatives. EL equations on the triadic objects (coefficient functions) should be formulated in the lowest order derivatives, in correspondence to the all the imposed principles and *irreducibility* demands.

9. *Existence of a conservative supercurrent* in general case is an intrinsic property of an EAP-based dual-covariant differential system as considered in section 5.7. Its nullification at possible particular structuring of Lagrangian would be equivalent to posing a specific degenerating restriction on dynamics of SFT which cannot be motivated from a physical or mathematical point of view. In contrary to such degeneration, existence of conservative supercurrent should be regarded as one of necessary dynamical properties of SFT as approach to UFT.

10. *Scale invariance* principle: Lagrangian form should be invariant relative of introduction of arbitrary real constants as multipliers at its items. The definition "invariant" implies that structure of Lagrangian does possess immunity to such deformation, by scaling of objects $X^a$ and variables $\breve{\psi}^k$. Note that, the *scale invariance* of a unified theory can be realized as a consistent principle *only at use of EAP* as a way to derive the basic equations. In this way, EAP becomes an inseparable and indispensable recipe for deriving basic equations of a unified theory.

11. *Mini-max* principle: while under press of all the above exhibited demands, Lagrangian should include a *maximum* variety of possible scalar items.

The exhibited set of principles can be considered as specification for implementation of a general and universal requirement that should be asserted to a fundamental field theory - the *irreducibility* principle, in accordance with legacy of W. Pauli [8].

### 6.2. Preliminary Matter Lagrangian and EL equations on DSV

As pointed above, the Lagrangian of UFT should include the first-only derivatives of objects. It follows from principle of *differential irreducibility* of equations for DSV that Lagrangian should also be a form *linear* in the derivatives of DSV; otherwise, the derived EL equations will include the second order derivatives of DSV. Together with the principle of general covariance, this requirement greatly reduces selection of possible structuring of the Lagrangian on DSV.

*General Lagrangian form linear on the derivatives*

For a Lagrangian form linear in the derivatives of objects, we can write a general expression as follows:

$$\mathcal{L}(X, \partial_k X) = \mathcal{L}_0(X) + \mathcal{L}_a^k(X)\partial_k X^a . \tag{6.2.1}$$

Substitution of this form into the general EL equations (6.1.29) results in the following general form of the *first only derivatives* differential system of a field theory:

$$\left(\frac{\partial \mathcal{L}_a^k}{\partial X^b} - \frac{\partial \mathcal{L}_b^k}{\partial X^a}\right)\partial_k X^b = \frac{\partial \mathcal{L}_0}{\partial X^a} . \tag{6.2.2}$$

It is obvious, that, if Lagrangian form contains the second and a higher power of the first order derivatives, the EL equations include the second order derivatives. Such extension contradicts to the principles of structural symmetry and irreducibility of a differential law for DSVderived based on the variation principle. On the other hand, there is no approach to formulation of the fundamental field theory equations that universal, clear, and consistent other than the Extreme Action, taken as principle of the *dynamic balance*.

*Covariant form of the matter scalar*

Now we will consider a possible scalar Lagrangian form that would lead to the initially proposed equations of type (3.3.35). In accordance to covariance requirements (6.2.4) we have to write the related Lagrangian as follows:

$$\mathcal{L}(X, \partial X) \implies \mathbb{L}_M(\Xi, \mathfrak{D}\Xi)\sqrt{w} . \tag{6.2.3}$$

Scalar Lagrangian $\mathbb{L}_M$ can be composed based on the scalar forms shown in part 5.1.:

$$\mathbb{L}_M = \Phi_\alpha \Psi^\alpha + \Lambda_\beta^{\alpha k}\Phi_\alpha \mathfrak{D}_k \Psi^\beta + \overline{\Lambda}_\beta^{\alpha k}\Psi^\beta \mathfrak{D}_k \Phi_\alpha , \tag{6.2.4}$$

where

$$\mathfrak{D}_k \Psi^\beta = \partial_k \Psi^\beta + \mathcal{A}_{\gamma k}^\beta \Psi^\gamma$$

$$\mathfrak{D}_k \Phi_\alpha = \partial_k \Phi_\alpha - \mathcal{A}_{\alpha k}^\gamma \Phi_\gamma \tag{6.2.5}$$

are covariant derivatives of DSV. Here we have introduced to the Lagrangian structure two h-tensors or *h-triads*, $\Lambda_\beta^{\alpha k}$ and $\overline{\Lambda}_\beta^{\alpha k}$, transformed according to definition (3.4.8), and an affine h-tensor $\mathcal{A}_{\beta k}^\alpha$ transformed according to (3.3.7). We consider these objects as the basic ones in the Lagrangian form along with the dual s-field $\Psi^\alpha, \Phi_\alpha$ i.e. subjects of independent variation when deriving the Euler-Lagrange equations. Matter scalar (6.2.4) arrives as a form structured on family of five basic objects, subjects of independent variations in the Extreme Action Principle:

$$\{X\} = \Psi^\alpha, \Phi_\alpha\ ;\quad \Lambda^{\alpha k}_\beta, \overline{\Lambda}^{\alpha k}_\beta\ ;\quad \mathcal{A}^\alpha_{\beta k}\,. \tag{6.2.6}$$

*Reduction of Grand Metric dependence on the basic objects*

Above in section 5.5., general structure of a covariant Lagrangian has been established at no assumptions about structure of GM tensor $w_{kl}$ as a form built on basic objects (unless this tensor itself is considered as one of the basic objects). Here we assume that its structure does not include DSV and its covariant derivatives. Structure of this tensor will be discussed and figured out below in our treat for *geometry* or *gauge Lagrangian* (section 6.4 and 6.5.). We also assume tensor $w_{kl}$ symmetric:

$$w_{kl} = w_{lk}\,. \tag{6.2.7}$$

*Initial EL equations on DSV*

Applying now the EL equations:

$$\partial_k \frac{\partial \mathcal{L}}{\partial(\partial_k \Phi_\alpha)} - \frac{\partial \mathcal{L}}{\partial \Phi_\alpha} = 0\,,$$

$$\partial_k \frac{\partial \mathcal{L}}{\partial(\partial_k \Psi^\alpha)} - \frac{\partial \mathcal{L}}{\partial \Psi^\alpha} = 0 \tag{6.2.8}$$

to Lagrangian (6.2.3), we can write these equations in the following form:

$$\frac{1}{\sqrt{w}} \partial_k \left[\sqrt{w}\, \frac{\partial \mathbb{L}_M}{\partial(\partial_k \Phi_\alpha)}\right] - \frac{\partial \mathbb{L}_M}{\partial \Phi_\alpha} = 0\,;$$

$$\frac{1}{\sqrt{w}} \partial_k \left[\sqrt{w}\, \frac{\partial \mathbb{L}_M}{\partial(\partial_k \Psi^\alpha)}\right] - \frac{\partial \mathbb{L}_M}{\partial \Psi^\alpha} = 0\,. \tag{6.2.9}$$

For convenience of the further derivations let us rewrite form (6.2.1) as follows:

$$\mathbb{L}_M = \Phi_\alpha \Psi^\alpha + \Lambda^{\alpha k}_\beta \Phi_\alpha\, \partial_k \Psi^\beta + \overline{\Lambda}^{\alpha k}_\beta \Psi^\beta\, \partial_k \Phi_\alpha) + \mathcal{A}^\alpha_\beta \Phi_\alpha \Psi^\beta \tag{6.2.10}$$

where

$$\mathcal{A}^\alpha_\beta \equiv \Lambda^{\alpha k}_\gamma \mathcal{A}^\gamma_{\beta k} - \overline{\Lambda}^{\gamma k}_\beta \mathcal{A}^\alpha_{\gamma k}\,. \tag{6.2.11}$$

Performing the variation derivatives as prescribed by equations (6.2.9), we obtain the following dual pair of equations:

$$(\overline{\Lambda}^{\alpha k}_\beta - \Lambda^{\alpha k}_\beta)\, \partial_k \Psi^\beta + \left[\left(\partial_k + \frac{\partial_k w}{2w}\right) \overline{\Lambda}^{\alpha k}_\beta - \mathcal{A}^\alpha_\beta\right] \Psi^\beta - \Psi^\alpha = 0\,; \tag{6.2.12}$$

$$(\Lambda^{\beta k}_\alpha - \overline{\Lambda}^{\beta k}_\alpha)\, \partial_k \Phi_\beta + \left[\left(\partial_k + \frac{\partial_k w}{2w}\right) \Lambda^{\beta k}_\alpha - \mathcal{A}^\beta_\alpha\right] \Phi_\beta - \Phi_\alpha = 0. \tag{6.2.13}$$

Derived based on EAP, equations for co- and contra-variant s-vectors have the h-tensor coefficients before the derivatives of s-vector fields (compared after transposition of one of them on indices $\alpha, \beta$) of same-magnitudes but opposite-sign. It will be shown below that, these equations are equivalent to those initially profiled as in (3.4.11).

*Covariant form of matter equations*

We have to show the general covariance of dual EL equations (6.2.12), (6.2.13) in accordance to assumed transformation law for object $\mathcal{A}^{\alpha}_{\beta k}$ (3.3.7). At first, using the relation (see equation (4.1.31)) :

$$\frac{\partial_m w}{2w} = \Gamma^n_{nm}, \qquad (6.2.14)$$

we replace derivative $\partial_k w$ in equations (6.2.12) and (6.2.13) :

$$(\bar{\Lambda}^{\alpha k}_{\beta} - \Lambda^{\alpha k}_{\beta}) \partial_k \Psi^{\beta} + \left[(\partial_k + \Gamma^n_{nk})\bar{\Lambda}^{\alpha k}_{\beta} - \mathcal{A}^{\alpha}_{\beta}\right]\Psi^{\beta} - \Psi^{\alpha} = 0;$$

$$(\Lambda^{\beta k}_{\alpha} - \bar{\Lambda}^{\beta k}_{\alpha}) \partial_k \Phi_{\beta} + \left[(\partial_k + \Gamma^n_{nk})\Lambda^{\beta k}_{\alpha} - \mathcal{A}^{\beta}_{\alpha}\right]\Phi_{\beta} - \Phi_{\alpha} = 0. \qquad (6.2.15)$$

Next, we substitute the ordinary derivatives in these equations via the covariant ones according to equations (6.3.2), (6.3.3) and definition of covariant derivatives of h-tensors according to equation (4.1.38). Then the dual equations acquire the following explicit covariant form:

$$(\bar{\Lambda}^{\alpha k}_{\beta} - \Lambda^{\alpha k}_{\beta})\mathfrak{D}_k \Psi^{\beta} + (\mathfrak{D}_k \bar{\Lambda}^{\alpha k}_{\beta})\Psi^{\beta} - \Psi^{\alpha} = 0 ;$$

$$(\Lambda^{\beta k}_{\alpha} - \bar{\Lambda}^{\beta k}_{\alpha})\mathfrak{D}_k \Phi_{\beta} + (\mathfrak{D}_k \Lambda^{\beta k}_{\alpha})\Phi_{\beta} - \Phi_{\alpha} = 0 . \qquad (6.2.16)$$

Here terms $\mathfrak{D}_k \Lambda^{\beta k}_{\alpha}$ and $\mathfrak{D}_k \bar{\Lambda}^{\alpha k}_{\beta}$ are *covariant divergences* h-tensors $\Lambda^{\beta k}_{\alpha}$ and $\bar{\Lambda}^{\alpha k}_{\beta}$ obtained from *versified* covariant derivatives of these objects according to general definition (4.1.38), but contracted on vector (Roman) indices ($l = k$):

$$\mathfrak{D}_k \Lambda^{\alpha k}_{\beta} \equiv \partial_k \Lambda^{\alpha k}_{\beta} + \Gamma^k_{mk}\Lambda^{\alpha m}_{\beta} + \mathcal{A}^{\alpha}_{\gamma k}\Lambda^{\gamma k}_{\beta} - \mathcal{A}^{\gamma}_{\beta k}\Lambda^{\alpha k}_{\gamma} ;$$

$$\mathfrak{D}_k \bar{\Lambda}^{\alpha k}_{\beta} \equiv \partial_k \bar{\Lambda}^{\alpha k}_{\beta} + \Gamma^k_{mk}\bar{\Lambda}^{\alpha m}_{\beta} + \mathcal{A}^{\alpha}_{\gamma k}\bar{\Lambda}^{\gamma k}_{\beta} - \mathcal{A}^{\gamma}_{\beta k}\bar{\Lambda}^{\alpha k}_{\gamma} . \qquad (6.2.17)$$

These objects are s-tensors, as well as $w_k \Lambda^{\beta k}_{\alpha}$ and $w_k \bar{\Lambda}^{\beta k}_{\alpha}$. Equations (6.2.16) are generally covariant under condition that, object $\mathcal{A}^{\alpha}_{\beta k}$ is *in fact* transformed according to equation (3.3.7). Transformation law for $\mathcal{A}^{\alpha}_{\beta k}$ to be proved below.

*Equivalence of EL equations to the pre-viewed DSV equations*

Equations (6.2.16) can be transformed to a view of equations for DSV initially derived in section 3.4. (see (3.4.11)). To show this, let us write the derived covariant equations (6.2.16) in the following view:

$$(\Lambda_\beta^{\alpha k} - \overline{\Lambda}_\beta^{\alpha k})\mathfrak{D}_k\Psi^\beta = -\overline{M}_\beta^\alpha\Psi^\beta \, ;$$

$$(\Lambda_\alpha^{\beta k} - \overline{\Lambda}_\alpha^{\beta k})\mathfrak{D}_k\Phi_\beta = M_\alpha^\beta\Phi_\beta \, ,$$

(6.2.18)

where

$$\overline{M}_\beta^\alpha \equiv \delta_\beta^\alpha - \mathfrak{D}_k\,\overline{\Lambda}_\beta^{\alpha k} \, ;$$

$$M_\alpha^\beta \equiv \delta_\alpha^\beta - \mathfrak{D}_k\Lambda_\alpha^{\beta k} \, .$$

(6.2.19)

Now, multiplying these equations by matrices inverse to matrices $\overline{M}_\beta^\alpha$ and $M_\alpha^\beta$, respectively, we obtain our equations in a form which is an explicit equivalence to equations (3.4.11) with the following relations:

$$P_\beta^{\alpha k} = -(\overline{M}_\gamma^\alpha)^{-1}(\Lambda_\beta^{\gamma k} - \overline{\Lambda}_\beta^{\gamma k}) \, ;$$

$$\overline{P}_\beta^{\alpha k} = (M_\gamma^\alpha)^{-1}(\Lambda_\beta^{\gamma k} - \overline{\Lambda}_\beta^{\gamma k}) \, .$$

(6.2.20)

We conclude with that, based on the Extreme Action principle complemented with the set of the above exhibited principles, we have attained a "special realization" of initially derived dual equations (3.4.11).

### 6.3. Reduction of Split Metric and matter Lagrangian

In paper [1*] matrices $\boldsymbol{\Lambda}^k$ and $\overline{\boldsymbol{\Lambda}}^k$ were considered as subjects of independent variation in the Lagrange formalism, so the total number of basic objects in the action integral was five. In this paper we will explore a SFT concept with only four basic objects, reducing Split Metric to a set of $N$ matrices, in formal terms posing a condition $\overline{\boldsymbol{\Lambda}}^k = -\boldsymbol{\Lambda}^k$. Substantiation of such a reduction is in the following consideration. EL equations (6.2.12) and (6.2.13) can be written in the following view::

$$(\overline{\Lambda}_\beta^{\alpha k} - \Lambda_\beta^{\alpha k})\mathfrak{D}_k\Psi^\beta + \frac{1}{2}[\mathfrak{D}_k(\overline{\Lambda}_\beta^{\alpha k} - \Lambda_\beta^{\alpha k})]\Psi^\beta + \frac{1}{2}[\mathfrak{D}_k(\overline{\Lambda}_\beta^{\alpha k} + \Lambda_\beta^{\alpha k})]\Psi^\beta - \Psi^\alpha = 0 \, ;$$

$$(\Lambda_\alpha^{\beta k} - \overline{\Lambda}_\alpha^{\beta k})\mathfrak{D}_k\Phi_\beta + \frac{1}{2}[\mathfrak{D}_k(\Lambda_\alpha^{\beta k} - \overline{\Lambda}_\alpha^{\beta k})]\Phi_\beta + \frac{1}{2}[\mathfrak{D}_k(\Lambda_\alpha^{\beta k} + \overline{\Lambda}_\alpha^{\beta k})]\Phi_\beta - \Phi_\alpha = 0 \, ,$$

(6.3.1)

with matter scalar written correspondently as follows:

$$\mathbb{L}_M = \Phi_\alpha\Psi^\alpha + \frac{1}{2}(\Lambda_\beta^{\alpha k} - \overline{\Lambda}_\beta^{\alpha k})(\Phi_\alpha\mathfrak{D}_k\Psi^\beta - \Psi^\beta\mathfrak{D}_k\Phi_\alpha) + \frac{1}{2}(\overline{\Lambda}_\beta^{\alpha k} + \Lambda_\beta^{\alpha k})\mathfrak{D}_k(\Psi^\beta\Phi_\alpha). \quad (6.3.2)$$

Observing structure of equations (6.3.1) and Lagrangian in form (6.3.2), we recognize that, presence of matrices $\Lambda_\alpha^{\beta k} - \overline{\Lambda}_\alpha^{\beta k}$ is of a critical meaning, since they are necessary for utilization of connection between DSV and its derivatives in principle according to equations (6.3.1), while matrices $\Lambda_\beta^{\alpha k} + \overline{\Lambda}_\beta^{\alpha k}$ do not play such a critical role. So, one can consider a theory with reduced system of DSV equations, removing the sum matrices and simply using notation $\Lambda_\alpha^{\beta k}$ for the difference matrices:

$$\Lambda_\beta^{\alpha k} \mathfrak{D}_k \Psi^\beta + \frac{1}{2}(\mathfrak{D}_k \Lambda_\beta^{\alpha k})\Psi^\beta + \Psi^\alpha = 0 ;$$

$$\Lambda_\alpha^{\beta k} \mathfrak{D}_k \Phi_\beta + \frac{1}{2}(\mathfrak{D}_k \Lambda_\alpha^{\beta k})\Phi_\beta - \Phi_\alpha = 0 .$$

(6.3.3.)

with reduced matter scalar:

$$\mathbb{L}_M \Longrightarrow \Phi_\alpha \Psi^\alpha + \frac{1}{2}\Lambda_\beta^{\alpha k}(\Phi_\alpha \mathfrak{D}_k \Psi^\beta - \Psi^\beta \mathfrak{D}_k \Phi_\alpha) . \tag{6.3.4}$$

Thus, association of five basic objects of EAP can be reduced to four while leaving EL equations on DSV non-degenerated, formally assuming $\bar{\Lambda}_\beta^{\alpha k} \Longrightarrow -\Lambda_\beta^{\alpha k}$.

Introducing notation $\mathfrak{D}_{\beta k}^\alpha$ (*matter matrix*, MM) for matrix function:

$$\mathfrak{D}_{\beta k}^\alpha \equiv \frac{1}{2}(\Phi_\beta \mathfrak{D}_k \Psi^\alpha - \Psi^\alpha \mathfrak{D}_k \Phi_\beta) , \tag{6.3.5}$$

we can write matter Lagrangian in the following symbolic view:

$$\mathbb{L}_M = \mathbb{N} + \mathbb{D}; \qquad \mathbb{N} \equiv \Phi_\alpha \Psi^\alpha; \qquad \mathbb{D} \equiv \Lambda_\beta^{\alpha k} \mathfrak{D}_{\alpha k}^\beta . \tag{6.3.6}$$

### 6.4. Structuring of the Gauge Scalar

Total scalar Lagrangian $\mathbb{L}$ should include in its structure covariant derivative of unified gauge field $\mathcal{A}_{\beta k}^\alpha$, the hybrid curvature form (HCF) $\mathfrak{R}_{\beta k l}^\alpha$, generally, in contraction with some dual h-tensor $\Sigma_\beta^{\alpha k l}$. Such form as scalar

$$\mathbb{L}_G = \Sigma_\beta^{\alpha k l} \mathfrak{R}_{\alpha k l}^\beta \tag{6.4.1}$$

with matrices $\Sigma_\beta^{\alpha k l}$ independent of DSV should be introduced to scalar Lagrangian in additive way:

$$\mathbb{L} \Longrightarrow \mathbb{L}_M + \mathbb{L}_G , \tag{6.4.2}$$

in order to attain a disposition in which DSV can play a role of an inducing object with respect to the triadic basic object of the theory, especially with respect to UGF. Two minimal options can be considered.

1. H-tensor $\Sigma_\beta^{\alpha k l}$ composed based on *split metric* h-tensor $\Lambda_\beta^{\alpha k}$ (gauge scalar of the first power on HCF):

$$\Sigma_\beta^{\alpha kl} \Rightarrow \hat{\Sigma}_\beta^{\alpha kl} = \frac{1}{2}(\Lambda_\gamma^{\alpha k}\Lambda_\beta^{\gamma l} - \Lambda_\gamma^{\alpha l}\Lambda_\beta^{\gamma k}). \tag{6.4.3}$$

2. Other possibility is the following: build h-tensor $\Sigma_\beta^{\alpha kl}$ by use GM to lift Roman indices of HCF itself (GS of the second power on HCF):

$$\Sigma_\beta^{\alpha kl} \Rightarrow \frac{1}{4}\mathfrak{R}_\beta^{\alpha kl} \equiv \frac{1}{4}w^{km}w^{ln}\mathfrak{R}_{\beta mn}^\alpha, \tag{6.4.4}$$

so then:

$$\mathbb{L}_G \Rightarrow \frac{1}{4}w^{km}w^{ln}\mathfrak{R}_{\beta kl}^\alpha \mathfrak{R}_{\alpha mn}^\beta. \tag{6.4.5}$$

Let us consider properties of the first option. For convenience of deriving EL equations on UGF, scalar Lagrangian (6.4.2) can be written in the following view:

$$\mathbb{L} = \Phi_\alpha \Psi^\alpha + \frac{1}{2}\Lambda_\beta^{\alpha k}(\Phi_\alpha \partial_k \Psi^\beta - \Psi^\beta \partial_k \Phi_\alpha) + \mathcal{A}_{\beta k}^\alpha \mathcal{J}_\alpha^{\beta k} + \frac{1}{2}\hat{\Sigma}_\beta^{\alpha kl}\mathfrak{R}_{\beta mn}^\alpha ; \tag{6.4.6}$$

here $\mathcal{J}_\alpha^{\beta k}$ denotes *supercurrent matrix*, a hybrid triadic tensor :

$$\mathcal{J}_\alpha^{\beta k} \equiv \frac{1}{2}(\Lambda_\alpha^{\gamma k}\Phi_\gamma\Psi^\beta + \Lambda_\gamma^{\beta k}\Phi_\alpha\Psi^\gamma). \tag{6.4.7}$$

Performing variation derivatives in EL equations (5.6.4), we obtain the following differential equations for *unified gauge field* $\mathcal{A}_{\beta k}^\alpha$ :

$$(\partial_l + \frac{\partial_l w}{2w})\hat{\Sigma}_\beta^{\alpha kl} + \hat{\Sigma}_\beta^{\gamma kl}\mathcal{A}_{\gamma l}^\alpha - \hat{\Sigma}_\gamma^{\alpha kl}\mathcal{A}_{\beta l}^\gamma = \mathcal{J}_\beta^{\alpha k}. \tag{6.4.8}$$

We thus have derived a system of equations that connect affine h-tensor, *unified gauge field* $\mathcal{A}_{\beta k}^\alpha$ to DSV and Split Metric $\Lambda_\beta^{\alpha k}$. Equations (6.4.8) can be reduced to an explicit covariant view:

$$\mathfrak{D}_l\hat{\Sigma}_\beta^{\alpha kl} = \mathcal{J}_\beta^{\alpha k}, \tag{6.4.9}$$

with a complete covariant divergence of h-tensor $\mathfrak{R}_\beta^{\alpha kl}$ on the left hand side:

$$\mathfrak{D}_l\hat{\Sigma}_\beta^{\alpha kl} \equiv \frac{1}{\sqrt{w}}\partial_l(\sqrt{w}\hat{\Sigma}_\beta^{\alpha kl}) + \mathcal{A}_{\gamma l}^\alpha\hat{\Sigma}_\beta^{\gamma kl} - \mathcal{A}_{\beta l}^\gamma\hat{\Sigma}_\gamma^{\alpha kl}. \tag{6.4.10}$$

Contracting equation (6.4.9) on Greek indices results in equations:

$$\mathcal{J}^k = 0$$

for vector of supercurrent

$$\mathcal{J}^k \equiv \mathcal{J}^{\alpha k}_\alpha = \Lambda^{\beta k}_\alpha \Phi_\beta \Psi^\alpha, \tag{6.4.11}$$

since $\Sigma^{\alpha k l}_\alpha \equiv 0$ (compare (6.4.3)). Considering meaning of this degradation from an external logical point of view, we note that, choice of option (6.4.3) in structuring the gauge scalar is equivalent to posing the specific requirements $\mathcal{J}^k = 0$ to the dynamics. However, such requirements cannot be motivated by a substantial reason. Moreover, it is contrary to the *existence of conservative supercurrent*, pointed out in general treat of variation principle for an isomorphic field theory in section 5.6 and posed as one of the principal physical requirement to an irreducible theory in section 6.1. Therefore, option (6.4.3) has to be denied, and we have to consider option (6.4.4), delivering the following scalar Lagrangian:

$$\mathbb{L} = \Phi_\alpha \Psi^\alpha + \frac{1}{2}\Lambda^{\alpha k}_\beta (\Phi_\alpha \partial_k \Psi^\beta - \Psi^\beta \partial_k \Phi_\alpha) + \mathcal{A}^\alpha_{\beta k}\mathcal{J}^{\beta k}_\alpha + \frac{1}{4} w^{km} w^{ln} \mathfrak{R}^\alpha_{\beta k l} \mathfrak{R}^\alpha_{\beta m n} \tag{6.4.12}$$

and subsequent EL equations on $\mathcal{A}^\alpha_{\beta k}$:

$$(\partial_l + \frac{\partial_l w}{2w})\mathfrak{R}^{\alpha k l}_\beta + \mathfrak{R}^{\gamma k l}_\beta \mathcal{A}^\alpha_{\gamma l} - \mathfrak{R}^{\alpha k l}_\gamma \mathcal{A}^\gamma_{\beta l} = \mathcal{J}^{\alpha k}_\beta. \tag{6.4.13}$$

Equations (6.4.13) can be reduced to an explicit covariant view:

$$\mathfrak{D}_l \mathfrak{R}^{\alpha k l}_\beta = \mathcal{J}^{\alpha k}_\beta \tag{6.4.14}$$

with a complete covariant divergence of h-tensor $\mathfrak{R}^{\alpha k l}_\beta$ on the left hand side:

$$\mathfrak{D}_l \mathfrak{R}^{\alpha k l}_\beta \equiv \frac{1}{\sqrt{w}} \partial_l (\sqrt{w}\mathfrak{R}^{\alpha k l}_\beta) + \mathcal{A}^\alpha_{\gamma l}\mathfrak{R}^{\gamma k l}_\beta - \mathcal{A}^\gamma_{\beta l}\mathfrak{R}^{\alpha k l}_\gamma. \tag{6.4.15}$$

Contraction of this equations on Greek indices leads to Maxwell type equations with supercurrent (6.4.11) for odd-symmetric tensor $\mathfrak{R}^{kl} \equiv \mathfrak{R}^{\alpha k l}_\alpha$:

$$\frac{1}{\sqrt{w}} \partial_l(\sqrt{w}\mathfrak{R}^{kl}) = \mathcal{J}^k. \tag{6.4.16}$$

### 6.5. Structuring of Grand Metric

Next, we have to consider constraints of introduction of a symmetric tensor $w_{kl}$ to the DSV-based covariant theory. There are two possible different approaches to this issue in the EAP-based approach to establishing differential system of a covariant field theory.

A) Introduce such tensor as one of basic objects of a field theory, a subject of independent variations in the applied EAP principle. This approach was used by D. Hilbert when he resorted to EAP to derive equations of relativistic theory of gravitation [8].

B) Not to include tensor $w_{kl}$ in collection of basic objects but compose such tensor on these objects, once such possibility can be realized in a way consistent with the imposed irreducibility demands. Such approach was utilized in paper [1*] and specified in the present paper with a reduction. This way was inspired by observation of Dirac's relativistic theory of electrons and guided by the *irreducibility* demand of W. Pauli to a unified field theory [8].

As it was treated in paper [1*], in the DSV-based approach to irreducible theory there is a possibility of structuring tensor $w_{kl}$ on triadic h-tensors $\Lambda_\beta^{\alpha k}$ and $\overline{\Lambda}_\beta^{\alpha k}$:

$$w_{kl} \Longrightarrow \Lambda_\beta^{\alpha k}\overline{\Lambda}_\alpha^{\beta l} + \Lambda_\beta^{\alpha l}\overline{\Lambda}_\alpha^{\beta k}. \tag{6.5.1}$$

In the present paper we produce reduction of basic objection collection from 5 objects to 4, replacing $\overline{\Lambda}_\alpha^{\beta k}$ by $-\Lambda_\beta^{\alpha k}$ in order to access more degree of irreducibility if possible. Then structure (6.4.1) is reduced to the simplest one based on single triad $\Lambda_\beta^{\alpha k}$: introduce contravariant tensor $w^{kl}$ as binary composition on matrices $\Lambda_\beta^{\alpha k}$:

$$w^{kl} \Longrightarrow \Lambda_\beta^{\alpha k}\Lambda_\alpha^{\beta l} \equiv \Lambda^{kl} \tag{6.5.2}$$

and covariant tensor $\Lambda_{kl}$ (*grand metric*, GM) as inverse to $w^{kl}$:

$$\Lambda^{km}\Lambda_{lm} = \Delta_l^k. \tag{6.5.3}$$

In this way, just one tensor form (6.5.2) serves three duties in structuring the unified covariant field theory:
- utilizing general covariance of the differential system;
- providing coupling of triadic h-tensor $\Lambda_\beta^{\alpha k}$ to DSV and UGF;
- utilizing gauge scalar form in Lagrangian based on triad $\Lambda_\beta^{\alpha k}$.

Concerning possibility of alternative options of structuring GM based on other hybrid objects in frame of the DSV-based theory, the following to be noted.
1. GM structure should not include derivatives of DSV, otherwise the *second* derivatives of DSV arrive in EL equations on DSV, in contrary to the demand of differential irreducibility of the DSV equations.
2. GM cannot be built on gauge matrices $\mathcal{A}_{\beta k}^\alpha$ since this object is not transformed as an h-tensor.
3. GM cannot be built on HCF $\mathfrak{R}_{\beta kl}^\alpha$ because of the fundamental skew-symmetry of this object on Roman indices $, l$.

Structuring of tensor $w_{kl}$ on triadic h-tensor $\Lambda_\beta^{\alpha k}$ as above shown appears to be the only possibility consistent with the all posed principles.

*Covariant Λ-matrices*

Using *Grand Metric* $\Lambda_{kl}$ as tensor inverse to $\Lambda^{kl}$, one can also introduce *covariant* Λ-matrices as

$$\Lambda_{\beta k}^\alpha \equiv \Lambda_\beta^{\alpha m}\Lambda_{km}; \tag{6.5.4}$$

note that,

$$\Lambda_\beta^{\alpha k}\Lambda_{\alpha l}^\beta = \Lambda^{km}\Lambda_{lm} = \Delta_l^k; \qquad \Lambda_{\beta k}^\alpha \Lambda_{\alpha l}^\beta = \Lambda_{kl}. \tag{6.5.5}$$

*Determination of Matched Connection*

Matched Connection $\Gamma^k_{lm}$ (see section 4.2.) now can be defined as

$$\Gamma^k_{lm} = \frac{1}{2}\Lambda^{kn}(\partial_l \Lambda_{mn} + \partial_m \Lambda_{ln} - \partial_n \Lambda_{lm}) . \tag{6.5.6}$$

Then

$$\nabla_m \Lambda_{kl} \equiv 0 ; \qquad \nabla_m \Lambda^{kl} \equiv 0 . \tag{6.5.7}$$

### 6.6. Final Lagrangian

By the above derivations we have completed our endeavor of structuring a minimal scalar Lagrangian form $\mathbb{L}$ and Grand Metric $w_{kl}$ of a self-contained field theory. The total Lagrangian $\mathcal{L}$ is determined as

$$\mathcal{L} = \mathbb{L}\sqrt{\Lambda} ; \tag{6.6.1}$$

here:

$$\mathbb{L} = \mathbb{M} + \mathbb{G} = \mathbb{N} + \mathbb{D} + \mathbb{G}; \tag{6.6.2}$$

$$\mathbb{N} = \Phi_\alpha \Psi^\alpha; \qquad \mathbb{D} = \Lambda^{\alpha k}_\beta \mathfrak{D}^\beta_{\alpha k} ; \tag{6.6.3}$$

$$\mathfrak{D}^\beta_{\alpha k} \equiv \frac{1}{2}(\Phi_\alpha \mathfrak{D}_k \Psi^\beta - \Psi^\beta \mathfrak{D}_k \Phi_\alpha) ; \tag{6.6.4}$$

$$\mathfrak{D}_k \Psi^\beta \equiv \partial_k \Psi^\beta + \mathcal{A}^\beta_{\gamma k} \Psi^\gamma; \qquad \mathfrak{D}_k \Phi_\alpha \equiv \partial_k \Phi_\alpha - \mathcal{A}^\gamma_{\alpha k} \Phi_\gamma ; \tag{6.6.5}$$

$$\mathbb{G} \equiv \frac{1}{4}\Lambda^{km}\Lambda^{ln}\mathfrak{R}^\beta_{\alpha kl}\mathfrak{R}^\alpha_{\beta mn} ; \tag{6.6.6}$$

$$\mathfrak{R}^\alpha_{\beta kl} = \breve{\mathfrak{R}}^\alpha_{\beta kl} - \breve{\mathfrak{R}}^\alpha_{\beta lk} ; \qquad \breve{\mathfrak{R}}^\alpha_{\beta kl} \equiv \partial_k \mathcal{A}^\alpha_{\beta l} + \mathcal{A}^\alpha_{\gamma k}\mathcal{A}^\gamma_{\beta l}; \tag{6.6.7}$$

$$\Lambda \equiv |det\Lambda_{kl}| = \frac{1}{|det\Lambda^{kl}|} ; \qquad \Lambda^{kl} \equiv \Lambda^{\alpha k}_\beta \Lambda^{\beta l}_\alpha ; \qquad \Lambda_{km}\Lambda^{lm} = \Delta^m_l . \tag{6.6.8}$$

*Gauge scalar* $\mathbb{G}$ can also be represented as:

$$\mathbb{G} = \frac{1}{4}\Lambda^{kl}\Lambda^{mn}\mathbb{G}_{km;ln} , \tag{6.6.9}$$

where notation $\mathbb{G}_{km;ln}$ is for *gauge 4-tensor* :

$$\mathbb{G}_{km;ln} \equiv \mathfrak{R}^\alpha_{\beta km}\mathfrak{R}^\beta_{\alpha ln} . \tag{6.6.10}$$

It should be noted that, all introduced tensor and scalar forms are unambiguous i.e. unique, relative the contraction procedures, since contractions between Roman and Greek indices are not legitimate in the differential theory under treat.

*Resume of Chapter 6*

In Chapter 6 a system of principles: *homomorphism*, *duality, homogeneity, uniformity, reality, covariance, differential irreducibility, existence of supercurrent, scale invariance and mini-max principle* has been set as guidance for forming the Lagrangian. The system of four basic objects, subjects to vary in the Extreme Action Principle, includes *Dual State Vector* (DSV) field $\Xi = (\Psi^\alpha, \Phi_\alpha)$, triadic hybrid tensor, *Split Metric* $\Lambda^{k\alpha}_\beta$, and affine hybrid tensor, *unified gauge field* $\mathcal{A}^\beta_{k\alpha}$. Lagrangian form does not include the conventional Christoffel symbols and their covariant derivatives i.e. Riemann-Christoffel curvature tensor, since these objects are not inquired by the necessity of covariant structuring the matter scalar and building a closed differential system. Metric tensor (Grand Metric, GM) is introduced for invariant integration of scalar forms in space of Unified Manifold and in order to compose the *geometry* or *gauge* scalar form based on the hybrid curvature tensor. GM itself is structured on SM, thus not playing a role of an independent basic object complement to the *family of four*.

Based on Lagrangian (6.6.1), Euler-Lagrange equations are straightforward to be derived as a closed system of differential equations for *four* real objects. This is performed in Chapter 8. Before doing this we have to answer the following critical question: is the presented Lagrangian something unique, or there still a room for more general forms which will not contradict to the imposed *irreducibility* requirements?

# 7. The Scale Invariance and Mini-max principles

## 7.1. Scale invariance of Lagrangian

There is an ultimate requirement or principle which was not invited yet to the consideration: *scale invariance,* number 9 in the list of demands to a fundamental differential system exhibited in section 6.1. . The requirement implies two questions.

1. Does the introduction of real constant numbers as multipliers to the invariant parts of scalar Lagrangian (6.7.2) influence the properties of the differential system and quantities that may come from the solutions?
2. Can the derived Lagrangian of the DSV-based covariant field theory be modified to other forms or extended to more general ones if scaling invariance is posed?

Based on the analysis exposed below, answer to both these questions is: *no.*

It seems at the first glance at the structure of the scalar Lagrangian (6.7.2), that, it is not generally defined, since one may insert three arbitrary real constants $g_1, g_2$, in the following way:

$$\mathcal{L} \Longrightarrow (g_1 \mathbb{N} + g_2 \Lambda_\beta^{\alpha k} \mathfrak{D}_{\alpha k}^\beta + \frac{1}{4} \mathfrak{R}_\beta^{\alpha k l} \mathfrak{R}_{\alpha k l}^\beta)\sqrt{\Lambda}. \tag{7.1.1}$$

We have to notice that, the introduced constants should be real (either positive or negative), to be consistent with our concept of a real manifold and all real objects. Under this condition, the scalar Lagrangian form (7.1.1) can be returned to form (6.7.1) by a simple change of scaling of the involved objects: vector $\Psi^\alpha$ (or $\Phi_\alpha$) and h-tensor $\Lambda_\beta^{\alpha k}$. Namely let us define "renormalized" objects $\acute{\Psi}^\alpha$, $\acute{\Lambda}_\beta^{\alpha k}$ and $\acute{\bar\Lambda}_\beta^{\alpha k}$ as

$$\Psi^\alpha = p\acute{\Psi}^\alpha, \quad \Lambda_\beta^{\alpha k} = q\acute{\Lambda}_\beta^{\alpha k}, \tag{7.1.2}$$

with two undefined real constants $p, q$, and re-write form (7.1.1) as a function of the "new" objects (we omit symbol (´)):

$$\mathcal{L} \Longrightarrow (g_1 p\mathbb{N} + g_2 pq \Lambda_\beta^{\alpha k} \mathfrak{D}_{\alpha k}^\beta + \frac{1}{4} q^4 \mathfrak{R}_\beta^{\alpha k l} \mathfrak{R}_{\alpha k l}^\beta)\sqrt{\Lambda/q^{2N}}, \tag{7.1.3}$$

Now, we impose conditions on the scaling parameters $p, q$, as follows:

$$g_1 p = g_2 pq = q^4 \tag{7.1.4}$$

to return the scalar Lagrangian to its "initial" form:

$$\mathbb{L} \longrightarrow \mathbb{L}(g) \Longrightarrow \mathbb{L} \cdot const\,; \tag{7.1.5}$$

a common constant factor as a multiplier at the Lagrangian is insignificant. Equations (7.1.4) imply:

$$q = \frac{g_1}{g_2}; \quad p = \frac{q^4}{g_1} = \frac{1}{g_1}(\frac{g_1}{g_2})^4; \tag{7.1.6}$$

i.e. scale "renormalization" is always real for arbitrary real constants $g_1, g_2$.

Thus, introduction of arbitrary constants to the composition of the scalar Lagrangian (6.7.2) is equivalent to scaling renormalization of DSV and Split Metric. We call this feature of the theory the *scale invariance*.

Note in particular that, this scale invariance also covers the structural ambiguity in the sign of the geometry scalar obtained by bundling products of odd-symmetric tensors $\Sigma_\beta^{\alpha k l}$ and $\mathfrak{R}_{\alpha k l}^\beta$: the sign of this scalar can be referred to arbitrary real constants $g_1, g_2$. In other words, the sign of the gauge scalar can be chosen *arbitrarily*. We will clarify the meaning of this point in the discussion of the derived EL equations.

## 7.2. Confirmation of the derived Lagrangian

We have proved scaling invariance of Lagrangian (6.7.1). There still be a question, however, is this form the only possible one to satisfy all requirements exhibited in section 6.1., including the *scale invariance* ?

First, let us consider an addition $\widetilde{N}$ to matter scalar $\mathbb{L}_M$ (or replacement to norm $N = \Phi_\alpha \Psi^\alpha$) which does not include derivatives of dual s-field but also does not invite new basic objects:

$$\widetilde{N} = g_3 \Phi_\alpha \Psi^\beta \mathfrak{D}_k \Lambda_\beta^{\alpha k}, \tag{7.2.1}$$

with parameter $g_3$ as a constant. It should be noted that, introduction of form (7.2.1) in Lagrangian would next invite the affine tensor $G_{nk}^m$ of differential geometry and Riemann-Christoffel curvature form $\mathcal{R}_{nkl}^m$ (see section 4.1) with the related geometry scalars. The scaling in basic objects as shown in equation (7.1.3) then cannot satisfy requirement of scale invariance, since number of conditions of invariance would then exceed number of scale parameters. Therefore, such extension cannot be executed without breaking of SIP.

*Incompatibility of scaling invariance with introduction of high order terms*

Finally, we have to answer a question, whether the Lagrangian could be extended by introduction of terms higher order on norm N or, more generally, in dual s-field. Apparently, the necessity of introduction of the higher order terms cannot be logically motivated (as well as a necessity to extend the family of basic objects of EAP). Note in this connection that, the presented system of EL equations already is an essentially non-linear differential system – due to the requirements of *homogeneity* and *general covariance*. On the other hand, the composed system does not include any constants – due to requirement of *scaling invariance*. On the contrary, high order additions in scalar Lagrangian lead to introduction of higher rank hybrid tensors (either new independent ones or structured on the triads) and/or weighting constants. These constants cannot be "dissolved" in the structure by a scaling transformation.

Structuring of metric tensor $w_{kl}$ on *Spilt Metric* h-tensor $\Lambda_\beta^{\alpha k}$ according to formula (6.5.2) is an option the simplest one and the most corresponding to requirement of *irreducibility* compare to any other possibilities. It should be noted that, the even-symmetric tensor $w_{kl}$ cannot be built on tensor $\mathfrak{R}_{kl} \equiv \mathfrak{R}^{\alpha}_{\alpha kl}$ because of the background skew-symmetry of form $\mathfrak{R}^{\alpha}_{\beta kl}$ on indices $k, l$. Rest possible options would include (covariant) derivatives of DSV, this would lead to arriving of the higher order derivatives in equations on DSV, in contrary to demand of *differential irreducibility* of these equations (as well as the whole differential system of SFT) – the master equations of the explored approach to UFT.

## *Resume of Chapter 7*

We have shown that, the presented DSV-based approach to UFT does possess two important features:
1) The derived differential system does satisfy the requirement of *scale invariance*, i.e. it is insensitive relative of introduction of arbitrary real constants (multipliers) before invariant scalar blocks of Lagrangian structure.
2) Extension of the derived Lagrangian with higher order non-linear terms is not compatible with the scale invariance. Thus, Lagrangian (6.7.1) is the *maximally possible* composition of the basic objects and their derivatives agreeable with principles of the presented dual-covariant differential approach to UFT, - in accordance with the *mini-max principle*.

We thus have confirmed the structuring of the Lagrangian in our approach to UFT. We can conclude this Chapter with a statement that, Lagrangian form (6.7.1) is the unique one to satisfy principles *1* through *11* exhibited in section 6.1.

# 8. Euler-Lagrange Equations of SFT

Here we will finally derive EL equations on Dual State Vector field $(\Psi^\alpha, \Phi_\alpha)$, hybrid affine tensor or *unified gauge field* $\mathcal{A}^\alpha_{\beta k}$, and hybrid tensor or *split metric* $\Lambda^{k\alpha}_\beta$, taking into account the above established structure of Grand Metric and geometry scalar .

## 8.1. Equations on Dual State Vector Field

*Final equations on DSV in the ordinary derivatives*

For convenience of deriving EL equations on DSV, Lagrangian (6.7.1) can be written in the following view:

$$\mathcal{L} = \{\tfrac{1}{2}\left[\Lambda^{\alpha k}_\beta (\Phi_\alpha \partial_k \Psi^\beta - \Psi^\beta \partial_k \Phi_\alpha) + \mathcal{A}^\alpha_\beta \Phi_\alpha \Psi^\beta\right] + \Phi_\alpha \Psi^\alpha + \mathbb{G}\}\sqrt{\Lambda} \qquad (8.1.1)$$

here

$$\mathcal{A}^\alpha_\beta \equiv \Lambda^{\alpha k}_\gamma \mathcal{A}^\gamma_{\beta k} + \Lambda^{\gamma k}_\beta \mathcal{A}^\alpha_{\gamma k} . \qquad (8.1.2)$$

Taking into account definition of $\Lambda$ according to (6.7.7), we receive EL equations on DSV in the following form:

$$\Lambda^{\alpha k}_\beta \partial_k \Psi^\beta + \frac{1}{2}\left[\left(\partial_k - \frac{1}{2}\Lambda_{lm}\partial_k\Lambda^{lm}\right)\Lambda^{\alpha k}_\beta + \mathcal{A}^\alpha_\beta \Psi^\beta\right] + \Psi^\alpha = 0;$$

$$\Lambda^{\beta k}_\alpha \partial_k \Phi_\beta + \frac{1}{2}\left[\left(\partial_k - \frac{1}{2}\Lambda_{lm}\partial_k\Lambda^{lm}\right)\Lambda^{\beta k}_\alpha - \mathcal{A}^\beta_\alpha \Phi_\beta\right] - \Phi_\alpha = 0 . \qquad (8.1.3)$$

*DSV equations in terms of covariant derivatives*

The even symmetry of Grand Metric tensor structured according to equation (6.6.4) leads to nullification of form $w_k$ shown in equation (4.1.31). Covariant equations for DSV (6.3.20) then

acquire a compact form:

$$\Lambda^{\alpha k}_\beta \mathfrak{D}_k \Psi^\beta + \frac{1}{2} \Psi^\beta \mathfrak{D}_k \Lambda^{\alpha k}_\beta + \Psi^\alpha = 0 ;$$

$$\Lambda^{\beta k}_\alpha \mathfrak{D}_k \Phi_\beta + \frac{1}{2} \Phi_\beta \mathfrak{D}_k \Lambda^{\beta k}_\alpha - \Phi_\alpha = 0 . \qquad (8.1.4)$$

Here

$$\mathfrak{D}_k \Lambda^{\alpha k}_\beta \equiv \frac{1}{\sqrt{\Lambda}} \partial_k(\sqrt{\Lambda}\Lambda^{\alpha k}_\beta) + \mathcal{A}^\alpha_{\gamma k}\Lambda^{\gamma k}_\beta - \mathcal{A}^\gamma_{\beta k}\Lambda^{\alpha k}_\gamma \qquad (8.1.5)$$

Note that, object $\mathfrak{D}_k \Lambda_\beta^{\alpha k}$ is an s-tensor, since it can be represented as covariant derivative $\mathfrak{D}_l \Lambda_\beta^{\alpha k}$ of h-tensor $\Lambda_\beta^{\alpha k}$ contracted on Roman indices $l = k$:

$$\mathfrak{D}_l \Lambda_\beta^{\alpha k} \equiv \partial_l \Lambda_\beta^{\alpha k} + \Gamma_{ml}^k \Lambda_\beta^{\alpha m} + \mathcal{A}_{\gamma l}^\alpha \Lambda_\beta^{\gamma k} - \mathcal{A}_{\beta l}^\gamma \Lambda_\gamma^{\alpha k} ; \tag{8.1.6}$$

here $\Gamma_{ml}^k$ are the conventional *Christoffel symbols* or *matched connection* form [8,9]:

$$\Gamma_{ml}^k = \frac{1}{2} \Lambda^{kn} (\partial_m \Lambda_{ln} + \partial_l \Lambda_{mn} - \partial_n \Lambda_{lm}) . \tag{8.1.7}$$

Object $\mathfrak{D}_k \Lambda_\beta^{\alpha k}$ can be characterized as *covariant divergence* of Split Metric $\Lambda_\beta^{\alpha k}$.

EL equations (8.1.4) coincide with the preliminary derived pair of equations (3.6.6).

Note that, DSV equations in the covariant form (8.1.4) include covariant derivatives of Split Metric as a hybrid tensor, which involve Matched Connection (MC) (8.1.7). It should be noted that, MC arrives in DSV equations only via derivatives of weigh factor $\sqrt{\Lambda}$ in the total Lagrangian $\mathcal{L} = \mathbb{L}\sqrt{\Lambda}$.

## 8.2. Gauge equations

Structure of scalar $\mathbb{M}$ does not include derivatives of $\mathcal{A}_{\beta k}^\alpha$, and also the structure of the grand metric $w_{kl}$ does not include both $\mathcal{A}_{\beta k}^\alpha$ and its derivatives. EL equations on $\mathcal{A}_{\beta k}^\alpha$ can initially be written in a reduced form as follows:

$$\frac{1}{\sqrt{\Lambda}} \partial_l \left[ \sqrt{\Lambda} \frac{\partial \mathbb{G}}{\partial (\partial_l \mathcal{A}_{\beta k}^\alpha)} \right] - \frac{\partial (\mathbb{M} + \mathbb{G})}{\partial \mathcal{A}_{\beta k}^\alpha} = 0 . \tag{8.2.1}$$

For convenience of deriving EL equations on UGF, scalar Lagrangian (6.6.2) can be written in the following view:

$$\mathbb{L} = \Phi_\alpha \Psi^\alpha + \frac{1}{2} \Lambda_\beta^{\alpha k} (\Phi_\alpha \partial_k \Psi^\beta - \Psi^\beta \partial_k \Phi_\alpha) + \mathcal{A}_{\beta k}^\alpha \mathcal{J}_\alpha^{\beta k} + \frac{1}{4} \Lambda^{km} \Lambda^{ln} \mathfrak{R}_{\beta kl}^\alpha \mathfrak{R}_{\beta mn}^\alpha ; \tag{8.2.2}$$

here $\mathcal{J}_\alpha^{\beta k}$ denotes *supercurrent matrix*, a hybrid triadic tensor :

$$\mathcal{J}_\alpha^{\beta k} \equiv \frac{1}{2} (\Lambda_\alpha^{\gamma k} \Phi_\gamma \Psi^\beta + \Lambda_\gamma^{\beta k} \Phi_\alpha \Psi^\gamma \tag{8.2.3}$$

Performing variation derivatives in EL equations (8.2.1), we obtain the following differential equations for *unified gauge field* $\mathcal{A}_{\beta k}^\alpha$ :

$$(\partial_l + \frac{\partial_l \Lambda}{2\Lambda}) \mathfrak{R}_\beta^{\alpha k l} + \mathfrak{R}_\beta^{\gamma k l} \mathcal{A}_{\gamma l}^\alpha - \mathfrak{R}_\gamma^{\alpha k l} \mathcal{A}_{\beta l}^\gamma = \mathcal{J}_\beta^{\alpha k} , \tag{8.2.4}$$

Equations (7.2.4) can be represented in an explicit covariant view:

$$\mathfrak{D}_l \mathfrak{R}_\beta^{\alpha kl} = \mathcal{J}_\beta^{\alpha k}, \tag{8.2.5}$$

with a complete *covariant divergence* of h-tensor $\mathfrak{R}_\beta^{\alpha kl}$ on the left hand side:

$$\mathfrak{D}_l \mathfrak{R}_\beta^{\alpha kl} \equiv \frac{1}{\sqrt{\Lambda}} \partial_k (\sqrt{\Lambda} \mathfrak{R}_\beta^{\alpha kl}) + \mathcal{A}_{\gamma l}^{\alpha} \mathfrak{R}_\beta^{\gamma kl} - \mathcal{A}_{\beta l}^{\gamma} \mathfrak{R}_\gamma^{\alpha kl}. \tag{8.2.6}$$

Object $\mathfrak{D}_l \mathfrak{R}_\beta^{\alpha kl}$ is an h-tensor, since it is a covariant derivative of h-tensor $\mathfrak{D}_m \mathfrak{R}_\beta^{\alpha kl}$:

$$\mathfrak{D}_m \mathfrak{R}_\beta^{\alpha kl} \equiv \partial_m \mathfrak{R}_\beta^{\alpha kl} + \Gamma_{nm}^{k} \mathfrak{R}_\beta^{\alpha nl} + \Gamma_{nm}^{l} \mathfrak{R}_\beta^{\alpha kn} + \mathcal{A}_{\gamma m}^{\alpha} \mathfrak{R}_\beta^{\gamma kl} - \mathcal{A}_{\beta m}^{\gamma} \mathfrak{R}_\gamma^{\alpha kl} \tag{8.2.7}$$

contracted on Roman indices $m = l$ (taking into account the even-symmetry of Matched Connection ($\Gamma_{nm}^{k} = \Gamma_{mn}^{k}$) against the skew-symmetry of HCF ($\mathfrak{R}_\beta^{\alpha kl} = -\mathfrak{R}_\beta^{\alpha lk}$)).

We thus have derived a system of equations that connect the affine h-tensor, *unified gauge field* $\mathcal{A}_{\beta k}^{\alpha}$ to DSV and Split Metric $\Lambda_\beta^{\alpha k}$. The total number of these equations is $N^3$, as it should be for a triadic hybrid object. When considering them as equations for $\mathcal{A}_{\beta k}^{\alpha}$ as matrices on Greek indices, there are $N$ such equations.

### 8.3. Metric equations

Finally, we will derive EL equations on Split Metric by taking variation of the action integral on h-tensors $\Lambda_\beta^{\alpha k}$ and $\overline{\Lambda}_\beta^{\alpha k}$. Lagrangian form (6.7.1) does not include the derivatives of these objects, but we have to take into account that determinant $\Lambda$ is a form built on SM. The related EL equations take a simple general form as follows:

$$\frac{\partial \mathbb{L}}{\partial \Lambda_\beta^{\alpha k}} + \frac{\mathbb{L}}{2\Lambda} \frac{\partial \Lambda}{\partial \Lambda_\beta^{\alpha k}} = 0. \tag{8.3.1}$$

Using an elementary formula [9, 21]:

$$\partial \Lambda = -\Lambda \Lambda_{kl} \partial \Lambda^{kl}, \tag{8.3.2}$$

equations (8.3.1) can be written in an explicit covariant form:

$$\frac{\partial \mathbb{L}}{\partial \Lambda_\beta^{\alpha k}} - \frac{1}{2} \mathbb{L} \frac{\partial \Lambda^{lm}}{\partial \Lambda_\beta^{\alpha k}} \Lambda_{lm} = 0; \tag{8.3.3}$$

For convenience of calculating the variation derivatives, let us write scalar Lagrangian and Grand Metric explicit as forms structured on Split Metric:

$$\mathbb{L} = \Phi_\alpha \Psi^\alpha + \Lambda_\beta^{\alpha k} \mathfrak{D}_{\alpha k}^{\beta} + \frac{1}{4} \mathbb{G}_{kl;mn} \Lambda^{km} \Lambda^{ln}; \tag{8.3.4}$$

Performing the variation derivatives in equations (7.3.3), we then obtain the EL equations on Split Metric as follows:

$$\Lambda_\beta^{\alpha l}(\Lambda^{mn}\mathbb{G}_{km;ln} - \mathbb{L}\Lambda_{kl}) = -\mathfrak{D}_{\beta k}^\alpha \qquad (8.3.5)$$

or

$$\Lambda_\beta^{\alpha l}(\mathbb{G}_{kl} - \mathbb{L}\Lambda_{kl}) = -\mathfrak{D}_{\beta k}^\alpha; \qquad (8.3.6)$$

here notation $\mathbb{G}_{kl}$ is for symmetric *gauge tensor*:

$$\mathbb{G}_{kl} \equiv \Lambda^{mn}\mathbb{G}_{km;ln}. \qquad (8.3.7)$$

### 8.4. EL equations in matrix notations

*Lagrangian*

Lagrangian $\mathcal{L}$ of the theory is composed on four basic objects: *dual state vector* field (DSV) as alliance of two vector fields, contravariant $\mathbf{\Psi} \equiv \Psi^\alpha$ (*column* of functions) and covariant $\mathbf{\Phi} \equiv \Phi_\alpha$ (*row*); *split metric* (SM) matrices $\mathbf{\Lambda}^k \equiv \Lambda_\beta^{\alpha k}$ and *unified gauge field* (UGF) matrices $\mathcal{A}_k \equiv \mathcal{A}_{\beta k}^\alpha$, all functions of UM variables and subjects of independent varying in Lagrangian form:

$$\mathcal{L} = \mathbb{L}\sqrt{\Lambda} = (\mathbb{M} + \mathbb{G})\sqrt{\Lambda}; \qquad (8.4.3)$$

where $\mathbb{M}$ is the *matter scalar* form:

$$\mathbb{M} = \mathbb{N} + \mathbb{D}$$

$$\mathbb{N} = \mathbf{\Phi}\mathbf{\Psi} = \Phi_\alpha \Psi^\alpha \qquad (8.4.4)$$

$$\mathbb{D} \equiv \frac{1}{2}[\mathbf{\Phi}\mathbf{\Lambda}^k\mathfrak{D}_k\mathbf{\Psi} - (\mathfrak{D}_k\mathbf{\Phi})\mathbf{\Lambda}^k\mathbf{\Psi}]$$

$$\mathfrak{D}_k\mathbf{\Psi} \equiv \partial_k\mathbf{\Psi} + \mathcal{A}_k\mathbf{\Psi};$$

$$\mathfrak{D}_k\mathbf{\Phi} \equiv \partial_k\mathbf{\Phi} - \mathbf{\Phi}\mathcal{A}_k, \qquad (8.4.5)$$

and $\mathbb{G}$ is *gauge scalar* form:

$$\mathbb{G} = Tr(\mathfrak{R}^{kl}\mathfrak{R}_{kl}); \qquad (8.4.6)$$

$$\mathfrak{R}_{kl} \equiv \mathfrak{R}_{\beta kl}^\alpha \equiv \partial_k\mathcal{A}_l - \partial_l\mathcal{A}_k + [\mathcal{A}_k, \mathcal{A}_l]; \qquad \mathfrak{R}^{kl} = \Lambda^{km}\Lambda^{ln}\mathfrak{R}_{mn} \qquad (8.4.7)$$

symbols [ ; ] denote commutator of two matrices. Further,

$$\Lambda \equiv |det\Lambda^{kl}|^{-1}; \qquad (8.4.8)$$

$$\Lambda^{kl} \equiv Tr(\boldsymbol{\Lambda}^k \boldsymbol{\Lambda}^l) . \tag{8.4.9}$$

Objects $\mathfrak{D}_k \boldsymbol{\Psi}$ and $\mathfrak{D}_k \boldsymbol{\Phi}$ are the *covariant derivatives* of $\boldsymbol{\Psi}$ and $\boldsymbol{\Phi}$, respectively; they also can be treated as valence 2 hybrid tensors. Object $\mathfrak{R}_{kl}$ (*hybrid curvature form,* HCF) is uniquely recognized as *covariant derivative of Dynamic Connection* $\mathcal{A}_k$ . We call tensor $\Lambda^{kl}$ *grand metric* (GM).

Lagrangian form (8.4.3) is found *unique* to satisfy all above mentioned the *irreducibility* requirements.

*Euler-Lagrange equations*

Euler-Lagrange equations on basic objects with Lagrangian (8.4.3) have been derived. They constitute the background differential system in the presented approach to UFT. In matrix notations they acquire the following forms.

1. Euler-Lagrange equations on Dual State Vector field (*DSV equations*).
   Equations on $\Psi^\alpha, \Phi_\alpha$ acquire the following view:

$$\boldsymbol{\Lambda}^k \partial_k \boldsymbol{\Psi} + \frac{1}{2}[\frac{1}{\sqrt{\Lambda}}\partial_k(\sqrt{\Lambda}\boldsymbol{\Lambda}^k) + \boldsymbol{\Lambda}^k \mathcal{A}_k + \mathcal{A}_k \boldsymbol{\Lambda}^k]\boldsymbol{\Psi} + \boldsymbol{\Psi} = 0 . \tag{8.4.12}$$

$$(\partial_k \boldsymbol{\Phi})\boldsymbol{\Lambda}^k + \frac{1}{2}\boldsymbol{\Phi}[\frac{1}{\sqrt{\Lambda}}\partial_k(\sqrt{\Lambda}\boldsymbol{\Lambda}^k) - \boldsymbol{\Lambda}^k \mathcal{A}_k - \mathcal{A}_k \boldsymbol{\Lambda}^k] - \boldsymbol{\Phi} = 0 . \tag{8.4.13}$$

These equations being written in terms of covariant derivatives have the following view:

$$\boldsymbol{\Lambda}^k \mathfrak{D}_k \boldsymbol{\Psi} + (\frac{1}{2}\mathfrak{D}_k \boldsymbol{\Lambda}^k + 1)\boldsymbol{\Psi} = 0 ;$$

$$(\mathfrak{D}_k \boldsymbol{\Phi})\boldsymbol{\Lambda}^k + \boldsymbol{\Phi}(\frac{1}{2}\mathfrak{D}_k \boldsymbol{\Lambda}^k - 1) = 0 . \tag{8.4.14}$$

Here

$$\mathfrak{D}_k \boldsymbol{\Lambda}^k \equiv \frac{1}{\sqrt{\Lambda}}\partial_k(\sqrt{\Lambda}\boldsymbol{\Lambda}^k) + [\mathcal{A}_k, \boldsymbol{\Lambda}^k] \tag{8.4.15}$$

is the covariant divergence of Split Metric (SM) as a hybrid triadic tensor.

Note that, equations (8.4.14) coincide with the preliminary derived DSV equations (3.6.6), with specification $w^{kl} \Longrightarrow \Lambda^{kl}$.

2. EL equations on Unified Gauge Field $\mathcal{A}_k$ (*gauge equations*) can be written as follows:

$$\mathfrak{D}_l \mathfrak{R}^{kl} = \boldsymbol{\mathcal{J}}^k ; \tag{8.4.16}$$

here

$$\mathfrak{D}_l \mathfrak{R}^{kl} \equiv \frac{1}{\sqrt{\Lambda}}\partial_l(\sqrt{\Lambda}\mathfrak{R}^{kl}) + [\mathcal{A}_l, \mathfrak{R}^{kl}] \tag{8.4.17}$$

is covariant divergence of the odd-symmetric *hybrid tensor* $\mathfrak{R}^{kl}$, and $\boldsymbol{\mathcal{J}}^k$ is the *supercurrent*

*matrix*:

$$\boldsymbol{J}^k \equiv \boldsymbol{J}^{\alpha k}_\beta \equiv \frac{1}{2}(\Lambda^{\alpha k}_\gamma \Phi_\beta \Psi^\gamma + \Lambda^{\gamma k}_\beta \Phi_\gamma \Psi^\alpha) = \frac{1}{2}(\boldsymbol{\Lambda}^k \mathbf{N} + \mathbf{N}\boldsymbol{\Lambda}^k) = \frac{1}{2}\{\boldsymbol{\Lambda}^k \mathbf{N}\}; \qquad (8.4.18)$$

$$\mathbf{N} \equiv \mathrm{N}^\alpha_\beta = \Phi_\beta \Psi^\alpha = \boldsymbol{\Phi} \times \boldsymbol{\Psi}.$$

3. EL equations on Split Metric matrices $\boldsymbol{\Lambda}^k$ (*SM equations*) in matrix notations acquire the following view:

$$\mathbb{L}\boldsymbol{\Lambda}_k - \boldsymbol{\Lambda}_l Tr(\mathfrak{R}^l_n \mathfrak{R}^n_k) = \mathfrak{D}_k \qquad (8.4.19)$$

or

$$[\mathbb{L}\delta^l_k - Tr(\mathfrak{R}_{kn}\mathfrak{R}^{ln})]\boldsymbol{\Lambda}_l = \mathfrak{D}_k ; \qquad (8.4.20)$$

here $\mathfrak{D}_k$ is notation for the associate matrices:

$$\mathfrak{D}_k \equiv \frac{1}{2}(\Phi_\beta \mathfrak{D}_k \Psi^\alpha - \Psi^\alpha \mathfrak{D}_k \Phi_\beta) . \qquad (8.4.21)$$

These equations connect the dual metric object, *Split Metric* matrices $\boldsymbol{\Lambda}^k$ to the primary object, *Dual State Vector* $\Psi^\alpha, \Phi_\alpha$ and *Dynamic Connection* matrices $\mathcal{A}_k$ . From point of view of an external comparison, SM equations correspond to equation for metric tensor in theories of gravitation field of Einstein-Hilbert-Weyl [8].

### 8.5. General covariance as an attribute of the Extreme Action principle

Considering transformation properties of basic objects coupled in dynamics by the EL equations, we first have to note that, structural form of EL equations *is not* and *cannot be* thought as referred to a certain (i.e. specific) "frame of coordinates". Even statement of such a question does not have sense with respect to the background differential system of an irreducible field theory. Realization of this circumstance is the essence of the *general covariance* paradigm in the context of the UFT foundations. The only legitimate question in this context is: how the basic objects are transformed at transformations of UM (*free*) variables ? At this stage of profiling dynamic properties of SFT, there is no a complete unambiguous answer of this question. Namely, if in accordance with the above stated invariance of the homomorphic differential law () assume that the contravariant state vector is transformed with a matrix *B*, then connections between basic objects in the system of EL equations determine transformations of covariant state vector $\boldsymbol{\Phi}$, unified gauge field $\mathcal{A}$ and split metric $\boldsymbol{\Lambda}$ in accordance with equations (1.18), (1.27), and (1.29), respectively. This follows from the corresponding transformation properties of the covariant derivatives forms constituting structure of the derived Lagrangian and EL equations: , $\mathfrak{D}_k \Psi^\beta$, $\mathfrak{D}_k \Phi_\alpha$ , and $\mathfrak{R}^\alpha_{\beta kl}$. Taken together, the derived EL equations obviously validate the presumed transformation properties of basic objects. However, in the exposed treat of SFT is still not answered a question about transformation of DSV itself, i.e. about connection of matrix *B* to matrix *A*.

As pointed above, matrix *B* is different from *A* but connected to the last one via the dynamic law of SFT. Connection $B(A)$ can be considered as aspect of the invariance property of the dynamic law based on the derived EL equations. Consideration of this issue, however, goes beyond the scope of this paper.

*Resume of Chapter 8*

By derivations in this Chapter, formulation of basic equations in the exposed dual-vector differential approach to unified field theory is accomplished, that was in essence the aim of this paper. Three systems of Euler-Lagrange equations for the basic objects have been derived:
- equations on *Dual State Vector* (DSV) (8.1.3);
- equations the *Hybrid Christoffel symbols* or *unified gauge field* (UGF) (8.2.4);
- equations on *Split Metric* (SM) (8.3.2).

These three systems of equations can also be written in matrix forms as shown in section 8.4., representing SM and UGF as matrices associated with the Matter Function space.

All equations are generally covariant in form, i.e. they do not change their tensor structure at arbitrary transformations of variables. The differential system also is *scale invariant* i.e. it does not include any constants and is insensitive in its dynamic properties to introduction of arbitrary constants in Lagrangian (6.7.1) of the system.

The DSV equations are the primary, pilot equations of the theory. Equations on UGF and SM, being equations for the *coefficient functions* of DSV equations, connect these objects to DSV.

Discussion of the derived system in more detail with some references to the existing field theories is held in the summarizing Chapters 10 and 11. Here we mark one general circumstance in the context of references to the existing concepts of *Quantum Fields*: *breaking the superposition principle* (SP) of QFT. Namely, the DSV equations are viewed as playing role of the Schrödinger-Dirac equation for Matter Function in the Unified Manifold. They could be viewed as an analog to *Schrödinger equation* for *State Vector* of QFT. However, once the *coefficient functions* of DSV equations *have to be connected to DSV*, the differential system becomes *essentially non-linear on DSV*. This is contrary to the basic QFT concepts, where all *operators* associated with various "elementary particles" are defined as *independent* of the *State Vector*. Breaking of SP seems to be an unavoidable logical aspect of the efforts to reach a consistent unified theory.

# 9. Contracted Equations

## 9.1. Extended Faraday-Maxwell equations

*Tensor identities of the contracted Hybrid Curvature Form*

By contracting HCF form $\mathfrak{R}^\alpha_{\beta kl}$ on Greek indices we obtain an odd-symmetric covariant tensor

$$\mathfrak{R}_{kl} \equiv \mathfrak{R}^\alpha_{\alpha kl} = \partial_k \mathcal{A}^\alpha_{\alpha l} - \partial_l \mathcal{A}^\alpha_{\alpha k} \equiv \mathfrak{R}^\alpha_{\alpha kl} = \partial_k \mathcal{A}_l - \partial_l \mathcal{A}_k \ ; \tag{9.1.1}$$

$$\mathcal{A}_k \equiv \mathcal{A}^\alpha_{\alpha k}$$

Note that, this object is tensor despite the fact that $\mathcal{A}^\alpha_{\alpha k}$ is not a vector. This tensor satisfies the identity equations similar to the *first pair* of Maxwell equations:

$$\partial_m \mathfrak{R}_{kl} + \partial_l \mathfrak{R}_{mk} + \partial_k \mathfrak{R}_{lm} = 0 \ . \tag{9.1.2}$$

These equations are generally covariant, since they can be re-written in a form with covariant derivatives:

$$\nabla_m \mathfrak{R}_{kl} + \nabla_l \mathfrak{R}_{mk} + \nabla_k \mathfrak{R}_{lm} = 0 \ . \tag{9.1.3}$$

*Equations for dual image of contracted HCF*

The left-hand side of equations on UGF (8.4.12) is commutator of two matrices, $\mathfrak{R}^{\alpha kl}_\beta$ and $\mathcal{A}^\alpha_{\beta k}$. By taking trace of these equations on indices $\beta = \alpha$ we obtain the following $N$ equations:

$$(\partial_l + \frac{\partial_l \Lambda}{2\Lambda})\mathfrak{R}^{\alpha kl}_\alpha - J^{\alpha k}_\alpha = 0 \ , \tag{9.1.4}$$

or

$$\check{\partial}_l \mathfrak{R}^{kl} = J^k, \tag{9.1.5}$$

or

$$\frac{1}{\sqrt{\Lambda}} \partial_l (\mathfrak{R}^{kl} \sqrt{\Lambda}) = J^k. \tag{9.1.6}$$

$$\mathfrak{R}^{kl} \equiv \Lambda^{km} \Lambda^{ln} \mathfrak{R}_{mn} \ ; \qquad \mathfrak{R}_{mn} \equiv \mathfrak{R}^\alpha_{\alpha mn} \ . \tag{9.1.7}$$

These $N$ equations connect two contravariant objects: an *odd-symmetric tensor* field

$$\mathfrak{R}^{\alpha kl}_\alpha \equiv \mathfrak{R}^{kl} = -\mathfrak{R}^{lk} \tag{9.1.8}$$

and a contravariant *vector* field, the *supercurrent* introduced above in section 6.3. :

$$\mathcal{J}^k \equiv \mathcal{J}_\alpha^{\alpha k} = \Lambda_\alpha^{\beta k}\Phi_\beta\Psi^\alpha. \tag{9.1.9}$$

## 9.2. Scalar equations derived from EL equations on DSV

Let us bundle equations (8.1.3) with $\Phi_\alpha$ and $\Psi^\alpha$ as follows:

$$\Phi_\alpha\Lambda_\beta^{\alpha k}\mathfrak{D}_k\Psi^\beta + \frac{1}{2}\Phi_\alpha\Psi^\beta\mathfrak{D}_k\Lambda_\beta^{\alpha k} + \Phi_\alpha\Psi^\alpha = 0,$$

$$\Psi^\alpha\Lambda_\alpha^{\beta k}\mathfrak{D}_k\Phi_\beta + \frac{1}{2}\Psi^\alpha\Phi_\beta\mathfrak{D}_k\Lambda_\alpha^{\beta k} - \Phi_\alpha\Psi^\alpha = 0,$$

and take sum and difference of these equations, then we obtain two scalar equations as follows:

$$\Lambda_\beta^{\alpha k}(\Phi_\alpha\mathfrak{D}_k\Psi^\beta + \Psi^\beta\mathfrak{D}_k\Phi_\alpha) + \Phi_\alpha\Psi^\beta\mathfrak{D}_k\Lambda_\beta^{\alpha k} = 0 \tag{9.2.1}$$

$$\Lambda_\beta^{\alpha k}(\Phi_\alpha\mathfrak{D}_k\Psi^\beta - \Psi^\beta\mathfrak{D}_k\Phi_\alpha) + 2\Phi_\alpha\Psi^\alpha = 0. \tag{9.2.2}$$

Two important properties result from these equations.

*Conservation of supercurrent*

The first scalar equation can be written as

$$\mathfrak{D}_k(\Lambda_\beta^{\alpha k}\Phi_\alpha\Psi^\beta) = 0 \tag{9.2.3}$$

or

$$\nabla_k\mathcal{J}^k \equiv (\partial_k + \Gamma_{kl}^l)\mathcal{J}^k \equiv \frac{1}{\sqrt{\Lambda}}\partial_k(\sqrt{\Lambda}\mathcal{J}^k) = 0, \tag{9.2.4}$$

where

$$\mathcal{J}^k \equiv \Lambda_\beta^{\alpha k}\Phi_\alpha\Psi^\beta \tag{9.2.5}$$

is above introduced vector field, the *supercurrent* (see (9.1.9)).

Note that, equation (9.2.4) is in an identical consistence with equation (9.1.5), since tensor $\mathfrak{R}^{kl}$ possesses property $\nabla_k(\nabla_l\mathfrak{R}^{kl}) \equiv 0$ due to its skew symmetry:

$$\nabla_k(\nabla_l\mathfrak{R}^{kl}) \equiv \frac{1}{\sqrt{\Lambda}}\partial_k[\sqrt{\Lambda} \cdot \frac{1}{\sqrt{\Lambda}}\partial_l(\sqrt{\Lambda}\mathfrak{R}^{kl})] \equiv \frac{1}{\sqrt{\Lambda}}\partial_k\partial_l(\sqrt{\Lambda}\mathfrak{R}^{kl}) \equiv 0.$$

*Nullification of Matter Scalar along the dynamics*

Expression on the left hand side of equation (9.2.2) is twice as *Matter Scalar* form (6.3.4). Thus, Matter Scalar is nullified in dynamics:

$$\mathbb{M} \Rightarrow 0. \tag{9.2.6}$$

In essence, this nullification means the following scalar *dynamic connection*:

$$\frac{1}{2}\Lambda_\beta^{\alpha k}(\Phi_\alpha \mathfrak{D}_k \Psi^\beta - \Psi^\beta \mathfrak{D}_k \Phi_\alpha) = -\Phi_\alpha \Psi^\alpha, \tag{9.2.7}$$

or

$$\mathbb{D} = -\mathbb{N}. \tag{9.2.8}$$

### 9.3. Contracted metric equations

#### 9.3.1. Equation for gauge scalar

Multiplying equations (8.4.21) by $\Lambda_\alpha^{\beta k}$, producing contraction on all indices and taking into account dynamic identity (9.2.8), we find the following dynamic relation:

$$(N-4)\mathbb{G} = \mathbb{D}. \tag{9.3.1}$$

So at $N \neq 4$ we find that gauge scalar $\mathbb{G}$ in dynamics is a proportion to state norm $\mathbb{N}$:

$$\mathbb{G} = \frac{\mathbb{D}}{N-4} = -\frac{\mathbb{N}}{N-4}. \tag{9.3.2}$$

When considering case $N = 4$ in equation (9.3.1), we have to accept dynamic condition $\mathbb{D} = 0$, hence, $\mathbb{N} = 0$, instead of relation of a proportion between $\mathbb{R}$ and $\mathbb{N}$, as in cases $N \neq 4$. It should be noted, however, that such condition of *mathematical consistence* of the theory as $\mathbb{N} = 0$ at $N = 4$ is contrary to the foundation of the *autonomic duality* of *state vector* as represented by the two *independent* vector fields in the *matter function* space, contravariant $\Psi^\alpha$ and covariant $\Phi_\alpha$. Therefore, this "option" should be regarded as *standing out* of the frame of the treated superdimensional field theory (meaning *inconsistent*; see, by the way, the related comments in section 11.2.).

#### 9.3.2. Reduction of metric equations

*First reduction*

It follows from dynamic identity (9.2.8) that, scalar Lagrangian $\mathbb{L}$ in *metrics equations* (8.4.22) can be replaced by gauge scalar $\mathbb{G}$:

$$\mathbb{L} \Rightarrow \mathbb{G}, \tag{9.3.3}$$

Using this dynamic reduction, we can write metric equations (8.4.22) in the following view:

$$\mathbb{H}_{kl}\Lambda_\beta^{\alpha l} = -\mathfrak{D}_{\beta k}^\alpha; \tag{9.3.4}$$

here

$$\mathbb{H}_{kl} \equiv \mathbb{G}_{kl} - \mathbb{G}\Lambda_{kl} = \mathbb{H}_{lk}. \tag{9.3.5}$$

*Second reduction*

At $N \neq 4$ scalar $\mathbb{G}$ can be replaced by its *dynamic identity* according to relation (9.3.2):

$$\Lambda_\beta^{\alpha l}(\mathbb{G}_{kl} + \frac{\mathbb{N}}{N-4}\Lambda_{kl}) = -\mathfrak{D}_{\beta k}^{\alpha} \qquad (9.3.6)$$

There are two interesting outcomes resulting from this equation.

### 9.3.3. Solution for Split Metric as function of DSV, UGF and GM

Metrics equations (9.3.6) can be directly solved relative Split Metrics $\Lambda_\beta^{\alpha k}$ considered as function of DSV, UGF and GM:

$$\Lambda_\beta^{\alpha k} = -\breve{\mathbb{H}}^{kl}\mathfrak{D}_{\beta l}^{\alpha} ; \qquad (9.3.7)$$

here $\breve{\mathbb{H}}^{kl}$ is tensor inverse to $\mathbb{H}_{kl}$, i.e. :

$$\breve{\mathbb{H}}^{kn}\mathbb{H}_{ln} = \Delta_l^k . \qquad (9.3.8)$$

### 9.3.4. Algebraic equations for GM as function of DSV and UGF

Contracting on Greek indices product of two equations (9.3.6) with Roman indices $k$ and $l$:

$$\mathbb{H}_{kn}\Lambda_\beta^{\alpha n} = -\mathfrak{D}_{\beta k}^{\alpha} ,$$

$$\mathbb{H}_{lm}\Lambda_\beta^{\alpha m} = -\mathfrak{D}_{\beta l}^{\alpha} ,$$

we obtain algebraic equations for Grand Metric $\Lambda^{kl}$ as function of DSV and UGF:

$$\mathbb{H}_{km}\mathbb{H}_{ln}\Lambda^{mn} = \mathfrak{D}_{\beta k}^{\alpha}\mathfrak{D}_{\alpha l}^{\beta} , \qquad (9.3.9)$$

or

$$\mathbb{H}_{km}\mathbb{H}_{ln}\Lambda^{mn} = \mathbb{D}_{kl} ; \qquad (9.3.10)$$

here notation $\mathbb{D}_{kl}$ is for *matter tensor* defined as follows:

$$\mathbb{D}_{kl} \equiv \mathfrak{D}_{\beta k}^{\alpha}\mathfrak{D}_{\alpha l}^{\beta} . \qquad (9.3.11)$$

Equations (9.3.15) can also be written in the following view:

$$\mathbb{H}_n^l \mathbb{H}_k^n = \mathbb{D}_k^l ; \qquad (9.3.12)$$

here:

$$\mathbb{H}_k^l \equiv \mathbb{G}_k^l - \mathbb{G}\delta_k^l ; \quad \mathbb{G}_k^l \equiv \Lambda^{lm}\mathbb{G}_{km} = \Lambda^{lm}\Lambda^{pq}\mathbb{G}_{kp;mq}; \quad \mathbb{G} = \frac{1}{4}\mathbb{G}_k^k ; \qquad (9.3.13)$$

$$\mathbb{D}_k^l \equiv \Lambda^{lm}\mathbb{D}_{km} . \qquad (9.3.14)$$

As one can see, equations (9.3.12) are a system of algebraic equations of the fourth power for *grand metric* tensor $\Lambda^{kl}$ as function of *matter tensor* $\mathbb{D}_{kl}$ and *gauge 4-tensor* $\mathbb{G}_{km;ln}$.

Once GM arrives from equations (9.3.12) as function of DSV and UGF, so is about SM given by formula (9.3.7).

### 9.4. Hamilton-Nöther equations

Considering the derivatives of Lagrangian $\mathcal{L}$ while taking into account general Euler-Lagrange equations (5.4.8), one finds the following generic *dynamic identities*:

$$\partial_k \mathcal{L} = \partial_l \left[ \frac{\partial \mathcal{L}}{\partial(\partial_l X^a)} \partial_k X^a \right]. \tag{9.4.1}$$

Introducing an object:

$$\mathcal{H}_k^l \equiv \frac{\partial \mathcal{L}}{\partial(\partial_l X^a)} \partial_k X^a - \delta_k^l \mathcal{L}, \tag{9.4.2}$$

one can rewrite dynamical identities (9.4.1) as a general law following from the EL equations:

$$\partial_l \mathcal{H}_k^l = 0. \tag{9.4.3}$$

For Lagrangian $\mathcal{L} = \mathbb{L}\sqrt{\Lambda}$ with $\mathbb{L}$ as a scalar form, and tensor form $\Lambda_{kl}$ independent of the derivatives of the basic objects, equations (9.4.3) can be written in a view as follows:

$$\partial_l \left( \sqrt{\Lambda} \mathcal{T}_k^l \right) = 0, \tag{9.4.4}$$

where $\mathcal{T}_k^l$ is a mix valence 2 *pseudo-tensor* object:

$$\mathcal{T}_k^l \equiv \frac{\partial \mathbb{L}}{\partial(\partial_l X^a)} \partial_k X^a - \delta_k^l \mathbb{L}. \tag{9.4.5}$$

In our case of Lagrangian (6.6.1), taking into account dynamic reduction (9.3.2) and (9.3.3), we obtain the following expression for object $\mathcal{T}_k^l$:

$$\mathcal{T}_k^l = \frac{1}{2} \Lambda_\beta^{\alpha l} (\Phi_\alpha \partial_k \Psi^\beta - \Psi^\beta \partial_k \Phi_\alpha) + \mathfrak{R}_\beta^{\alpha l n} \partial_k \mathcal{A}_{\alpha n}^\beta + \frac{\mathbb{N}}{N-4} \Delta_k^l. \tag{9.4.6}$$

Apparently, equations (9.4.4) for object (9.4.6) can be considered as special type of contracted equations. It should be noted that, object $\mathcal{T}_k^l$ is not tensor since ordinary derivatives of DSV and UGF are not the covariant ones i.e. do not transform as tensors or h-tensors. Form (9.4.6) can be called *pseudo-tensor*, following a tradition established in GTR [8, 9, 15]). It is essential that, this form as well as equations (9.4.4) maintain their general form at arbitrary transformation of UM variables and related transformations of objects and their derivatives, i.e. they are *generally covariant.* Note that, object $\mathcal{T}_k^l$ and equations (9.4.4) do not have an independent background status in a covariant EAP-based field theory, since these equations are the consequences of EL equations (5.4.8) (as well as so are all the contracted equations). However, equations (9.4.4) may play a significant role in analysis of the (asymptotic) solutions, since these equations express a general dynamic invariance of the EL differential system.

## *Resume of Chapter 9*

C*ontracted equations* derived in this Chapter reveal *connections in dynamics* between basic objects as *equations* for scalar, vector, tensor and pseudo-tensor structural forms obtained by contraction of matrix forms and equations of the DSV-based theory on Greek indices. These connections in most are immediate consequences of the derived system of the Euler-Lagrange equations. An exemption is tensor equations for the odd-symmetric tensor $\mathfrak{R}_{kl} \equiv \mathfrak{R}^{\alpha}_{\alpha kl} = -\mathfrak{R}_{lk}$ , an external analog of the electromagnetic tensor $F_{kl}$ in the intelligible space-time manifold.

Discussion of the meaning and significance of the contracted equations in the context of possible relevance to QFT and GTR is included in Chapter 10 devoted to general discussion of the content and results of this paper.

# 10. Discussion

## 10.1. Relations to Classical Field Theories

*Lagrangian*

    *Matter part* $\mathbb{M}$ of scalar Lagrangian $\mathbb{L}$ has an external structural similarity to the "Dirac's part" of Lagrangian of a *spinor electrodynamics* and also to that of the QCD Lagrangian in the Standard Model of QFT. But there are the following distinctions: spinor $\hat{\psi}$-fields of QFT are complex while DSV object is implied real; Dirac's $\gamma$-matrices are constant, while SM object $\Lambda_\beta^{\alpha k}$ is variable as function of *unified manifold variables*, being connected to DSV and UGF; object $\mathcal{A}_{\beta k}^\alpha$ is a *hybrid affine tensor* compared to *vector potential* of EM field. Furthermore, *geometry scalar* $\mathbb{G}$, in general, corresponds to Lagrangians of the *gauge fields* of QFT: electromagnetic field, gluon fields of QCD, and non-Abelian gauge fields by Yang-Mills. On the other hand, this form, when taken with the weigh factor $\sqrt{\Lambda}$ for invariant integration, corresponds to the gravitation Lagrangian of the Weyl's theory of gravitation.

    There are important features of introducing *interaction* in the considered approach to UFT, in distinction to utilization of the *gauge principle* in QFT. First, covariant extension of the derivatives is produced not specifically with respect to the particular fields associated with certain "elementary particles", but with respect the *universal wave function*, Dual State Vector field as the primary, autonomic object of the theory, representing *matter*. Second, introduction of gauge fields $\mathcal{A}_{\beta k}^\alpha$ has produced *uniformly* over all degrees of freedom (components) of DSV, based on the general requirement of the covariant extension of the DSV derivatives as a vector object in the *matter function* space. The third, it is produced in all *real* terms, in accordance with the general presumption of *reality* of all objects and the whole covariant differential system of a unified field theory. It should be stressed that, the presence of such an invariable object as the symbol of *imaginary unit i* in the structure of the dual equations for DSV would contradict to the *general covariance* principle, – even relative to the linear transformations of the manifold variables. The fourth, the rank of matrices $\mathcal{A}_{\beta k}^\alpha$ on Greek indices, according to their background definition as a *hybrid affine tensor*, is equal to the rank of Roman index $k$, which is $N > 4$. In our view, the general extended utilization of the gauge fields as hybrid Christoffels $\mathcal{A}_{\beta k}^\alpha$ might cover all particular types of gauge fields or interactions of QFT, including the charge interactions of electrodynamics, and unite them in one uniform *real* matrix structure of each of the two autonomic (but connected) DSV equations.

    Finally, Lagrangians of Standard Model do not possess the property of *scale invariance*, due to, in particular, the non-analytical structure of Lagrangians composed on the complex functions.

    Considering relations to GTR, one may note that, the hybrid curvature form (HCF) $\mathfrak{R}_{\beta kl}^\alpha$ corresponds to Riemann-Christoffel curvature form (RCF) $R_{mkl}^n$ in differential geometry of GTR. Scalar Lagrangian $\mathbb{G}$ is of the second power in HCF corresponding to Weyl's theory [8] unlike the *geometry scalar* $\mathbb{L}_g$ of Einstein − Hilbert theory linear in Riemann-Christoffel tensor [8, 9, 13-16, 20, 21].

    But there are also background distinctions from both Einstein-Hilbert and Weyl theories. Primary object of "geometry" in the presented approach to UFT is not the metric tensor or Grand Metric (GM) tensor $\Lambda^{kl}$ but the *Split Metric* $\Lambda_\beta^{\alpha k}$, a *hybrid triadic tensor* (being a triadic coefficient function of equations for DSV, it can be considered as an extended analog of Dirac matrices as discussed above). Tensor $\Lambda^{kl}$ is structured on SM as a binary form; so GM cannot be considered as one of the basic

objects to be varied independently when applying the Extreme Action principle (EAP). Correspondently and naturally, EAP is formulated and EL equations are derived for SM but not directly for GM. Note that, there are no specific requirements in advance to the properties of GM tensor as metric tensor of UM (besides that it should not be degenerated); in particular, its signature is not specified.

And, there is another critical difference. In Einstein equations as well as in Hilbert's Lagrangian of GTR, Riemann-Christoffel tensor $R_{mkl}^n$ is simplified to Ricci tensor $R_{kl} = R_{kml}^m$, while the *hybrid curvature form* (HCF) $\mathfrak{R}_{\beta kl}^\alpha$ in geometry scalar $\mathbb{L}_G$ is structured on *hybrid affine tensor* $\mathcal{A}_{\beta k}^\alpha$ (*Unified Gauge Field*, UGF) and uniquely recognized as *covariant derivative* of UGF; it cannot be simplified to a rank 2 even-symmetric tensor as in case of Riemann-Christoffel tensor, since such procedure is not applicable to HCF.

The derived Lagrangian as form structured on four basic objects is unique due to the *irreducibility demands*.

The correspondences to -and differences from- QFT and GTR are discussed below in more detail in terms of the derived Euler-Lagrange equations.

*Master equations*

Equations (8.1.3) or (8.1.4) for the primary object, dual couple $\Xi = (\Psi^\alpha, \Phi_\alpha)$ are the autonomic relations between $\Xi$ and its first derivatives, therefore these equations can be viewed as a covariant Dirac-type differential law for the *dual state vector of matter,* DSV, representing matter in the $N$-dimensional Unified Manifold. EL equations for DSV include triadic object $\Lambda_\beta^{\alpha k}$ as coefficients multiplying the derivatives of DSV, and triadic object $\mathcal{A}_{\beta k}^\alpha$ introduced for *covariant extension* of the DSV derivatives. These objects are connected to DSV by the correspondent EL equations.

In the context of comparison in general structural properties, DSV equations are playing in UM a role similar to Dirac equation in the Electrodynamics as *classical field theory* in the 4-dimensional space-time manifold. Due to the interaction term in Lagrangian, this theory is effectively non-linear in Dirac's dual object, and so is the theory under consideration relative to DSV.

Besides the duality and pilot role of DSV in the differential system, DSV equations have other properties corresponding to features of Dirac equations:
- Coefficient functions $\Lambda_\beta^{\alpha k}$ (*Split Metric*, SM) in the DSV equations can be viewed as matrices corresponding to Dirac matrices.
- In correspondence to Dirac equation where $\gamma$-matrices are connected to Minkowski metric tensor $g^{kl}$, a connection is established between the metric tensor of UM (Grand Metric $\Lambda_{kl}$) and SM as derived in section 6.6..
- In correspondence to Dirac equation, it follows from the DSV equations that, there exists a conservative vector current, the *supercurrent*.

On the other hand, the DSV equations distinguish from Dirac equations in several important properties:
- DSV equations are formulated for a dual $\mu$-components object as function of variables of an *extended* rank $N > 4$ *Unified Manifold*; presumably, $\mu > N$.
- In correspondence, rank $\mu$ of matrices $\Lambda_\beta^{\alpha k}$ is not supposed to be equal to the dimensionality of UM but exceeds it, in distinct to the case of Dirac matrices (4 matrices of rank 4).
- DSV equations do not include *imaginary unit*, as well as the rest of the EL equations.

- Dirac matrices are constants, while Split Metric matrices $\Lambda_\beta^{\alpha k}$ may vary with all variables in UM. Their derivatives arrive in the DSV equations as multipliers at DSV.
- DSV equations are generally covariant, and the covariance is provided by the introduction of matrices (on Greek indices) $\mathcal{A}_{\beta k}^{\alpha}$ (*Unified Gauge Field*) as an addition to derivative symbol $\partial_k$. UGF is connected to DSV and SM by the correspondent EL equations.
- Split Metric matrices are connected to DSV and UGF by Euler-Lagrange equations on SM.
- In Dirac equations, algebraic properties of $\gamma$-matrices are determined by their connection to Minkowski metric tensor of the pseudo-Euclidian space-time manifold, based on the requirement that $\psi$-function should satisfy Klein-Gordon equation for "free electron". In distinction to that, Grand Metric tensor of the DSV-based theory is determined not as a fixed background object but as structured on the Split Metric matrices. Due to this inverse arrow of determination, GM varies with SM in the space of UM.
- DSV equations do not include any constants like charge or mass parameters. Moreover, the equations (hence, the solutions) cannot be affected at all by an introduction of constant multipliers to scalar blocks in Lagrangian, because of the property of *scaling invariance* of the Lagrangian imposed on its structure as one of the principal requirements to UFT. Note that, Lagrangians of Standard Model do not possess such a property.
- A special aspect of the comparison is *duality* of DSV equations with respect to the complex duality of Dirac equations. Namely, the duality of DSV as an alliance of two real independent objects (contra- and co-variant vector fields in the *matter function* space) is not equivalent to the complex duality. A duality equivalent to the complex one may appear as a particular (yet approximate) internal property of each of the two DSV components associated with specific fragments of the group structure of matrices $\mathcal{A}_{\beta k}^{\alpha}$.

*Gauge equations*

EL equations on UGF (section 8.2.) connect matrices $\mathcal{A}_k$ to the primary object, *Dual State Vector* $\Psi^\alpha, \Phi_\alpha$, and metric object, *Split Metric* matrices $\boldsymbol{\Lambda}^k$.

In the presented approach to UFT, *interaction* in the *matter* Lagrangian arrives totally as an attribute of covariance or tensor invariance principle, in general correspondence to the *gauge principle* of introducing interactions in QFT. The "agent" of interaction is the *hybrid affine tensor* $\mathcal{A}_{\beta k}^{\alpha}$ (*unified gauge field*, UGF), introduced to the *matter* scalar Lagrangian for a covariant extension of the DSV derivatives. Implementation of the extended general covariance in the differential equations is based on EAP as a fundamental dynamic principle that connects SM and UGF to DSV. EAP naturally invites derivatives of UGF in the Lagrangian structure, but requires a covariant extension of these derivatives, thus leading to the appearance of the *hybrid curvature form,* HCF in Lagrangian to replace the ordinary curl-derivatives of UGF. EAP then results in equations that couple UGF to SM and DSV.

*Metric equations*

EL equations for *Split Metric* (8.3.2) or (8.4.22) connect this object to DSV and UGF:

$$(\mathbb{G}_{kl} - \mathbb{G}\Lambda_{kl})\boldsymbol{\Lambda}^l = -\mathbf{D}_k . \qquad (10.1.1)$$

These equations are in general symbolic correspondence to Einstein-Hilbert (EH) equations for metric tensor $g_{kl}$ [8, 9, 13-16]:

$$R_{kl} - \frac{1}{2} g_{kl} R = \kappa T_{kl}. \tag{10.1.2}$$

Term $\mathbb{G}_{kl}$ in equations (10.1.1) corresponds to Ricci tensor term $R_{kl}$ in EH equations. Term with geometry scalar $\mathbb{G}$ correspond to the term with scalar curvature $R$ in EH equations. And, the expressions on the right-hand side of metric equations (10.1.1) correspond to the energy-momentum tensor of matter in EH equations. It is obvious that, this correspondence is not an accidental one but, namely, is due to structuring of GM tensor (6.5.2) based on SM. This structuring also is not voluntary but determined logically by the two principal demands: the necessity to connect SM and UGF to DSV, on one hand, and the requirement of *irreducibility* of the Euler-Lagrange differential system, on the other hand.

As the Einstein-Hilbert and Weyl equations, resulting equations for GM are essentially non-linear.

But here the similarity ends. Equations (8.4.22) have been derived not for Grand Metric tensor $\Lambda^{kl}$ but for *Split Metric* matrices $\Lambda^{\alpha k}_{\beta}$, an extended analog of Dirac matrices. They are introduced as a necessary object connecting the covariant DSV derivatives to DSV itself in the equations of *matter* (8.4.13). SM has no analog in GTR which is directly a differential theory of metric tensor $g_{kl}$ in the 4-dimensions space-time manifold (more accurately, a theory of $g_{kl}$ deviations from Minkowski metric tensor in the "presence of matter"). Tensor $\Lambda_{kl}$ (*Grand Metric,* GM) in the DSV-based theory is introduced for invariant integration of scalar forms in the Unified Manifold; in this aspect, it corresponds to metric tensor $g_{kl}$. The difference is that, tensor $\Lambda_{kl}$ is built on the Split Metric matrices as shown in equation (6.6.5) and does not play a primary role in the considered theory. Furthermore, *hybrid curvature form* (HCF) $\mathfrak{R}^{\alpha}_{\beta kl}$ in equations (8.4.22) is structured on UGF matrices $\mathcal{A}^{\alpha}_{\beta k}$, coupled to SM and DSV by the EL equations on UGF. Riemann-Christoffel curvature form $R^{n}_{mkl}$ (RCF) in EH equations (10.1.6) is structured on *matched connection* $\Gamma^{n}_{mk}$ expressed through the derivatives of $g_{kl}$. Yet RCF in EH equations is simplified to Ricci tensor, while one cannot apply such contraction to HCF. Finally, the attributes of *matter* are presented in EH equations by a symmetric *valence 2 energy-momentum* tensor $T_{kl}$, which, in principle, can be associated in GTR with variation derivatives of *matter Lagrangian* $L_m \sqrt{g}$ on $g^{kl}$ [13, 9], while the right-hand side terms in equations (10.1.1) are the variation derivatives of matter Lagrangian $\mathbb{M}\sqrt{\Lambda}$ on SM.

It should be noted that, Einstein-Hilbert equations of *gravitational field* cannot be derived by a direct contraction of the complete system of EL equations derived in this paper. Also, they cannot be considered as equations that could be added as a complementary to the derived system of the EL equations, if not to violate the *scale invariance* principle. The derived system is *locked* by this demand for such introduction.

*Matched Connection in UM of SFT*

An immediate representative of *gravitational field* in GTR is *matched connection* (MC) or Christoffel symbols [8, 9, 14, 15, 16, 21, 22]. In SFT, MC $\Gamma^{k}_{lm}$ of extended *Unified Manifold* is given by similar definition based on the Grand Metric $\Lambda^{kl}$, similar to the correspondent background establishment in Riemannian geometry, and can be interpreted as *extended* (or *super-*) *gravitational field,* referring to the definition of such field in GTR. Though MC is not included in collection of basic objects of "Lagrange formalism" (being replaced by UGF as a more general object corresponding to DSV), it arrives with derivatives of the determinant of Grand Metric $\Lambda_{kl}$ in the EL equations on DSV and UGF. This appearance naturally provides for covariant extension of derivatives of the hybrid tensors as well as *vector* and *tensor* fields (objects obtained by contraction of the *hybrid tensors*). Also, MC can be used to define *geodesics* in the $N$-dimensions space of the Unified Manifold.

## 10.2. Relations to the *Quantum Field Theory*

Here we will discuss briefly relations of the presented *Superdimensional Field Theory* (SFT) principles and equations to methodology of QFT.

*Homogeneity principle of SFT versus the Superposition Principle of QFT*

Investigation of possibilities to reach a consistent approach to UFT by a covariant differential method in an $N$-dimensional manifold undertaken in this work has resulted in a system of Euler-Lagrange equations essentially non-linear on the principal or *pilot* object of the system, *Dual State Vector* (DSV) field. This non-linearity is due to that the coefficient functions (associated with *interactions*) in these equations are connected to DSV by the correspondent *non-autonomic* equations in which DSV plays a role of an "inducing force". It is important that the system is space-homogenous i.e. it does not include any explicit i.e. *given* functions of the manifold variables, in accordance to the imposed the *homogeneity* principle in Unified Manifold. It is noteworthy that, the presented conservative approach to UFT is naturally inheriting non-linearity types of electrodynamics and theory of gravity together – but with extension to $N$-dimensional manifold and suited generalizations of the gauge principle and connections to objects of "geometry".

In the context of a comparison with equations and methodology of QFT, the SFT corresponds in general to a principal object of QFT, the *secondary quantization wave function* (SQF) or *state vector* $\Upsilon(\hat{x}, Q)$, where $Q$ denotes a collection of individual *fields* as degrees of freedom (*fermions* and *bosons*). One might perceive a correspondence between $\Upsilon$ and $\Xi$ in parallel to correspondence between $(\hat{x}, Q)$ and $\check{\psi}$ manifolds. However, in contrary to SFT, dynamic algorithm of QFT is linear on SQF, since the principles and methodology of *quantum theories* are essentially based on assumption that "*operators of physical values*" are *not affected* by the state vector $\Upsilon$. Equation for SQF, a principal law of QFT can be written in its general symbolic form as follows:

$$i\frac{\partial}{\partial t}\Upsilon = \mathcal{H}\left(Q, i\frac{\partial}{\partial Q}\right)\Upsilon,$$

where differential form $\widehat{\mathcal{H}}$ is *quantum Hamiltonian*, the *energy operator* [1,2]. Property of linearity of the principal equation (10.2.1) on *state vector* $\Upsilon$ is known as *superposition principle*, a *quantum postulate*. Such concept does not meet principle of the *homogeneity* of a unified theory in space of variables – one of the requirements in the list of the *irreducibility principles* of SFT exhibited in this paper.

In approach to UFT as a differential law for a real irreducible *State Vector* field and related *coefficient functions* in a real extended manifold, not the *superposition principle* but *homogeneity* of the differential system is considered as one of the background properties of a unified theory.

In the context of SFT relation to methodology of QFT, it can be noted that, it is similar to relation of Dirac-Maxwell electrodynamics as classical *field theory* to Schrödinger-Dirac quantum mechanics of particles (represented by particular *wave functions*) in a *given external field*.